\documentclass{article}
\PassOptionsToPackage{numbers, compress}{natbib}

\usepackage[final]{neurips_2025}

\usepackage[utf8]{inputenc}
\usepackage[T1]{fontenc}
\usepackage[pdftex,bookmarksnumbered,hidelinks,breaklinks]{hyperref}
\usepackage{url}
\usepackage{booktabs}
\usepackage{amsfonts}
\usepackage{nicefrac}
\usepackage{microtype}
\usepackage{xcolor}

\usepackage{graphicx}
\usepackage{subcaption}

\usepackage{amsmath}
\usepackage{amssymb}
\usepackage{mathtools}
\usepackage{amsthm}
\usepackage[capitalize,noabbrev]{cleveref}
\usepackage[textsize=tiny]{todonotes}
\usepackage{silence}
\usepackage{enumitem}
\usepackage{adjustbox}
\usepackage{xstring}
\usepackage{multirow}
\usepackage{float}
\usepackage{tikz}
\usetikzlibrary{calc}
\usepackage{algorithm}
\usepackage{algorithmic}
\usepackage{wrapfig}
\usepackage{stackengine}
\usepackage{bbm}
\usepackage{mathabx}
\setstackgap{S}{-1.5pt}

\theoremstyle{plain}
\newtheorem{theorem}{Theorem}[section]

\theoremstyle{definition}

\theoremstyle{remark}

\newcommand{\tikzmark}[1]{\tikz[overlay,remember picture,baseline] \node [anchor=base] (#1) {};}
\newcommand{\commentout}[1]{}
\newcommand{\aref}[1]{{Appendix~\ref{#1}}}

\usepackage{amsmath,amsfonts,bm}

\def\eqref#1{equation~\ref{#1}}

\def\floor#1{\lfloor #1 \rfloor}
\def\1{\bm{1}}

\def\vf{{\bm{f}}}
\def\vg{{\bm{g}}}

\def\ws{{\bm{s}}}

\def\vu{{\bm{u}}}

\def\vw{{\bm{w}}}
\def\vx{{\bm{x}}}

\def\vz{{\bm{z}}}

\DeclareMathAlphabet{\mathsfit}{\encodingdefault}{\sfdefault}{m}{sl}
\SetMathAlphabet{\mathsfit}{bold}{\encodingdefault}{\sfdefault}{bx}{n}

\DeclareMathOperator*{\argmin}{arg\,min}

\newcommand{\sysname}{\textbf{DiffGrad}}
\newcommand{\libname}{\textit{DiffBreak}}

\newenvironment{SUBENVccomment}[2]{\color{#1}[#2:]~}{\color{black}}

\definecolor{author1}{rgb}     {0.9,0.5,0.0}
\definecolor{author2}{rgb}     {0.6,0.0,0.8}
\definecolor{author3}{rgb}     {0.0,0.5,0.0}
\definecolor{author4}{rgb}     {0.9,0.2,0.2}

\newcommand{\RETURN}{\STATE \textbf{return} }
\newcommand{\CCOMMENT}[2][]{\hfill {\color{#1}/* #2 */}}

\newcommand{\MULTILINECOMMENT}[2][]{%
    \hfill \raisebox{-1.5ex}{\color{#1}/*} 
    \begin{minipage}[t]{6cm}%
        \raggedleft \color{#1} #2 \hspace{0pt} */%
    \end{minipage}
}

\definecolor{red}{rgb}{1.0, 0.0, 0.0}
\definecolor{blue}{rgb}{0.0, 0.0, 1.0}
\definecolor{gray}{rgb}{0.5, 0.5, 0.5}
\definecolor{black}{rgb}{0.0, 0.0, 0.0}
\makeatletter\newcommand{\printAffiliations}[1]{%
\stepcounter{@affiliationcounter}%
{\let\thefootnote\relax\footnotetext{\hspace*{-\footnotesep}\ifdefined\isaccepted #1\fi%
\forloop{@affilnum}{1}{\value{@affilnum} < \value{@affiliationcounter}}{
\textsuperscript{\arabic{@affilnum}}\ifcsname @affilname\the@affilnum\endcsname%
\csname @affilname\the@affilnum\endcsname%
\else
{\bf AUTHORERR: Missing \textbackslash{}icmlaffiliation.}
\fi
}.
\ifdefined\icmlcorrespondingauthor@text
Correspondence to: \icmlcorrespondingauthor@text.
\else
{\bf AUTHORERR: Missing \textbackslash{}icmlcorrespondingauthor.}
\fi

}
}
}

\title{DiffBreak: Is Diffusion-Based Purification Robust?}

\author{
  Andre Kassis, Urs Hengartner, Yaoliang Yu \\
  Cheriton School of Computer Science, University of Waterloo\\
  Waterloo, Ontario, Canada \\
 \texttt{\{akassis, urs.hengartner, yaoliang.yu\}@uwaterloo.ca}
}

\begin{document}

\if@neuripsfinal
\maketitle
\else
\if@preprint
\maketitle
\else
\vbox{%
    \hsize\textwidth
    \linewidth\hsize
    \vskip 0.1in
    \@toptitlebar
    \centering
    {\LARGE\bf \@title\par}
    \@bottomtitlebar
    \vskip 0.3in \@minus 0.1in
}
\fi
\fi

\if@neuripsfinal
    \newcommand{\codeurl}{\url{https://github.com/andrekassis/DiffBreak}}
\else
\if@preprint
    \newcommand{\codeurl}{\url{https://github.com/andrekassis/DiffBreak}}
\else
    \newcommand{\codeurl}{\url{https://anonymous.4open.science/r/DiffBreak-A768}}
\fi
\fi

\begin{abstract}
Diffusion-based purification (\textit{DBP}) has become a cornerstone defense against adversarial examples (\textit{AE}s), regarded as robust due to its use of diffusion models (\textit{DM}s) that project \textit{AE}s onto the natural data manifold. We refute this core claim, theoretically proving that gradient-based attacks effectively target the \textit{DM} rather than the classifier, causing \textit{DBP}'s outputs to align with adversarial distributions. This prompts a reassessment of \textit{DBP}'s robustness, accrediting it two critical factors: inaccurate gradients and improper evaluation protocols that test only a single random purification of the \textit{AE}. We show that when accounting for stochasticity and resubmission risk, \textit{DBP} collapses. To support this, we introduce \textit{DiffBreak}, the first reliable toolkit for differentiation through \textit{DBP}, eliminating gradient mismatches that previously further inflated robustness estimates. We also analyze the current defense scheme used for \textit{DBP} where classification relies on a single purification, pinpointing its inherent invalidity. We provide a statistically grounded majority-vote (\textit{MV}) alternative that aggregates predictions across multiple purified copies, showing partial but meaningful robustness gain. We then propose a novel adaptation of an optimization method against deepfake watermarking, crafting systemic perturbations that defeat \textit{DBP} even under \textit{MV}, challenging \textit{DBP}'s viability.
\end{abstract}

\section{Introduction}
\label{sec:intro}
\textit{ML} classifiers are vulnerable to \textit{Adversarial Examples (AEs)}—imperceptible perturbations that induce misclassification~\citep{DBLP:journals/corr/SzegedyZSBEGF13,DBLP:journals/corr/GoodfellowSS14}. \textit{Adversarial Training}~\citep{DBLP:conf/iclr/MadryMSTV18,DBLP:conf/icml/ZhangYJXGJ19} is attack-specific and costly~\citep{DBLP:conf/iclr/WongRK20}, while other defenses~\citep{countering,stoc,DBLP:conf/iclr/BuckmanRRG18,Xiao2020Enhancing,xie2018mitigating,pang2019mixup,yang2019me,DBLP:conf/icml/CohenRK19} are vulnerable to adaptive attacks~\citep{DBLP:conf/nips/TramerCBM20}. \textit{Diffusion-Based Purification (DBP)}~\citep{diffpure,gmdp}, which leverages \textit{diffusion models (DMs)}~\citep{dmint,DiffusionModels}, has emerged as a promising defense~\citep{DBLP:journals/corr/abs-2402-02316,DBLP:journals/corr/abs-2206-10875,chen2024robust,DBLP:journals/corr/abs-2310-18762,10.5555/3618408.3619519,DBLP:conf/icml/WangPDL0Y23}. \textit{DBP} purifies inputs by solving the reverse stochastic differential equation (\textit{SDE}) associated with \textit{DM}s, diffusing the input with random noise to dilute adversarial changes and iteratively denoising it. Given \textit{DMs'} data modeling abilities~\citep{DBLP:conf/nips/DhariwalN21,DBLP:conf/nips/VahdatKK21}, \textit{DBP} operates under the assumption that each denoising (reverse) step's output belongs to a corresponding \textit{marginal natural distribution} (a distribution obtained by introducing Gaussian noise into natural inputs, with decreasing variance proportional to that step). 

This is the foundation for \textit{DBP}’s \textit{robustness} as it ensures its outputs are from the natural distribution, while \textit{AE}s lie outside this manifold and are unlikely to be preserved. Recent works treat \textit{DBP} as a static pre-processor that standard gradient-based attacks cannot directly manipulate but rather only attempt to evade. Thus, attackers are assumed to either fail due to randomness that masks the classifier's vulnerabilities~\cite{liu2024towards}, or be forced to rely on surrogate losses and heuristic approximations~\cite{kang2024diffattack,diffhammer}. We theoretically refute \textit{DBP}'s robustness rooted in its ability to perform projection onto the natural manifold. This is paradoxical as it assumes correct behavior of the score model $\ws_\theta$ used by the \textit{DM} to mimic the gradients of the marginal distributions. Yet, $\ws_\theta$ is an \textit{ML} system with exploitable imperfections. In~\S\ref{subsec:path}, we prove that \textbf{standard} adaptive attacks that simply backpropagate the classifier's loss gradients through \textit{DBP} \textbf{effectively target $\ws_\theta$ rather than the classifier}, forcing it to generate samples from an adversarial distribution. Hence, \textit{DBP}'s theory no longer holds. 

\begin{figure*}[t]
\includegraphics[width=0.99\textwidth,height=4.2cm]
{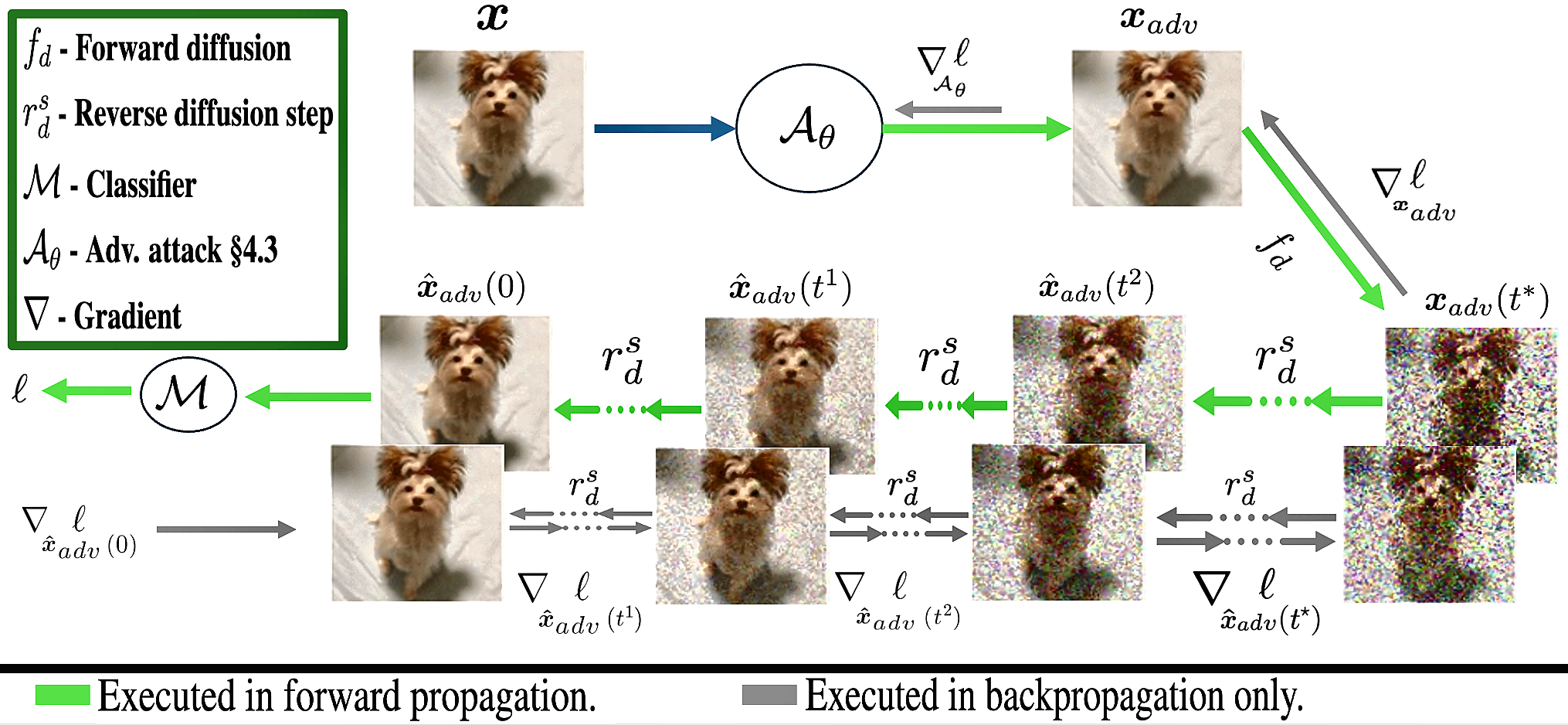}
\caption{
\sysname{}. \mbox{\small$\vx$} is given to the iterative attack algorithm \mbox{\small$\mathcal{A}_\theta$} to generate \mbox{\small$\vx_{adv}$}. At each iteration, \mbox{\small$\vx_{adv}$} is propagated through \textit{DBP}, yielding \mbox{\small$\hat{\vx}_{\!_{adv}}\!(0)$} that is given to \mbox{\small$\mathcal{M}$} while storing each intermediate \mbox{\small$\hat{\vx}_{\!_{adv}}\!(t)$} (bottom replicas) from \textit{DBP}'s reverse pass (see~\S\ref{sec:Background}) but without saving any graph dependencies. Backpropagation uses the stored samples: Starting from \mbox{\small$t\!=\!-dt$} ($dt<0$---see~\S\ref{sec:Background}), each \mbox{\small$\hat{\vx}_{\!_{adv}}\!(t)$} is used to recompute  \mbox{\small$\hat{\vx}_{_{adv}}\!(t\!+\!dt)$}, retrieving the required dependencies. Then, we recursively obtain the gradient w.r.t. \mbox{\small$\hat{\vx}_{_{adv}}\!(t)$} from the gradients w.r.t. \mbox{\small$\hat{\vx}_{_{adv}}\!(t\!+\!dt)$} using this recovered sub-graph (see~\S\ref{subsec:precise_grads}). Finally, gradients are backpropagated in a standard manner from \mbox{\small$\vx_{\!_{adv}}\!(t^*)$} to \mbox{\small$\mathcal{A}_\theta$} to update \mbox{\small$\vx_{adv}$}.
}
\label{fig:diagram}
\vspace{-5mm}
\end{figure*}

In~\S\ref{subsec:precise_grads} and~\S\ref{subsec:flaws}, we revisit \textit{DBP}'s previous robustness, attributing it to backpropagation issues and improper evaluation protocols. Most works~\cite{kang2024diffattack,diffpure,liu2024towards,gmdp} judge attacks based on a single purification of the final adversarial input. While this seemingly mirrors \textit{DBP}’s intended ``purify once and classify'' deployment~\cite{diffpure,gmdp}, it is statistically invalid: since \textit{DBP} samples noise randomly, one evaluation of the \textit{AE} fails to capture true \textbf{single} misclassification probability across possible purifications. Moreover, due to \textit{DBP}’s memory-intensive gradients, prior works implement exact backpropagation~\cite{kang2024diffattack,liu2024towards,diffhammer} via checkpointing~\cite{checkpoint}. We discover and fix issues in all previous implementations, introducing \sysname{}—the first reliable module for exact backpropagation through \textit{DBP} (see~Fig.\ref{fig:diagram}). In~\S\ref{sec:exp}, we show that even under the one-sample evaluation protocol, \textit{AutoAttack}~\cite{autoattack} significantly outperforms prior works, exposing their gradient fallacies. We further use a multi-sample evaluation that captures \textit{DBP}’s stochasticity and resubmission risk—aligned with \textbf{DiffHammer}~\cite{diffhammer}, which also identified the evaluation gap, but retained problematic gradients. By fixing both, we prove \textit{DBP} even more vulnerable than they report, with robustness dropping below $17.19\%$. Finally, we show prior attack enhancements tailored to \textit{DBP} (e.g., \textbf{DiffHammer}, \textbf{DiffAttack}~\cite{kang2024diffattack}), being oblivious to the ability of standard gradient-based attacks to reshape its behavior, in fact, disrupt optimization, lowering attack performance, and confirming our theoretical scrutiny.

To replace the current inherently vulnerable single-evaluation \textit{DBP} defense scheme, we propose a more robust majority-vote (\textit{MV}) setup, where multiple purified copies are classified and the final decision is based on the majority label. Yet, even under \textit{MV}, \textit{DBP} retains only \textbf{partial} robustness against norm-bounded attacks—see~\S\ref{subsec:exp_mv}, supporting our theoretical findings. As we established the inefficacy of intermediate projections, this resistance is due to \textit{DBP}'s vast stochasticity: common \textit{AE}s that introduce large changes to \textit{individual} pixels are diluted and may not impact most paths. Instead, we require stealthy modifications that affect many pixels. In~\S\ref{subsec:low_freq}, we propose a novel adaptation of a recent strategy~\citep{unmarker} from image watermarking that crafts low-frequency perturbations, accounting for links between neighboring pixels. This technique defeats \textit{DBP} even under \textit{MV}

\textbf{Contributions.} (i) Analytically scrutinizing adaptive attacks on \textit{DBP}, proving they nullify its theoretical robustness claims. (ii) Addressing protocol and gradient issues in prior attacks, enabling reliable evaluations, and demonstrating degraded performance of \textit{DBP} and the ineffectiveness of recent attack enhancement strategies~\citep{kang2024diffattack,diffhammer}. (iii) Introducing and evaluating a statistically grounded \textit{MV} defense protocol for \textit{DBP}. (iv) Proposing and adapting low-frequency (\textit{LF}) attack optimizations to \textit{DBP}, achieving unprecedented success even under \textit{MV}. (v) Availability: aside from scalability and backpropagation issues, existing \textit{DBP} implementations and attacks lack generalizability. We provide \libname{}\footnote{\codeurl{}}---the first toolkit for evaluating any classifier with \textit{DBP} under various optimization methods, including our novel \textit{LF}, using our reliable \sysname{} module for backpropagation. (vi) Extensive evaluations on \textit{\textbf{ImageNet}} and \textit{\textbf{CIFAR-10}} prove our methods defeat \textit{DBP}, bringing its robustness to $\sim\!0\%$, outperforming previous works by a large margin.
\section{Background \& Related Work}
\label{sec:Background}
\label{sec:threat_model}
\textbf{Adversarial Attacks.} Given $\vx \!\in\! \mathbb{R}^d$ with true label $y$, classifier $\mathcal{M}$, and preprocessing defense $G$ ($G\!\equiv\!\!Id$ if no defense), attackers aim to generate a \textit{similar} $\vx_{adv}$ s.t. $\mathcal{M}(G(\vx_{adv}))\!\neq\!\! y$. Formally:
\begin{equation*}
\begin{aligned}
    &\vx_{adv} \!=\! \underset{\boldsymbol{\mathcal{D}}(\vx^\prime,\; \vx) \leq \mbox{\large$\epsilon$}_{_{\boldsymbol{\mathcal{D}}}}}{\argmin} \mathbb{E}[\ell_G^\mathcal{M}(\vx^\prime,\; y)]
\end{aligned}
\end{equation*}
for loss $\ell_G^\mathcal{M}$. Typically, $\ell_G^\mathcal{M}(\vx,y)=\ell(\mathcal{M}(G(\vx)),y)$, where $\ell$ is a loss over $\mathcal{M}$'s output that captures the desired outcome. For instance, $\ell(\mathcal{M}(G(\vx)),y)$ can be chosen as the probability that the classifier's output label is $y$, which we strive to minimize. $\boldsymbol{\mathcal{D}}$ is a distance metric that ensures similarity if kept below some $\mbox{\large$\epsilon$}_{_{\boldsymbol{\mathcal{D}}}}$. These \textit{untargeted} attacks are the focus of many works~\citep{diffpure,gmdp,pang2019mixup,meng2017magnet}. The expected value accounts for potential stochasticity in $G$ (e.g., \textit{DBP} below).

\textbf{Diffusion models (DMs)}~\citep{DiffusionModels,dmint} learn to model a distribution $p$ on $\mathbb{R}^d$ by reversing the process that diffuses inputs into noise. \textit{DM}s involve two stochastic processes. The forward pass converts samples into pure Gaussians, and is governed by the following \textit{SDE} for an infinitesimal step $dt>0$:
\begin{equation}
d\vx = \vf(\vx,\, t)dt + g(t)d\vw\label{eq:forward_sde}.
\end{equation}
 \hyperref[eq:forward_sde]{\textit{eq.~(1)}} describes a stochastic integral whose solution up to $t^* \!\in \![0, 1]$ gives $\vx(t^*)$. Here, $\vf\!:\mathbb{R}^d \! \times \! \mathbb{R} \! \longrightarrow \! \mathbb{R}^d$ is the drift, $\vw\!:\mathbb{R} \longrightarrow \!\mathbb{R}^d$ is a Wiener process, and $g\!: \mathbb{R} \!\longrightarrow\! \mathbb{R}$ is the diffusion coefficient. We focus on \textbf{VP-SDE}, which is the most common \textit{DM} for \textit{DBP}~\citep{diffpure,densepure,gmdp}. Yet, our insights generalize to all \textit{DM}s (see~\citep{dmint} for a review). In \textbf{VP-SDE}, $\vf(\vx,\; t)\!=\!-\frac{1}{2}\beta(t)\vx(t)$ and $g(t)\!=\!\sqrt{\beta(t)}$, where $\beta (t)$ is a noise scheduler outputting small positive constants. These choices yield the solution:
\begin{equation}
\vx(t^*) = \sqrt{\alpha(t^*)} \vx + \sqrt{1-\alpha (t^*)} \boldsymbol{\epsilon}
    \label{eq:closed}
\end{equation}
for $\boldsymbol{\epsilon}\!\sim\! \mathcal{N}(\boldsymbol{0}, \boldsymbol{I}_d)$ and $\alpha(t)\!=\!e^{-\int_{0}^{t} \beta (s) \,ds}$. With proper parameters, we have $\vx(1) \!\sim \!\mathcal{N}(\boldsymbol{0}, \boldsymbol{I}_d)$. Thus, a process that inverts \hyperref[eq:forward_sde]{\textit{eq.~(1)}} from $t^*\!=\!1$ to $0$ allows generating samples in $p$ from random noise. Due to~\citet{anderson1982reverse}, the reverse pass is known to be a stochastic process with:
\begin{equation}
    d\hat{\vx} = [\vf(\hat{\vx}(t),\, t) - g^2(t)\nabla_{\hat{\vx}(t)} \log p_t (\hat{\vx}(t))]dt + g(t)d\bar{\vw}.
    \label{eq:reverse_raw}
\end{equation}
Defining $\hat{\vx}(t^*)\! :=\! \vx(t^*)$, the process evolves from $t^*$ to $0$ with a \textbf{negative} time step $dt$ and reverse-time Wiener process $\bar{\vw}(t)$. Let $p(\vx)$ be the probability of $\vx$ under $p$, and $p_{0t}(\Tilde{\vx}|\vx)$ the conditional density at $t$ given $\vx(0) = \vx$. Then, the marginal density is given by $p_t(\Tilde{\vx}) = \int p(\vx)p_{0t}(\Tilde{\vx}|\vx)d\vx, \text{where } p_0 \equiv p$. Solving \hyperref[eq:reverse_raw]{\textit{eq.~(3)}} requires the \textbf{score} $\nabla_{\hat{\vx}(t)} \log p_t (\hat{\vx}(t))$, which can be approximated via a trained model $\boldsymbol{s}_\theta$ s.t. $\boldsymbol{s}_\theta(\hat{\vx}(t), t) \approx \nabla_{\hat{\vx}(t)}\log p_t(\hat{\vx}(t))$~\citep{dmint}, yielding:
\begin{equation}
d\hat{\vx}= -\frac{1}{2}\beta(t) [\hat{\vx}(t) + 2 \ws_\theta (\hat{\vx}(t), t)] dt + \sqrt{\beta(t)} d\bar{\vw}.
\label{eq:reverse_sde}
\end{equation}
As no closed-form solution exists, the process runs iteratively over \textbf{discrete} negative steps $dt$. Starting from $t^*$, $d\hat{\vx}$ is calculated at each $i\!=\!|\frac{t}{dt}|$, until $t\!=\!0$. This \textit{continuous-time} \textit{DM} describes a stochastic integral (despite the discretized implementations). An alternative, \textit{Denoising diffusion probabilistic modeling} (\textbf{DDPM})~\citep{ho2020denoising,DiffusionModels}, considers a \textit{discrete-time} \textit{DM}. The two are equivalent (see \aref{app:ddpm}).

\noindent
\textbf{DBP}~\citep{diffpure,densepure,gmdp,adp} performs purification by diffusing each input $\vx$ until optimal time $t^*$ that preserves class semantics while still diminishing malicious interruptions. $\vx(t^*)$ is then given to the reverse pass, reconstructing a clean $\hat{\vx}(0)\!\approx\!\vx$ s.t. $\hat{\vx}(0)\!\sim \!p$ for correct classification.

\noindent
\textbf{Certified vs. Empirical Robustness.} While recent work has explored certified guarantees for \textit{DBP} via randomized smoothing (RS)~\cite{densepure,certified!!,287372,DBLP:journals/corr/abs-2402-02316,DBLP:journals/corr/abs-2206-10875}, these guarantees hold only under restrictive assumptions — e.g., small $\ell_2$ perturbations and thousands of Monte Carlo samples per input. In contrast, empirical defenses~\cite{gmdp,diffpure,10.5555/3618408.3619519,chen2024robust,DBLP:journals/corr/abs-2310-18762} aim for practical robustness under realistic threats~\cite{autoattack} and efficient inference. Like prior work~\cite{diffhammer,kang2024diffattack,diffpure,gmdp,lee2023robust}, we explicitly target these empirical defenses---not orthogonal certified variants whose guarantees do not hold under stronger practical perturbations or operational constraints. Certification protocols are computationally prohibitive, and their bounds fail to capture threats that remain imperceptible but exceed certified radii.
\section{Revisiting Diffusion-Based Purification}
\label{sec:theory}
We study gradient-based attacks. In~\S\ref{subsec:path}, we theoretically prove that adaptive attacks invalidate \textit{DBP}'s principles. Then, we reconcile this with previous findings of \textit{DBP}'s robustness, attributing them to backpropagation issues and improper evaluation protocols. In~\S\ref{subsec:precise_grads}, we analyze backpropagation mismatches in previous works and propose fixes. Finally, we present an improved evaluation protocol for \textit{DBP} in~\S\ref{subsec:flaws} to better measure its robustness.

\subsection{Why \textit{DBP} Fails: Theoretical Vulnerability to Adaptive Attacks} 
\label{subsec:path}
\textit{DBP}'s robustness is primarily attributed to its ability to project the purified $\hat{\vx}(0)$ onto the natural manifold~\cite{diffpure,densepure,kang2024diffattack}. 
This foundational assumption—rooted in the inherent behavior of the purification process—has remained unchallenged. In fact, the most advanced attacks have been explicitly tailored to exploit or circumvent it: \textbf{DiffAttack}~\cite{kang2024diffattack} introduces suboptimal per-step losses to perturb the purification trajectory (see~\S\ref{sec:exp}), while \textbf{DiffHammer}~\cite{diffhammer} constrains its optimization to feasible paths that evade detection by the \textit{DBP} process itself. \textit{DBP} is often justified through its marginal consistency: since $\boldsymbol{s}_\theta\!\!\approx\!\!\nabla\log p_t$, $\{\vx(t)\}_{t \in\! [0, 1]}$ and $\{{\hat{\vx}}(t)\}_{t \in\! [0, 1]}$ follow the same marginals~\citep{anderson1982reverse}, yielding $\hat{\vx}(0) \sim p$. Specifically,~\citet{densepure} show that:
\vspace{-4mm}
\begin{equation*}
    \Pr(\hat{\vx}(0)|\vx) ~\propto~ p(\hat{\vx}(0))\cdot e^{\mbox{\large${-\frac{\alpha(t^*)\|\hat{\vx}(0)-\vx\|_2^2}{2(1-\alpha(t^*))}}$}}
\end{equation*}
where $p$ is the density of the natural data distribution and $\alpha(t^*)$ is the variance schedule at time $t^*$. Thus, \textit{DBP} is expected to reject adversarial examples by construction: the probability of producing any $\hat{\vx}(0)$ that is both adversarial and lies off-manifold is exponentially suppressed by the score model. However, this reasoning assumes the reverse process remains faithful to $p$. In practice, the score function $\boldsymbol{s}_\theta$ is itself an ML model—differentiable and susceptible to adversarial manipulation. Our key insight is that traditional gradient-based attacks, when backpropagated through \textit{DBP}, do not simply bypass the purifier—they implicitly target it, steering the score model to generate samples from an adversarial distribution. This challenges the assumptions underlying prior attack strategies, many of which distort gradients or constrain optimization in ways that ignore the score model's vulnerability, undermining their own effectiveness. Below, we formally characterize this vulnerability.

\paragraph{Definitions.} \underline{\textit{Diffusion Process \& Score Model.}} For $t_1\geq t_2$, let $\hat{\vx}_{t_1:t_2}$ represent the joint reverse diffusion trajectory $\hat{\vx}(t_1),\hat{\vx}(t_1 + dt), \dots, \hat{\vx}(t_2)$. Let $\ws_\theta$ denote the score model used by \textit{DBP} to approximate the gradients of the natural data distribution at different time steps. $\ws_{{\theta^t}}$ is the abstract score model invoked at time step $t$ with parameters ${{\theta^t}}$, corresponding to $\ws_\theta(\cdot, t)$. Given that the score model $\ws_\theta$ is an ML model that interacts with adversarial input $\vx$, we denote the parameters at each reverse step as $\theta^t_{\vx}$, capturing their dependence on $\vx$. That is, $\theta^t_{\vx} \equiv \theta^t(\vx)$, which makes explicit that the purification process is \textbf{not immutable}, as adversarial modifications to $\vx$ can shape its behavior.

\underline{\textit{Classifier.}} For a classifier $\mathcal{M}$ and label $y$, let $\mathcal{M}^y(\vu)$ denote the probability that $\vu$ belongs to class $y$.

\underline{\textit{Adaptive Attack.}} A gradient-based algorithm that iteratively updates $\vx$ with learning rate $\eta>0$:
\vspace{-2mm}
\begin{equation*}
\vx = \vx - \eta \widetilde{\nabla}_{\vx}, \quad \text{where} \quad  
\widetilde{\nabla}_{\vx} = \frac{1}{N}\sum_{n=1}^N \nabla_{\vx}[\mathcal{M}^y(\hat{\vx}(0)_n)].
\end{equation*}
\vspace{-4mm}

Here, $\widetilde{\nabla}_{\vx}$ is the empirical gradient estimate over $N$ purified samples $\hat{\vx}(0)_n$, obtained via Monte Carlo approximation and backpropagated through \textit{DBP}’s stochastic process.

\begin{theorem}
\label{thm:dbp_attack}
The adaptive attack \textbf{optimizes the entire reverse diffusion process}, modifying the parameters $\{\theta^t_{\vx}\}_{t\leq t^*}$ such that the output distribution $\hat{\vx}(0) \sim \textit{DBP}^{\{\theta^t_{\vx}\}}(\vx)$, where $\textit{DBP}^{\{\theta^t_{\vx}\}}(\vx)$ is the \textit{DBP} pipeline with the score model's weights ${\{\theta^t_{\vx}\}_{t\leq t^*}}$ adversarially aligned. That is, the adversary \textbf{implicitly controls the purification path} and optimizes the weights ${\{\theta^t_{\vx}\}_{t\leq t^*}}$ to maximize:
\begin{equation*}
\underset{\{\theta^t_{\vx}\}_{t\leq t^*}}{\max}\mathbb{E}_{\hat{\vx}(0) \sim \textit{DBP}^{\{\theta^t_{\vx}\}}(\vx)} [\Pr(\neg y | \hat{\vx}(0))].
\end{equation*}
Since $\{\theta^t_{\vx}\}_{t\leq t^*}$ depend on the purification trajectory $\hat{\vx}_{t^*\!:0}$, optimizing $\vx$ under the adaptive attack \textbf{directly shapes the purification process itself}, forcing \textit{DBP} to generate adversarially structured samples that maximize misclassification.
\end{theorem}

\paragraph{Impact.} As \textit{DBP} is widely perceived as immutable, its robustness is often attributed to gradient masking (vanishing or explosion) due to its presumed ability to project inputs onto the natural manifold. However, this reasoning treats \textit{DBP} as a \textbf{static pre-processing step} rather than a \textbf{dynamic optimization target}. Our result \textbf{overturns this assumption}: rather than resisting adversarial influence, \textit{DBP} itself becomes an \textbf{active participant in adversarial generation}. Crucially, standard adaptive attacks that backpropagate accurate gradients through \textit{DBP} do not merely evade it—they implicitly \textbf{exploit and reshape} its behavior, forcing the purification process to align with the attacker's objective. The introduced perturbations \textbf{directly shape the purification trajectory}, causing \textit{DBP} to generate, rather than suppress, adversarial samples. Thus, \textit{DBP} \textbf{does not inherently neutralize adversarial perturbations} but instead shifts the optimization target from the classifier to the score model $\ws_\theta$, leaving this paradigm highly vulnerable.

\begin{proof}
(Sketch) The adaptive attack maximizes the probability that the classifier mislabels the purified output, which is expressed as an expectation over the stochastic purification process. Applying the law of the unconscious statistician (LOTUS), we reformulate this expectation over the joint distribution of the reverse diffusion trajectory, shifting the dependency away from the classifier’s decision at the final step. Leveraging smoothness, we interchange differentiation and expectation, revealing that the adaptive attack’s gradient corresponds to the expected gradient of the purification trajectory, which is governed by the score model’s parameters. Crucially, these parameters are optimized implicitly through perturbations to $\vx$, rather than being explicitly modified. This demonstrates that the adaptive attack does not merely navigate \textit{DBP}---it fundamentally shapes its behavior, exploiting the score model rather than directly targeting the classifier. The full proof is in \aref{app:proof}.
\end{proof}

\subsection{Precise \textit{DBP} Gradients with \sysname{}}
\label{subsec:precise_grads}

Building on our theoretical analysis, we examine why prior adaptive attacks~\cite{diffpure,gmdp,kang2024diffattack,liu2024towards,lee2023robust} underperform and show how \sysname{}, provided by our \libname{} toolkit, resolves these limitations. 

We reserve the terms \textbf{forward/reverse pass} for \textit{DBP} and \textbf{forward/backpropagation} for gradient computations. Let $\hat{\vx}(t^*) := \vx(t^*)$. The reverse process follows:
\begin{equation}
\hat{\vx}(t+dt) = \hat{\vx}(t) + d\hat{\vx}(t), \quad dt<0,
\label{eq:recurse}
\end{equation}
and by chain rule the gradient of any function $F$ of $\hat{\vx}(t+dt)$ w.r.t. $\hat{\vx}(t)$ is:
\begin{equation}
\nabla_{\hat{\vx}(t)} F = \nabla_{\hat{\vx}(t)} \langle \hat{\vx}(t+dt),  \nabla_{\hat{\vx}(t+dt)} F \rangle.
\label{eq:recurse_grad}
\end{equation}
Applying this recursively yields gradients w.r.t. $\hat{\vx}(t^*)$, and via \hyperref[eq:closed]{eq.~(2)}, the input $\vx$. Yet, standard automatic differentiation is impractical for \textit{DBP} due to excessive memory overhead from dependencies between all $\hat{\vx}(t)$. Prior work resorts to approximations such as the \textbf{adjoint} method~\cite{adjoint}, or checkpointing~\cite{checkpoint} to compute the exact gradients. However, approximations yield suboptimal attack performance~\cite{lee2023robust}. On the other hand, we identify critical issues in existing checkpointing-based implementations that previously led to inflated estimates of \textit{DBP}'s robustness (see \aref{app:mem_efficient_diff} for detailed analysis, \sysname{}'s pseudo-code, and empirical evaluations of backpropagation issues):

\textit{\underline{1) High-Variance Gradients:}} Gradient-based attacks require estimating expected gradients using $N$ Monte Carlo (EOT) samples. When $N$ is small, gradient variance is high, leading to unreliable updates. Prior works used small $N$ due to limitations in how EOT samples were purified—one at a time via serial loops in standard attack benchmarks (e.g., \textit{AutoAttack}~\cite{autoattack}). \sysname{} parallelizes EOT purifications. This integrates seamlessly with standard attacks. Coupled with our termination upon success protocols---see~\S\ref{subsec:exp_mv}, this enables us to use up to $128$ EOT samples (for \textbf{\textit{CIFAR-10}}) with a speedup of up to $41\times$ upon early termination, drastically reducing variance.

\textit{\underline{2) Time Consistency:}} \textit{torchsde}\footnote{https://github.com/google-research/torchsde} is the de facto standard library for \textit{SDE} solvers. Hence, several prior checkpointing approaches likely use \textit{torchsde} as their backend. Yet, \textit{torchsde} internally converts the integration interval into a \textit{PyTorch} tensor, which causes a discrepancy in the time steps on which the score model is invoked during both propagation phases due to rounding issues if the checkpointing module is oblivious to this detail. \textbf{DiffGrad} ensures time steps match in both phases.

\textit{\underline{3) Reproducing Stochasticity:}} Forward and backward propagation should reuse identical randomness to preserve fidelity. In vanilla \textit{DBP}, diffusion noise cancels in the gradient, so not preserving stochastic components has little effect. Yet, for guided schemes whose computations depend on randomness (some current ones and potential future variants), failing to reproduce the noise realizations can bias gradients and derail optimization. We thus introduce a structured \textit{Noise Sampler $\boldsymbol{NS}$} that records all noise during the forward pass and reuses it in backpropagation. We also enable deterministic CuDNN kernel selection to avoid backend nondeterminism---see \aref{app:stoc} for details.

\textit{\underline{4) Guidance Gradients:}} Guided \textit{DBP}~\cite{gmdp} uses guidance metrics along the reverse pass. This guidance is obtained by applying a guidance function $\boldsymbol{g_{fn}}$ to (potentially) both the original input $\vx$ and the reverse-diffused $\hat{\vx}(t)$. As such, it creates paths from $\vx$ to the loss that basic checkpointing fails to consider. Furthermore, the guidance itself is often in the form of a gradient, necessitating second-order differentiation during backpropagation, which is, by default, disabled by differentiation engines. We extend \textbf{DiffGrad} to include these ``guidance'' gradients. Critically, recent SOTA \textit{DBP} defenses generate purified outputs from pure noise, relying on $\vx$ only via guidance (see~\S\ref{sebsec:sota}). As such, the lack of support for guidance gradients in all previous implementations leads to a false sense of security. To our knowledge, \textbf{DiffGrad} is the first to account for this component.

\underline{5) Proper Use of The Surrogate Method~\cite{lee2023robust}:} Upon inspecting \textbf{DiffHammer}'s code, we find their checkpointing implementation effectively applies~\citet{lee2023robust}'s \textbf{surrogate} approximation to calculate the gradients, which slightly degrades gradient quality with no practical performance gains---see \aref{app:one_shot}. Yet, the real issue lies in \textbf{DiffHammer}'s implementation of the \textbf{surrogate}, which diverges from~\cite{lee2023robust}, further corrupting the gradients. We defer details to \aref{app:diffhammer_surrogate}.

\subsection{On The Issues in \textit{DBP}'s Evaluation Protocols.} 
\label{subsec:flaws}
\textit{DBP} is typically deployed in a single-purification (\textit{SP}) setting: an input $\vx$ is purified once via a stochastic reverse process and classified as $c = \mathcal{M}(\hat{\vx}(0))$~\cite{diffpure,gmdp}. This assumes $\hat{\vx}(0)$ lies on the natural manifold, yielding a correct label with high probability—see~\S\ref{subsec:path}. Yet, prior studies~\cite{kang2024diffattack,diffpure,gmdp} fail to assess this very property: they evaluate robustness by purifying only the final adversarial iterate, testing just a single stochastic path, yielding noisy, unrepresentative estimates.

\textbf{DiffHammer}~\cite{diffhammer} addressed this with worst-case robustness (Wor.Rob): the attack succeeds if any of $N$ purifications (evaluated at each attack step) leads to misclassification. The metric is $ \text{Wor.Rob} := 1 - \frac{1}{S} \sum_{j=1}^{S} \underset{i \in [\![N]\!]}{max} \mathcal{A}_i^{(j)}$, where $\mathcal{A}_i^{(j)} = 1$ if the $i$-th purification of sample $j$ fails, and $S$ is the dataset size. They justify this protocol by modeling resubmission attacks (e.g., in login attempts), thereby limiting $N$ to (typically) $N=10$ after which the attacker will be blocked. However, attackers can retry inputs arbitrarily in stateless settings like spam, CSAM, and phishing. Even in stateful systems, the defender has no control over the stochastic path. Thus, \textit{SP} evaluations require many purifications per input for statistically meaningful conclusions.

Yet, any stochastic defense fails given enough queries~\cite{DBLP:journals/corr/abs-2302-13464}. Even if \textit{DBP} nearly always projects to the manifold, \textit{SP} is only reliable if misclassification probability is negligible over all paths—rare in practice (\S\ref{sec:exp}). Robust predictions must hence aggregate over multiple purifications. We thus propose:

\textit{\underline{Majority Vote (MV).}} Given input $\vx$, generate $K$ purifications and predict by majority: $c=MV^K(\vx):=mode\{\mathcal{M}(\hat{\vx}^{(1)}(0)),\ldots,\mathcal{M}(\hat{\vx}^{(K)}(0))\}$. Unlike certified smoothing~\cite{densepure}, \textit{MV} requires no excessive sampling, yet offers a stable, variance-tolerant robustness estimate, albeit without formal guarantees. We use the majority-robustness metric: $\text{MV.Rob} := 1 - \frac{1}{S} \sum_{j=1}^{S} \mathbbm{1} \left[ {MV^K(\vx_j) \ne y_j} \right]$, where $y_j$ is the ground-truth label. \textbf{DiffHammer}'s additional $\text{Avg.Rob} := 1 - \frac{1}{NS} \sum_{j=1}^{S} \sum_{i=1}^{N} \mathcal{A}_i^{(j)}$ averages path-level failures across all samples, conflating individual copy failures with sample robustness. A few fragile samples can dominate this metric, making it unreliable for deployment.

\section{Experiments}
\label{sec:exp}
\label{sec:exp:settings} 
We reevaluate \textit{DBP}, demonstrating its degraded performance when evaluation and backpropagation issues are addressed, cementing our theoretical findings from~\S\ref{subsec:path}.

\textbf{Setup.} We evaluate on \textit{\textbf{CIFAR-10}}~\cite{cifar} and \textit{\textbf{ImageNet}}~\cite{imagenet} similar to previous work~\cite{diffhammer,kang2024diffattack,diffpure,gmdp}. We consider two foundational \textit{DBP} defenses: The \textbf{VP-SDE} \textit{DBP} (\textit{DiffPure})~\cite{diffpure} and the \textit{Guided}-\textbf{DDPM} (see~\S\ref{subsec:precise_grads}), \textit{GDMP}~\cite{gmdp}. We use the \textit{DM}s~\cite{DBLP:conf/nips/DhariwalN21,ho2020denoising,dmint} studied in the original works, adopting the same purification settings---See \aref{app:systems}. Following~\citet{diffpure}, we use WideResNet-28-10 and WideResNet-70-16~\cite{zagoruyko2016wide} for \textit{\textbf{CIFAR-10}}, and WideResNet-50-2 and DeiT-S~\cite{dosovitskiy2020image} for \textit{\textbf{ImageNet}}. As in previous work~\cite{gmdp,diffpure,kang2024diffattack,diffhammer,liu2024towards,lee2023robust}, we focus on the white-box setting and use \textit{AutoAttack}-$\ell_\infty$ (\textit{AA}--$\ell_\infty$)~\cite{autoattack} with $\epsilon_{\infty}\!=\!8/255$ for \textbf{\textit{CIFAR-10}} and $\epsilon_{\infty}\!=\!4/255$ for \textbf{\textit{ImageNet}}. Existing works focus on norm-bounded ($\ell_{\infty}$ and $\ell_2$). For \textit{DBP}, $\ell_{\infty}$ has repeatedly proven superior~\cite{kang2024diffattack,lee2023robust,liu2024towards}, making it the focus of our evaluations. We report \textit{clean accuracy} (\textit{Cl-Acc})---without attacks--- and \textit{robust accuracy} (\textit{Rob-Acc})---the fraction of correctly classified attack samples. We use $256$ random test samples per dataset, consistent with prior \textit{DBP} work~\cite{diffhammer,chen2024robust}. Because \textit{DBP} is a compute-intensive defense, we prioritize \textbf{breadth over scale}---two datasets (\textbf{\textit{CIFAR-10}}, \textbf{\textit{ImageNet}}), multiple classifiers and \textit{DBP} schemes, diverse attacks and baselines, and \textit{MV} ablations---so readers see the same qualitative conclusions across settings (see~\S\ref{sec:exp}-~\S\ref{sec:counter}; extended results in \aref{app:one_shot}-\aref{app:ablation}).

\subsection{Reassessing One-Shot \textit{DBP} Robustness with Accurate Gradients}
\label{subsec:one_shott}
As noted in~\S\ref{sec:theory}, \textit{DBP}'s robustness stems from two factors: inaccurate gradients and improper evaluation. As our work offers enhancements on both fronts, we evaluate each factor separately. Here, we isolate the gradient issue by re-running prior experiments under the same (problematic) single-evaluation protocol, but with accurate gradients via our \textbf{DiffGrad} module and compare the results to those from the literature. Despite using only 10 optimization steps (vs. up to 100 in prior work), our method significantly outperforms all gradient approximations and even recent full-gradient methods. For \textit{DiffPure}, we lower robust accuracy from $62.11\%$ (reported by~\citet{liu2024towards}) to $48.05\%$, and for \textit{GDMP}, from $24.53\%$ (Surrogate~\cite{lee2023robust}) to $19.53\%$. These results expose the issues in existing gradient implementations and invalidate claimed gains from enhancement strategies like \textbf{DiffAttack}. Full breakdowns and additional baselines are provided in \aref{app:one_shot}.

\textbf{Evaluations Under~\citet{liu2024towards}'s Protocol.} \citet{liu2024towards}, like \textbf{DiffHammer}~\cite{diffhammer} and our own analysis, note that evaluating a single purification at attack termination inflates robustness scores. To address this, they propose a refined 1-evaluation protocol, which tests 20 purifications of the final \textit{AE} and declares success if any fail. This offers a stricter assessment than earlier one-shot methods, though still weaker than Wor.Rob (see~\S\ref{subsec:flaws}). Accordingly,~\citet{liu2024towards} group their method with one-shot evaluations. Repeating their setup using \textbf{DiffGrad} (20 attack iterations, $N=10$ EOT), we observe a dramatic drop in robust accuracy: on WideResNet-28-10, we improve upon~\citet{liu2024towards}'s results by $25.39\%$ and $30.42\%$ for \textit{DiffPure} and \textit{GDMP}, respectively—bringing the Rob-Accs down to $30.86\%$ and $10.55\%$. These results confirm the superiority of \textbf{DiffGrad}'s gradients and expose \textit{DBP}’s realistic vulnerability. Detailed results are in \aref{app:liu}.

\subsection{\textit{DBP} Under Realistic Protocols}
\label{subsec:exp_mv}
\begin{wraptable}{r}{0.6\textwidth}
    \vspace{-4mm}
    \centering
    \caption{\textit{AA}-$\ell_{\infty}$ performance comparison on \textbf{\textbf{\textit{CIFAR-10}}} ($\epsilon_{\infty}\!\!=\!\!8/255$) under realistic threat models. Metrics include Wor.Rob and MV.Rob under a 10-evaluation protocol. $\dag$ indicates strategy is \textit{PGD}.}
    \label{tab:batch_auto}
\resizebox{0.95\hsize}{2.1cm}{
    \begin{tabular}{cl|c|c}
    \toprule
    Pur. & Gradient Method & \begin{tabular}{cc}
         \multicolumn{2}{c}{Wor.Rob \%}  \\
        Cl-Acc & Rob-Acc
    \end{tabular} & \begin{tabular}{cc}
         \multicolumn{2}{c}{MV.Rob \%}  \\
        Cl-Acc & Rob-Acc
    \end{tabular}\\
    \midrule
    \multirow{7}{*}{\textit{DiffPure}~\cite{diffpure}} & 
         {\textbf{BPDA}} & \begin{tabular}{lc}
        \multirow{6}{*}{89.06} & \phantom{0}32.81
    \end{tabular} & \begin{tabular}{lc}
        \multirow{6}{*}{91.02} & \phantom{0}72.27
    \end{tabular} \\
    \cline{2-2}
         & {\textbf{DiffAttack} (DiffHammer~\cite{diffhammer})} & \begin{tabular}{cr}
         & \phantom{000000}33.79
    \end{tabular} & \begin{tabular}{lc}
         & \phantom{00000000}NA
    \end{tabular} \\
         & {\textbf{DiffAttack}\textbf{-DiffGrad}} & \begin{tabular}{cl}
         & \phantom{000000}15.63
    \end{tabular}  & \begin{tabular}{lc}
         & \phantom{000000}39.45
    \end{tabular} \\
    \cline{2-2}
         & \textbf{{DiffHammer}} (DiffHammer~\cite{diffhammer}) & \begin{tabular}{cr}
         & \phantom{000000}22.66
    \end{tabular} & \begin{tabular}{lc}
         & \phantom{00000000}NA
    \end{tabular} \\
         & \textbf{{DiffHammer}}\textbf{-DiffGrad} & \begin{tabular}{cr}
         & \phantom{000000}10.16
    \end{tabular} & \begin{tabular}{lc}
         & \phantom{000000}38.28
    \end{tabular} \\
        \cline{2-2}
          & \textbf{Full} (DiffHammer~\cite{diffhammer}) $\dag$ & \begin{tabular}{lc}
         & \phantom{000000}36.91
    \end{tabular} & \begin{tabular}{lc}
         & \phantom{00000000}NA
    \end{tabular} \\
          & \textbf{Full-}\sysname{} & \begin{tabular}{cr}
         & \phantom{000000}\textbf{17.19}
    \end{tabular}  & \begin{tabular}{lc}
         &\phantom{000000}{\textbf{39.45}}
    \end{tabular}  \\
    \cline{1-4}
    \multirow{7}{*}{\textit{GDMP}~\cite{gmdp}} & 
         {\textbf{BPDA}} & \begin{tabular}{lc}
        \multirow{6}{*}{91.80} & \phantom{0}27.73
    \end{tabular} & \begin{tabular}{lc}
        \multirow{6}{*}{92.19} & \phantom{0}53.52
    \end{tabular} \\
        \cline{2-2}
         & {\textbf{DiffAttack} (DiffHammer~\cite{diffhammer})} & \begin{tabular}{lc}
         & \phantom{000000}37.7
    \end{tabular} & \begin{tabular}{lc}
         & \phantom{00000000}NA
    \end{tabular}\\
         & {\textbf{DiffAttack}\textbf{-DiffGrad}} & \begin{tabular}{lc}
         & \phantom{000000}3.91
    \end{tabular} & \begin{tabular}{lc}
         & \phantom{000000}14.45
    \end{tabular} \\
        \cline{2-2}
         & \textbf{DiffHammer} (DiffHammer~\cite{diffhammer}) & \begin{tabular}{lc}
         & \phantom{000000}27.54
    \end{tabular} & \begin{tabular}{lc}
         & \phantom{00000000}NA
    \end{tabular} \\
         & \textbf{DiffHammer}\textbf{-DiffGrad} & \begin{tabular}{lc}
         & \phantom{0000000}7.81
    \end{tabular} & \begin{tabular}{lc}
         & \phantom{000000}25.39
    \end{tabular} \\
        \cline{2-2}
          & \textbf{Full} (DiffHammer~\cite{diffhammer}) $\dag$ & \begin{tabular}{lc}
         & \phantom{000000}31.05
    \end{tabular} & \begin{tabular}{lc}
         & \phantom{00000000}NA
    \end{tabular} \\
          & \textbf{Full-}\sysname{} & \begin{tabular}{lc}
         & \phantom{0000000}\textbf{7.03}
    \end{tabular} & \begin{tabular}{lc}
         & \phantom{0000000}\textbf{16.8}
    \end{tabular} \\
    \bottomrule
    \end{tabular}
}
\vspace{-2mm}
\end{wraptable}
Here, we continue to highlight the shortcomings of one-shot evaluation and the strength of \textbf{DiffGrad}, while exposing the invalidity of previous attack enhancements over the standard gradient-based methodologies. To do so, we adopt a Wor.Rob protocol, similar to \textbf{DiffHammer}---see~\S\ref{subsec:flaws} and also include our majority-vote variant (MV.Rob). We reimplement \textbf{DiffHammer} and \textbf{DiffAttack} using \textbf{DiffGrad}, and compare them to their original versions from~\cite{diffhammer} and our standard \textit{AA}-$\ell_\infty$. All evaluations use \textbf{\textit{CIFAR-10}} with WideResNet-70-16; We also report \textbf{\textit{ImageNet}} results below. \textbf{Full} denotes the standard $AA$ attack that uses the full exact gradients (i.e., via checkpointing).

\underline{\textit{Choice of batch size ($N$):}} We set $N=10$ for \textbf{\textit{CIFAR-10}} to match \cite{diffhammer} (which uses $N=10$ for both EOT and evaluation copies) and to balance robustness/gradient accuracy with runtime and memory; these values reflect practical deployment and evaluation constraints. For \textbf{\textit{ImageNet}}, we choose $N=8$ for similar reasons. Ablations with larger $N$, latency measurements, and practical recommendations appear in \aref{app:ablation}; complexity details are in \aref{app:stoc} and \aref{app:mem_efficient_algs}.

\textbf{\underline{Results.}} Table~\ref{tab:batch_auto} reports Wor.Rob and MV.Rob scores. The attacks run for 100 iterations, terminating upon success. For all attacks—\textbf{Full}, \textbf{DiffAttack}, and \textbf{DiffHammer}—\textbf{DiffGrad} yields significantly lower Wor.Rob compared to~\cite{diffhammer}, confirming the gradient mismatches discussed in~\S\ref{subsec:precise_grads} (MV.Rob is unique to our work). This establishes the need for our reliable \textbf{DiffGrad} as an essential tool for future progress in the field, given the repeated problems that continue to surface in implementations of the checkpointing method. With correct gradients (i.e., \textbf{DiffGrad}), all three attacks yield Wor.Rob $<20\%$; for \textit{GDMP}, $<10\%$, exposing the failure of the single-purification defense and reinforcing the claim that a statistically resistant alternative (e.g., our MV.Rob) must be used with \textit{DBP}. Similarly, this highlights the issues in the attack evaluation protocols that consider a single sample (i.e., 1-evaluation) as they drastically inflate robustness estimates. 

Notably, minor Wor.Rob differences among attacks in this range where predictions are highly noisy reflect random variation, not real gains. Hence, we must focus on the MV.Rob when comparing the three: For the same N, our \textit{MV} is strictly harder as it takes the majority label meaning success requires misclassifying $> \floor{N/2}$ of the N purified copies, whereas Wor.Rob counts a single error. Thus, MV.Rob is an upper bound on Wor.Rob, which aligns with the empirical findings attained with \textbf{DiffGrad}, showing significant robustness gains. Violations arise only when our MV.Rob results are compared to the Wor.Rob values reported by \textbf{DiffHammer}~\cite{diffhammer} and reflect implementation differences in gradient computation. As \textbf{DiffGrad} removes backpropagation mismatches that can reduce gradient fidelity and inflate robustness (both Wor.Rob and MV.Rob), \textit{MV}-versus-worst-case comparisons should be made using a common, \textbf{DiffGrad} implementation of each attack.

Throughout~\S\ref{sec:exp}, we have thus far shown that \textit{DBP}’s reported robustness has been overstated due to gradient inaccuracies. Yet, some recent works introduce \textit{DBP}-specific attack augmentations to improve success rates, but these overlook our theoretical finding that accurate gradients alone can reshape \textit{DBP}’s output distribution---rendering such enhancements unnecessary. \textbf{DiffHammer} builds on the assumption that \textit{DBP}’s trajectories split into “vulnerable” and “non-vulnerable” groups, with misleading gradients from the latter, but our analysis (see \hyperref[thm:dbp_attack]{\textbf{Theorem 3.1}}) shows that correct differentiation allows the attack to steer \textit{DBP}’s stochastic process toward adversarial outcomes, contradicting that fixed partition. Likewise, \textbf{DiffAttack}’s per-step losses artificially strive to alter the output distribution, ignoring the standard gradient-based attack’s inherent ability to do so, which could lead to suboptimal optimization updates. Below, we show that both these methods offer no advantages over the standard (full-gradient) attack and can, in fact, be counter-productive.

Under MV.Rob, \textbf{DiffAttack-DiffGrad} and \textbf{Full-DiffGrad} are identical on \textit{DiffPure}, and \textbf{DiffHammer-DiffGrad} leads to MV.Rob lower by a mere $1.17\%$, which amounts to only 3 samples out of 256 and is thus statistically insignificant. On \textit{GDMP}, \textbf{DiffHammer-DiffGrad} performs significantly worse than both others. Hence, \textbf{DiffHammer} worsens attack performance in accordance with our theory. \textbf{DiffAttack-DiffGrad} matches \textbf{Full-DiffGrad} on \textit{DiffPure}; on \textit{GDMP}, it shows a $2.35\%$ edge, which despite the questionable statistical significance given the test set size, could indicate a potential advantage. Yet, our \textbf{\textit{ImageNet}} comparisons below refute this hypothesis.

\begin{wraptable}{r}{0.7\textwidth}
    \vspace{-4mm}
    \centering
    \caption{\textit{AA}-$\ell_{\infty}$ comparison on \textbf{\textit{ImageNet}} ($\epsilon_{\infty}\!=\!4/255$).}
    \label{tab:single_imagenet_auto}
\resizebox{\hsize}{1.cm}{
    \begin{tabular}{ccl|c|c}
    \toprule
    Models &  Pur. & Gradient Method & \begin{tabular}{cc}
         \multicolumn{2}{c}{Wor.Rob \%}  \\
        Cl-Acc & Rob-Acc
    \end{tabular} & \begin{tabular}{cc}
         \multicolumn{2}{c}{MV.Rob \%}  \\
        Cl-Acc & Rob-Acc
    \end{tabular} \\
    \midrule
    \multirow{1}{*}{WideResNet-50-2} & \multirow{1}{*}{\textit{DiffPure}~\cite{diffpure}} & \textbf{Full-}\sysname{} & \begin{tabular}{lc}
         74.22 & \textbf{12.11}
    \end{tabular} & \begin{tabular}{lc}
        77.02 & \textbf{29.69}
    \end{tabular} \\
    \hline
    \multirow{3}{*}{DeiT-S} & \multirow{2}{*}{\textit{DiffPure}~\cite{diffpure}} & \textbf{DiffAttack}\textbf{-DiffGrad} & \begin{tabular}{lc}
        \multirow{2}{*}{73.63} & \phantom{00}25
    \end{tabular} & \begin{tabular}{lc}
        \multirow{2}{*}{77.34} & 42.21
    \end{tabular} \\
          & & \textbf{Full-}\sysname{} & \begin{tabular}{lc}
         & \phantom{00000}\textbf{21.09}
    \end{tabular} & \begin{tabular}{lc}
         & \phantom{0000}\textbf{32.81}
    \end{tabular} \\
    \cline{2-5}
    & \textit{GDMP}~\cite{gmdp} & \textbf{Full-}\sysname{} & \begin{tabular}{lc}
         69.14 & \textbf{20.70}
    \end{tabular} & \begin{tabular}{lc}
        75.0 & \textbf{32.83}
    \end{tabular} \\
    \bottomrule
    \end{tabular}
}
\vspace{-4mm}
\end{wraptable}

\textbf{ImageNet.}
We evaluate $AA$-$\ell_\infty$ on \textbf{\textit{ImageNet}} using WideResNet-50-2 and DeiT-S classifiers under $\epsilon_\infty{=}4/255$ ($100$ iterations), following standard practice. For DeiT-S, we also reimplement \textbf{DiffAttack} via \textbf{DiffGrad} for comparison. We use $16$ EOT samples (two batches of $N=8$) and $8$ samples for prediction (Wor.Rob/MV.Rob). As with \textbf{\textit{CIFAR-10}}, robustness drops sharply: Wor.Rob ranges from just $12.11\%$ to $21.09\%$, while MV.Rob peaks at $32.83\%$. This confirms the vulnerability of single-purification and the strength of gradient-based attacks. \textbf{DiffAttack} underperforms our standard attack by $9.4\%$ MV.Rob on DeiT-S, reinforcing its inferiority.
\section{Defeating Increased Stochasticity}
\label{subsec:low_freq}
\textit{DBP}'s stochasticity boosts its robustness under \textit{MV} (see~\S\ref{subsec:exp_mv}). Typical adversarial strategies incur high-frequency changes as they directly operate on pixels, altering each significantly w.r.t. its neighbors. This leads to visual inconsistencies, limiting the distortion budget. Such modifications are also easily masked by \textit{DBP}'s noise. Instead, systemic, low-frequency (\textit{LF}) changes allow larger perturbations and resist randomness.

Our \textit{LF} method is inspired by a recent attack---\textbf{\textit{UnMarker}}~\citep{unmarker}---on image watermarking that employs novel units termed \textit{Optimizable Filters (OFs)}. In signal processing, a filter is a smoothing operation defined by a kernel $\boldsymbol{\mathcal{K}} \!\in\! \mathbb{R}_{+}^{M \! {\times} \! N}$ (with values that sum to $1$), with which the input is convolved. The output at each pixel is a weighted average of all pixels in its $M \! {\times} \! N$ vicinity, depending on the weights assigned by $\boldsymbol{\mathcal{K}}$. Hence, filters incur systemic changes. Yet, they apply the same operation universally, unable to produce stealthy \textit{AE}s, as the changes required to alter the label will be uniformly destructive. \textit{OFs} allow each pixel $(i,j)$ to have its own kernel $\boldsymbol{\mathcal{K}}^{i,j}$ to customize the filtering at each point. $\boldsymbol{\mathcal{K^*}}$ is the set of all per-pixel $\boldsymbol{\mathcal{K}}^{i,j}$s. The weights $\boldsymbol{\theta}_{\boldsymbol{\mathcal{K}}^*}$ are learned via feedback from a perceptual metric (\textit{\textbf{lpips}})~\citep{lpips}, leading to an assignment that ensures similarity while maximizing the destruction at visually non-critical regions to optimize a specific objective. Note that the \textit{\textbf{lpips}} constraint replaces the traditional norm constraint. To guarantee similarity, they also impose geometric constraints via \textit{color kernels} $\sigma_c$, similar to \textit{\textbf{guided}} filters (details in \aref{app:low_freq}).

We subject $\vx$ to a chain \mbox{\small$\mathrel{\raisebox{1pt}{$\mbox{\tiny$\overset{B}{\underset{\boldsymbol{OF}}{\prod}}$}$}}\! \equiv\!$} \mbox{\small${\text{\tiny \mbox{$\raisebox{2pt}{$\boldsymbol{OF}_{\boldsymbol{\mathcal{K}}^*\!\!\!_{_1}, \vx, {\sigma\!_{c_{_1}}}} \!\!\!\!\! \circ \cdot\cdot\cdot \!\!\circ \boldsymbol{OF}_{\boldsymbol{\mathcal{K}}^*\!\!\!_{_B}, \vx, {\sigma\!_{c_{_B}}}}$}$}}}\!\!$} of \textit{OFs} similar to \textit{\textbf{UnMarker}}, replacing the objective pertaining to watermark removal in the filters' weights' learning process with the loss over $\mathcal{M}$. Each \textit{OF} has a kernel set {\small \mbox{${\boldsymbol{\mathcal{K}}}^*\!\!\!_{_b}$}} (with wights {\small \mbox{$\boldsymbol{\theta}_{\boldsymbol{\mathcal{K}}^*\!\!\!_{_b}}$}} and shape {\small \mbox{$M_b \! {\times} \! N_b$}}), and ${\sigma\!_{c_{_b}}}\!$. We optimize:
\vspace{-2mm}
\begin{equation}
\resizebox{!}{0.65cm}{$
\vx_{adv} = \; \underset{\{\boldsymbol{\theta}_{\boldsymbol{\mathcal{K}}^*\!\!\!_{_b}}\}, \boldsymbol{\delta}}{argmin}\; 
    \left[
        \begin{array}{cc}
    \ell_G^\mathcal{M}(\text{\mbox{\tiny$\overset{B}{\underset{\boldsymbol{OF}}{\prod}}$}($\vx+\boldsymbol{\delta}$}),\; y) \\
    + c \cdot max\{\boldsymbol{lpips}(\vx,\; \text{\mbox{\tiny$\overset{B}{\underset{\boldsymbol{OF}}{\prod}}$}($\vx+\boldsymbol{\delta})$})-\tau_p, 0\}
    \end{array}
    \right]
\label{eq:struct}
$}
\end{equation}
$\ell_G^\mathcal{M}$ denotes any loss as defined in~\S\ref{sec:threat_model}. $\boldsymbol{\delta}$ is a modifier that directly optimizes $\vx$, similar to traditional attacks. \textit{AE}s are generated by manipulating $\vx$ via $\boldsymbol{\delta}$ and propagating the result through the filters. While direct modifications alone do not cause systemic changes, with \textit{OF}s, they are smoothed over neighbors of the receiving pixels. $\boldsymbol{\delta}$ allows disruptions beyond interpolations. Similar to \textit{\textbf{UnMarker}}, we chain several \textit{OFs} with different shapes to explore various interpolations. Optimization is iterative (code in \aref{app:pseudo_c}). {\small \mbox{$max\{\boldsymbol{lpips}(\vx,\! \text{$\mathrel{\raisebox{1pt}{$\mbox{\tiny$\overset{B}{\underset{\boldsymbol{OF}}{\prod}}$}$}}$($\vx\!+\!\boldsymbol{\delta})$})-\tau_p, 0\}$}} enforces similarity: If the distance exceeds $\tau_p$, the $\boldsymbol{lpips}$ gradients lower it in the next iteration. Otherwise, it returns $0$, minimizing $\ell_G^\mathcal{M}$ unconditionally. This gives a solution within the $\tau_p$ constraint (violating outputs are discarded), yielding optimal {\tiny \mbox{$\{\widehat{\boldsymbol{\theta}}_{\boldsymbol{\mathcal{K}}^*\!\!\!_{_b}}\}\!$}}, \mbox{\small$\mathrel{\raisebox{-1pt}{$\widehat{\boldsymbol{\delta}}$}}$} s.t. \mbox{\small$\mathrel{\raisebox{0pt}{$\vx_{adv}\!\!=\!\!\text{$\mathrel{\raisebox{2pt}{$\mbox{\tiny$\overset{B}{\underset{\widehat{\boldsymbol{OF}}}{\prod}}$}$}}$}(\vx\!+\!\widehat{\boldsymbol{\delta}})$}}$}, where \mbox{\small$\mathrel{\raisebox{2pt}{$\mbox{\tiny$\overset{B}{\underset{\widehat{\boldsymbol{OF}}}{\prod}}$}$}}$} are the filters with {\tiny \mbox{$\{\widehat{\boldsymbol{\theta}}_{\boldsymbol{\mathcal{K}}^*\!\!\!_{_b}}\}$}}.
\nocite{cw}

\subsection{\textit{DBP} Against Low-Frequency \textit{AE}s}
\label{subsec:exp_lf}

\begin{wraptable}{r}{0.55\textwidth}
    \vspace{-4mm}
    \centering
    \caption{Performance of \textit{LF} attack under \textit{MV}.}
    \label{tab:unmarker}
\resizebox{\hsize}{1.5cm}{
    \begin{tabular}{ccc|cc}
    \toprule
    Pur. & Dataset & Models & Cl-Acc \% & Rob-Acc \% \\
    \midrule
     \multirow{5}{*}{\textit{DiffPure}~\cite{diffpure}} & \multirow{3}{*}{\textbf{\textit{ImageNet}}} & ResNet-50 & 72.54 & \textbf{0.00} \\
     & & WideResNet-50-2 & 77.02 & \textbf{0.00} \\
    & & DeiT-S & 77.34 & \textbf{0.00} \\
     \cline{2-5}
     & \multirow{2}{*}{\textbf{\textbf{\textit{CIFAR-10}}}} & WideResNet-28-10 & 92.19 &  \textbf{2.73} \\
     & & WideResNet-70-16 & 92.19 &  \textbf{3.13} \\
     \hline
     \multirow{5}{*}{\textit{GDMP}~\cite{gmdp}} & \multirow{3}{*}{\textbf{\textit{ImageNet}}} & ResNet-50 &  73.05 & \textbf{0.39} \\
     & & WideResNet-50-2 & 71.88 & \textbf{0.00} \\
    & & DeiT-S & 75.00 & \textbf{0.39} \\
     \cline{2-5}
     & \multirow{2}{*}{\textbf{\textbf{\textit{CIFAR-10}}}} & WideResNet-28-10 & 93.36 &  \textbf{0.00} \\
     & & WideResNet-70-16 & 92.19 &  \textbf{0.39} \\
    \bottomrule
    \end{tabular}
}
\vspace{-2mm}
\end{wraptable}
Our final question is: Can \textit{DBP} be degraded further under \textit{MV}? Based on~\S\ref{subsec:low_freq}, this is possible with \textit{LF}, which we test in this section. For \textit{LF} (see~\S\ref{subsec:low_freq}), we use \textit{VGG-LPIPS}~\cite{lpips} as the distance metric with $\tau_p\!=\!0.05$, ensuring imperceptibility~\cite{looks,unmarker}. Remaining parameters are similar to \textbf{\textit{UnMarker}}'s (see \aref{app:cifar_network_unmarker}). We use $128$ EOT samples for \textit{\textbf{CIFAR-10}} (same for label predictions since increasing the sample set size leads to enhanced robustness of \textit{DBP}---see \aref{app:ablation}). For \textbf{\textit{ImageNet}}, the numbers are identical to~\S\ref{subsec:exp_mv} (note that the dimensionality of \textit{\textbf{ImageNet}} makes larger sample sets prohibitive and our selected set sizes reflect realistic deployments---the largest number of samples that can fit into a modern GPU simultaneously).

\underline{\emph{Choice of Perceptual Constraint}}: We use \textbf{\textit{lpips}} with \textit{LF} because it is widely adopted, and prior work~\cite{unmarker,looks} provides known thresholds we can reuse for adversarial optimization. For \textit{AE}s, calibrated thresholds are essential; otherwise, an overly permissive constraint can label visibly altered adversarial images as successful, which defeats the point and inflates attack success (i.e., underestimates robustness). Both independently calibrating other non-norm-bounded metrics (e.g., via user studies) and re-running robustness evaluations are time-consuming and require additional resources. Thus, we limit our scope to \textbf{\textit{lpips}} but encourage future work to reproduce our experiments with alternative perceptual metrics for which calibrated thresholds may exist or become available.

\textbf{\underline{Results.}} The \textit{LF} results (using \textbf{Full-}\sysname{} for backpropagation) are in Table~\ref{tab:unmarker}. We also include a ResNet-50 classifier for \textit{\textbf{ImageNet}}. Not only does it defeat all classifiers completely, leaving the strongest with Rob-Acc of $3.13\%$, but it also does so in the \textit{MV} setting, where previous attacks fail. 

\textbf{Concluding Remarks.}  
Our findings highlight a limitation in current robustness evaluations, which focus heavily on norm-bounded attacks while overlooking powerful alternatives like low-frequency (\textit{LF}) perturbations. Though not norm-bounded, \textit{LF} is perceptually constrained—similar to \textit{StAdv}~\cite{stadv}, a longstanding benchmark against \textit{DBP}~\citet{diffpure}. \textit{LF}'s imperceptibility is guaranteed due to using a perceptual threshold $\tau_p=0.05$ that has previously been proven to guarantee stealthiness and quality~\cite{looks,unmarker}. We include qualitative results in \aref{app:attack_images} confirming this. One may question if existing attacks like \textit{StAdv} can also defeat \textit{DBP} under \textit{MV}, rendering \textit{LF} incremental. We address this in \aref{app:attack_images}, showing \textit{StAdv} fails to produce stealthy \textit{AE}s in this setting. While other techniques may also be adaptable to attacking certain \textit{DBP} variants in the future, \textit{LF} enjoys a solid theoretical foundation such alternatives may lack, limiting their generalizability to all \textit{DBP} defenses.

\textbf{Future Extensions.} While we focus on vision models similar to earlier \textit{DBP} work, both \hyperref[thm:dbp_attack]{\textbf{Theorem 3.1}} and \textbf{DiffGrad} are modality‑agnostic: they depend only on the diffusion dynamics and implementation, not structure. We thus expect the same vulnerabilities for, e.g., speech or video purification. The arguments in support of our \textit{LF} attack's efficacy due to its low-frequency nature also hold across various domains. Yet, the specific implementation that relies on spectral filter networks would require domain-based adaptations to apply it to different tasks (e.g., replacing \textit{2D} Fourier filters with their \textit{1D} counterparts for audio). We therefore urge future work to extend our methods to other modalities.
\section{Potential Countermeasures}
\label{sebsec:sota} 
\textbf{Adversarial Training (AT).}  
While AT remains a leading defense, it performs poorly on unseen threats. As expected, it fails against our \textit{LF}. On an adversarially trained WideResNet-28-10 for \textbf{\textit{CIFAR-10}} (\textit{DiffPure}), \textit{LF} reduces robust accuracy to just $0.78\%$, confirming this limitation.

\textbf{SOTA \textit{DBP}: \textit{MimicDiffusion}.}
\label{sec:counter}
One might ask whether recent variants offer improved robustness. Most build incrementally on the foundational defenses in~\S\ref{sec:exp}, meaning our results broadly generalize. One notable exception is \textit{MimicDiffusion}~\cite{mimic}, which generates outputs entirely from noise, using the input only as guidance (see~\S\ref{subsec:precise_grads}). Its goal is to preserve semantic content for classification while eliminating adversarial perturbations, and it reports SOTA robustness to adaptive attacks.

However, based on our theoretical analysis, we hypothesize this robustness is illusory—an artifact of improper evaluation and broken gradient flow. Unlike \textbf{GDMP}, which involves both direct and guidance paths from $\vx$ to the loss, \textit{MimicDiffusion} relies solely on guidance. Since guidance gradients require second-order derivatives (disabled by default), the original paper fails to compute meaningful gradients, severely overestimating robustness. We correct this using \textbf{DiffGrad}, the first method to properly differentiate through guidance paths. On \textbf{\textit{CIFAR-10}}, using \textit{AA}-$\ell_\infty$ (we exclude \textit{LF} for time limitations) with $N=128$ for both EOT and evaluation, we find that the original Rob-Acc scores of $92.67\%$ (WRN-28-10) and $92.26\%$ (WRN-70-16) collapse under proper evaluation: Wor.Rob drops to $2.73\%$ and $4.3\%$, respectively. Even under stricter MV.Rob, \textit{MimicDiffusion} performs worse than the standard \textit{DBP} defenses from~\S\ref{sec:exp}, with accuracies of just $25.78\%$ and $27.73\%$. This confirms that \textit{MimicDiffusion}'s robustness is not real, and collapses under proper gradients.

\textbf{\underline{Outlook.}} Our findings highlight the need for more robust defenses. \textit{DBP} remains promising as it improves clean accuracy compared to prior purification methods. Yet, \hyperref[thm:dbp_attack]{\textbf{Theorem 3.1}} shows it is vulnerable to adaptive attacks \textit{\textbf{when accurate gradients are available}}. We believe future work should explore \textit{DBP} variants that relax the assumptions of \hyperref[thm:dbp_attack]{\textbf{Theorem 3.1}}. One possible direction is to modify the purification process at inference so that the model follows private stochastic dynamics that are not directly exposed to adversaries. Such approaches, if carefully designed to avoid gradient masking~\cite{obfuscated}, could reduce the transferability of attacker-optimized gradients and increase robustness.

\section{Conclusion}
\label{sec:conclusion}
We scrutinized \textit{DBP}'s theoretical foundations, overturning its core assumptions. Our analysis of prior findings revealed their reliance on inaccurate gradients, which we corrected to enable reliable evaluations, exposing degraded performance under adaptive attacks. Finally, we evaluated \textit{DBP} in a stricter setup, wherein we found its increased stochasticity leaves it partially immune to norm-bounded \textit{AE}s. Yet, our novel low-frequency approach defeats this defense in both settings. We find current \textit{DBP} is not a viable response to \textit{AE}s, highlighting the need for improvements.
\if@neuripsfinal \commentout{
\section*{Acknowledgements}
This work was supported by the NSERC Discovery Grant RGPIN-2020-04722 and the Waterloo-Huawei Joint Innovation Laboratory.
}

\textbf{Acknowledgements.} This work was supported by NSERC Discovery Grant RGPIN-2020-04722. \fi
\if@preprint \commentout{
\section*{Acknowledgements}
This work was supported by the NSERC Discovery Grant RGPIN-2020-04722 and the Waterloo-Huawei Joint Innovation Laboratory.
}

\textbf{Acknowledgements.} This work was supported by NSERC Discovery Grant RGPIN-2020-04722. \fi
{
    \small
    \bibliographystyle{plainnat}
    \bibliography{main}
}

\newpage
\section*{NeurIPS Paper Checklist}

\begin{enumerate}

\item {\bf Claims}
    \item[] Question: Do the main claims made in the abstract and introduction accurately reflect the paper's contributions and scope?
    \item[] Answer: \answerYes{} 
    \item[] Justification: 

\item {\bf Limitations}
    \item[] Question: Does the paper discuss the limitations of the work performed by the authors?
    \item[] Answer: \answerYes{} 
    \item[] Justification: We provide a "Limitations" section in Appendix A.

\item {\bf Theory assumptions and proofs}
    \item[] Question: For each theoretical result, does the paper provide the full set of assumptions and a complete (and correct) proof?
    \item[] Answer: \answerYes{} 
    \item[] Justification: The assumptions underlying Theorem~\ref{thm:dbp_attack} are in Section~\ref{subsec:path} and the proof in Appendix~\ref{app:proof}.

    \item {\bf Experimental result reproducibility}
    \item[] Question: Does the paper fully disclose all the information needed to reproduce the main experimental results of the paper to the extent that it affects the main claims and/or conclusions of the paper (regardless of whether the code and data are provided or not)?
    \item[] Answer: \answerYes{} 
    \item[] Justification: Reproducibility information is given in Section~\ref{sec:exp:settings} and Appendices~\ref{app:cifar_network_unmarker} and~\ref{app:systems}.

\item {\bf Open access to data and code}
    \item[] Question: Does the paper provide open access to the data and code, with sufficient instructions to faithfully reproduce the main experimental results, as described in supplemental material?
    \item[] Answer: \answerYes{} 
    \item[] Justification: Our code is available here: \codeurl{}. We use public datasets CIFAR-10 and ImageNet.

\item {\bf Experimental setting/details}
    \item[] Question: Does the paper specify all the training and test details (e.g., data splits, hyperparameters, how they were chosen, type of optimizer, etc.) necessary to understand the results?
    \item[] Answer: \answerYes{} 
    \item[] Justification: Experiment settings and details are given in Section~\ref{sec:exp:settings} and Appendices~\ref{app:cifar_network_unmarker} and~\ref{app:systems}.

\item {\bf Experiment statistical significance}
    \item[] Question: Does the paper report error bars suitably and correctly defined or other appropriate information about the statistical significance of the experiments?
    \item[] Answer: \answerNo{} 
    \item[] Justification: Computing error bars would have been too computationally expensive given our limited computational resources and the extensive scope of our experiments, far exceeding typical settings in relevant prior work.

\item {\bf Experiments compute resources}
    \item[] Question: For each experiment, does the paper provide sufficient information on the computer resources (type of compute workers, memory, time of execution) needed to reproduce the experiments?
    \item[] Answer: \answerYes{} 
    \item[] Justification: Information about compute resources is given in Section~\ref{subsec:exp_mv}. Experiments are conducted on a 40GB NVIDIA A100 GPU.
    
\item {\bf Code of ethics}
    \item[] Question: Does the research conducted in the paper conform, in every respect, with the NeurIPS Code of Ethics \url{https://neurips.cc/public/EthicsGuidelines}?
    \item[] Answer: \answerYes{} 
    \item[] Justification: 

\item {\bf Broader impacts}
    \item[] Question: Does the paper discuss both potential positive societal impacts and negative societal impacts of the work performed?
    \item[] Answer: \answerYes{} 
    \item[] Justification: Broader impacts are discussed in Appendix~\ref{sec:impact}.
    
\item {\bf Safeguards}
    \item[] Question: Does the paper describe safeguards that have been put in place for responsible release of data or models that have a high risk for misuse (e.g., pretrained language models, image generators, or scraped datasets)?
    \item[] Answer: \answerNA{} 
    \item[] Justification: 

\item {\bf Licenses for existing assets}
    \item[] Question: Are the creators or original owners of assets (e.g., code, data, models), used in the paper, properly credited and are the license and terms of use explicitly mentioned and properly respected?
    \item[] Answer: \answerYes{} 
    \item[] Justification: We use public datasets CIFAR-10 and ImageNet and give corresponding citations.

\item {\bf New assets}
    \item[] Question: Are new assets introduced in the paper well documented and is the documentation provided alongside the assets?
    \item[] Answer: \answerYes{} 
    \item[] Justification: Our code repository (\codeurl{}) is well documented.

\item {\bf Crowdsourcing and research with human subjects}
    \item[] Question: For crowdsourcing experiments and research with human subjects, does the paper include the full text of instructions given to participants and screenshots, if applicable, as well as details about compensation (if any)? 
    \item[] Answer: \answerNA{} 
    \item[] Justification: 

\item {\bf Institutional review board (IRB) approvals or equivalent for research with human subjects}
    \item[] Question: Does the paper describe potential risks incurred by study participants, whether such risks were disclosed to the subjects, and whether Institutional Review Board (IRB) approvals (or an equivalent approval/review based on the requirements of your country or institution) were obtained?
    \item[] Answer: \answerNA{} 
    \item[] Justification: 

\item {\bf Declaration of LLM usage}
    \item[] Question: Does the paper describe the usage of LLMs if it is an important, original, or non-standard component of the core methods in this research? Note that if the LLM is used only for writing, editing, or formatting purposes and does not impact the core methodology, scientific rigorousness, or originality of the research, declaration is not required.
    \item[] Answer: \answerNA{} 
    \item[] Justification: 

\end{enumerate}
\clearpage
\appendix

\section{Broader Impact \& Limitations}
\label{sec:impact}
This work advances adversarial robustness by exposing a fundamental vulnerability in Diffusion-Based Purification (\textit{DBP}). We show that adaptive attacks do not merely circumvent \textit{DBP} but repurpose it as an adversarial generator, invalidating its theoretical guarantees. While this insight highlights a critical security risk, it also provides a foundation for designing more resilient purification strategies. To facilitate rigorous and reproducible evaluation, we introduce \libname{}, an open-source toolkit providing the first reliable gradient module for \textit{DBP}-based defenses, in addition to a wide array of attack implementations, including our low-frequency (\textit{LF}) strategy that outperforms existing approaches, seamlessly applicable to a broad range of classifiers. Similar to existing adversarial attack toolkits (e.g., \textit{Foolbox}~\cite{rauber2017foolbox}, \textit{AutoAttack}~\cite{autoattack}), our framework is intended for research and defensive purposes. While this tool enhances robustness assessments, we acknowledge its potential misuse in adversarial applications. To mitigate risks, we advocate its responsible use for research and defensive purposes only. Finally, we urge developers of \textit{ML} security systems to integrate our findings to design more resilient defenses against adaptive attacks. We encourage future research to explore fundamentally secure purification methods that are inherently resistant to manipulations.

\textbf{Limitations.}  Our work focuses on dissecting the robustness of \textit{DBP} defenses, both theoretically and empirically, with an emphasis on attack strategy. While we briefly consider existing mitigation techniques, future work should explore how diffusion models can be more effectively leveraged for defense. Classifiers could also be adversarially trained on \textit{LF} perturbations to improve robustness. Yet, the vast hyperparameter and architecture space of \textit{LF} potentially creates an effectively infinite threat surface, limiting the practicality of adversarial training. We leave this to future investigation.

 While we do not report formal error bars or confidence intervals, this decision aligns with standard practice in the literature (e.g.,~\cite{kang2024diffattack,liu2024towards}) due to the high computational cost of gradient-based \textit{DBP} attacks. Instead, we perform extensive experiments and also provide repeated empirical trials and variance analyses in
 \aref{app:practical_examples} that confirm our conclusions are robust, despite the absence of formal statistical intervals. 
 
 Although our main text focuses on conceptual and empirical issues in \textit{DBP}, \aref{app:stoc} and \aref{app:mem_efficient_algs}  provide theoretical and asymptotic cost analysis, and \aref{app:ablation} presents latency measurements and real-world speedup comparisons. Our module is efficient, scales linearly with diffusion steps, and outperforms prior implementations in both latency and throughput. Thus, while cost is not emphasized in the main paper, it is extensively addressed and poses no limitation in practice. Nonetheless, even with our efficient implementation, the dominant cost from \textit{DBP} itself is inherited by attacks and robustness evaluations, rendering them costly regardless. That said, \textit{DBP}'s practicality may improve with advanced hardware or more efficient schemes in the future.

\section{Equivalence of \textbf{DDPM} and \textbf{VP-SDE}}
\label{app:ddpm}
An alternative to the \textit{continuous-time} view described in~\S\ref{sec:Background} (i.e., \textbf{VP-SDE}) is \textit{Denoising diffusion probabilistic modeling} (\textbf{DDPM})~\citep{ho2020denoising,DiffusionModels}, which considers a \textit{discrete-time} framework (\textit{DM}) where the forward and reverse passes are characterized by a maximum number of steps $T$. Here, the forward pass is a Markov chain:
\vspace{-2mm}
\begin{equation*}
\vx_i = \sqrt{1-\beta_i} \vx_{i-1} + \sqrt{\beta_i} \vz_i
\vspace{-1mm}
\end{equation*}
where $\vz_i\sim \mathcal{N}(\boldsymbol{0},\; \boldsymbol{I}_d)$ and $\beta_i$ is a small positive noise scheduling constant. Defining $dt\!=\!\frac{1}{T}$, we know due to~\citet{dmint} that when $T \!\longrightarrow \!\infty$ (i.e, $dt\! \longrightarrow \!0$, which is the effective case of interest), this converges to the \textit{SDE} in \hyperref[eq:forward_sde]{\textit{eq.~(1)}} (with the drift and diffusion function $\boldsymbol{f}$ and $g$ described in~\S\ref{sec:Background}). The reverse pass is also a Markov chain given as:
\begin{equation}
d\hat{\vx}=\hat{\vx}_{i-1} \!-\! \hat{\vx}_i \!=\! \frac{1}{\sqrt{1-\beta_i}}((1 - \sqrt{1-\beta_i}) \hat{x}_i + \beta_i \ws_\theta (\hat{\vx}_i, i)) + \sqrt{\beta_i} \vz_i
    \label{eq:ddpm}
\end{equation}
When \mbox{\small$T \!\longrightarrow\! \infty$}, $d\hat{\vx}$ converges to \hyperref[eq:reverse_sde]{\textit{eq.~(4)}} (see~\cite{dmint}). Thus, the two views are effectively equivalent. Accordingly, we focus on the continuous view, which encompasses both frameworks. For discrete time, $\vx(t)$ will be used to denote $\vx_{|\frac{t}{dt}|}$.

\section{Proof of Theorem 3.1}
\label{app:proof}
\textbf{Theorem 3.1.} The adaptive attack \textbf{optimizes the entire reverse diffusion process}, modifying the parameters $\{\theta^t_{\vx}\}_{t\leq t^*}$ such that the output distribution $\hat{\vx}(0) \sim \textit{DBP}^{\{\theta^t_{\vx}\}}(\vx)$, where $\textit{DBP}^{\{\theta^t_{\vx}\}}(\vx)$ is the \textit{DBP} pipeline with the score model's weights ${\{\theta^t_{\vx}\}_{t\leq t^*}}$ adversarially aligned. That is, the adversary \textbf{implicitly controls the purification path} and optimizes the weights ${\{\theta^t_{\vx}\}_{t\leq t^*}}$ to maximize:
\begin{equation*}
\underset{\{\theta^t_{\vx}\}_{t\leq t^*}}{\max}\mathbb{E}_{\hat{\vx}(0) \sim \textit{DBP}^{\{\theta^t_{\vx}\}}(\vx)} [\Pr(\neg y | \hat{\vx}(0))].
\end{equation*}

\begin{proof}
The adversary aims to maximize:
\begin{equation*}
\begin{split}
\mathcal{Q}(\vx)\equiv\mathbb{E}_{\hat{\vx}(0) \sim \textit{DBP}^{{\{\theta^t\}}}(\vx)} [\Pr(\neg y | \hat{\vx}(0))]
\end{split}
\end{equation*}
Expanding this expectation yields the following alternative representations:
\begin{align*}
    &\mathcal{Q}(\vx) = \int \Pr(\neg y|\hat{\vx}(0))p(\hat{\vx}(0)) d\hat{\vx}(0) \\
    &= \int \Pr(\neg y|\hat{\vx}(0)) \int p(\hat{\vx}(0)|\hat{\vx}_{t^*\!:-dt}) p(\hat{\vx}_{t^*\!:-dt}) d\hat{\vx}_{t^*\!:-dt} d\hat{\vx}(0) \\
    &= \int \Pr(\neg y|\hat{\vx}(0)) p(\hat{\vx}_{t^*\!:0}) d\hat{\vx}_{t^*\!:0}.
\end{align*}
The first transition is due to the definition of expectation, the second follows from the definition of marginal probability, and the final transition replaces the joint density function with its compact form.

Since this formulation abstracts away \( \vx \) and \( \{\theta_{\vx}^t\}_{t\leq t^*} \), we explicitly include them as $p(\hat{\vx}_{t^*\!:0}) \equiv p(\hat{\vx}_{t^*\!:0} | \vx, \{\theta_{\vx}^t\}_{t\leq t^*})$. For notation simplicity, we define $p_{\theta_\vx}(\hat{\vx}_{t^*\!:0}|\vx) \equiv p(\hat{\vx}_{t^*\!:0}|\vx,\;\{\theta_{\vx}^t\}_{t\leq t^*})$. 
Thus, the final optimization objective is:
\begin{align*}
    &\mathcal{Q}(\vx) = \int \Pr(\neg y|\hat{\vx}(0)) p_{\theta_\vx}(\hat{\vx}_{t^*\!:0}|\vx) d\hat{\vx}_{t^*\!:0}.
\end{align*}
Since we assume smoothness (\( \mathcal{C}^2 \)), we interchange gradient and integral:
\begin{equation*}
\nabla_{\vx}[\mathcal{Q}(\vx)] = \int \Pr(\neg y|\hat{\vx}(0))\nabla_{\vx} p_{\theta_\vx}(\hat{\vx}_{t^*\!:0}|\vx) d\hat{\vx}_{t^*\!:0}.
\end{equation*}
where the last step is because $\Pr(\neg y|\hat{\vx}(0))$ does not depend on $\vx$, but only on $\hat{\vx}(0)$ which is independent of $\vx$ (though its probability of being the final output of \textit{DBP} is a function of $\vx$). Optimizing the objective via gradient ascent relies on altering \textit{DBP}'s output distribution alone, evident in its gradient where $\Pr(\neg y|\hat{\vx}(0))$ assigned by the classifier is a constant with its gradient ignored. While we lack direct access to the gradients of the probabilistic paths above, we may still attempt to solve the optimization problem.

Recall that by definition,  $\forall \vu\; \mathcal{M}^y(\vu)=\Pr(y|\vu)$. Thus, the required gradient can also be expressed as:
\begin{equation*}
\begin{aligned}
    &\nabla_{\vx}\big[\mathcal{Q}(\vx)\big] = \nabla_{\vx}\bigg[ \displaystyle\int\Pr(\neg y|\hat{\vx}(0))p_{\theta_\vx}(\hat{\vx}_{t^*\!:0}|\vx) d\mbox{\small${\hat{\vx}_{t^*\!:0}}$}\bigg]=\\
    &
    \nabla_{\vx}\bigg[ \displaystyle\mathop{\mathbb{E}}_{p_{\theta_\vx}(\hat{\vx}_{t^*\!:0}|\vx)}\!\!\!\!\!\!\![1-\mathcal{M}^y(\hat{\vx}(0))] \bigg] = -\nabla_{\vx}\bigg[\displaystyle\mathop{\mathbb{E}}_{p_{\theta_\vx}(\hat{\vx}_{t^*\!:0}|\vx)}\!\!\!\!\!\!\!\![\mathcal{M}^y(\hat{\vx}(0))]\bigg]
\end{aligned}
\end{equation*}

Yet, this expectation is over probabilistic paths whose randomness is due to the noise $\boldsymbol{\epsilon}$ of the forward pass and the Brownian motion $d\bar{\vw}_{t}$ at each reverse step $t$ that are independent. Hence, by the law of the unconscious statistician:
\vspace{-2mm}
\begin{equation*}
\begin{aligned}
&-\nabla_{\vx}\bigg[\displaystyle\mathop{\mathbb{E}}_{p_{\theta_\vx}(\hat{\vx}_{t^*\!:0}|\vx)}\!\!\!\!\!\!\!\!\!\![\mathcal{M}^y(\hat{\vx}(0))]\bigg]=-\nabla_{\vx}\bigg[\displaystyle\mathop{\mathbb{E}}_{p_r(\boldsymbol{\epsilon},d\bar{\vw}_{t^*},...,d\bar{\vw}_0)}\!\!\!\!\!\!\!\!\!\!\!\!\!\![\mathcal{M}^y(\hat{\vx}(0))]\bigg]
\end{aligned}
\end{equation*}
where $p_r(\boldsymbol{\epsilon},d\bar{\vw}_{t^*},..d\bar{\vw}_0)$ denotes the joint distribution of the noise $\boldsymbol{\epsilon}$ and the Brownian motion vectors in the reverse pass. Note that $\hat{\vx}(0)$ on the RHS denotes the output obtained by invoking the \textit{DBP} pipeline with some assignment of these random vectors on the sample $\vx$. As earlier, we can interchange the derivation and integration, obtaining:
\begin{equation*}
\begin{aligned}
&-\!\!\!\!\!\!\!\!\!\!\!\!\!\!\displaystyle\mathop{\mathbb{E}}_{p_r(\boldsymbol{\epsilon},d\bar{\vw}_{t^*},...,d\bar{\vw}_0)}\bigg[\nabla_{\vx}[\mathcal{M}^y(\hat{\vx}(0))]\bigg].
\end{aligned}
\end{equation*}

Let $\boldsymbol{G}^{\vx}$ denote the random variable that is assigned values from $\nabla_{\vx} \mathcal{M}^y(\hat{\vx}(0))$, where $\hat{\vx}(0)$ is as described above, and denote its covariance matrix by $\Sigma_{\boldsymbol{G}^{\vx}}$. Essentially, we are interested in $\mathbb{E}[\boldsymbol{G}^{\vx}]$. If we define $\widetilde{\nabla}_{\vx}$ as:
\begin{equation*}
 \widetilde{\nabla}_{\vx} = \frac{1}{N}\sum_{n=1}^N \nabla_{\vx}[\mathcal{M}^y(\hat{\vx}(0)_n)]
\end{equation*}
where each $\hat{\vx}(0)_n$ is the output of \textit{DBP} invoked with a certain random path $(\boldsymbol{\epsilon}^n,d\bar{\vw}_{t^*}^n,..d\bar{\vw}_0^n) \overset{i.i.d}{\sim} p(\boldsymbol{\epsilon},d\bar{\vw}_{t^*},..d\bar{\vw}_0)$ and $N$ is a sufficiently large number of samples. Then due to the central limit theorem, $\widetilde{\nabla}_{\vx} \longrightarrow \mathcal{N}(\mathbb{E}[\boldsymbol{G}^{\vx}], \frac{\Sigma_{\boldsymbol{G}^{\vx}}}{N})$. That is, propagating multiple copies through \textit{DBP} and then averaging the loss gradients computes the required gradient for forcing \textit{DBP} to alter its output distribution. The larger the number $N$ of samples is, the lower the variance of the error, as can be seen above. Note that the adaptive attack operates exactly in this manner, assuming it uses $\mathcal{M}^y$ as the loss it minimizes (but the soundness of the approach generalizes intuitively to any other loss with the same objective over the classifier's logits), proving our claim.
\end{proof}

\section{Details on Our \sysname{} Gradient Module}
\label{app:mem_efficient_diff}
In \aref{app:prev_issues}, we identify key implementation challenges in prior work, detailing how \sysname{} systematically resolves each one. Then, in \aref{app:practical_examples}, we empirically demonstrate their impact on \textit{DBP}'s gradients when left unaddressed, reinforcing their practical significance and validating the need for our reliable module. \aref{app:mem_efficient_algs} presents the pseudo-code for our memory-efficient checkpointing algorithms, offering a transparent blueprint for reproducible and correct gradient-enabled purification. Finally, in \aref{app:manual_verification}, we verify the correctness of our implementation, confirming that the gradients computed by \sysname{} match those produced by a true differentiable purification pipeline.

\subsection{The Challenges of Applying Checkpointing to \textit{DBP} and How \sysname{} Tackles Them}
\label{app:prev_issues}
\subsubsection{High-Variance Gradients}
\label{app:high_variance}
The challenge of high-variance gradients due to insufficient EOT samples is introduced in~\S\ref{subsec:precise_grads}. We defer its empirical analysis to \aref{app:practical_examples}, where we quantify the gradient variability under different sample counts. Practical implications for performance and throughput—enabled by \sysname{}'s design—are explored in the ablation study in \aref{app:ablation}. Accordingly, we focus below on the remaining implementation issues.

\subsubsection{Time Consistency}
\label{app:time_inconsistency}
Rounding issues may emerge when implementing checkpointing over \textit{torchsde} solvers: Given the starting time $t^*$, the solver internally converts it into a \textit{PyTorch} tensor, iteratively calculating the intermediate outputs by adding negative time increments to this tensor. On the other hand, checkpointing requires re-calculating the intermediate steps' samples during back-propagation. If this code is oblivious to \textit{torchsde}'s conversion of the initial time into a \textit{PyTorch} tensor, it will continue to treat it as a floating point number, updating it with increments of the time step to obtain each intermediate $t$ used to re-compute the dependencies. The \textit{PyTorch} implementation does not aim for $100\%$ accuracy, leading to minute discrepancies in the current value of $t$ compared to pure floating-point operations. These inaccuracies accumulate over the time horizon, potentially severely affecting the gradients. Instead, we ensure both the solver and checkpointing code use the same objects (either floating points or tensors).

\subsubsection{Reproducing Stochasticity}
\label{app:stoc}
\textbf{CuDNN-Induced Nondeterminism.} We found that even in deterministic purification pipelines, subtle nondeterminism from PyTorch’s backend can cause discrepancies between forward and backward propagations. Specifically, CuDNN may select different convolution kernels during the recomputation of intermediate states in checkpointing, leading to minute numerical drift.

To eliminate this, \libname{} explicitly forces CuDNN to behave deterministically, guaranteeing consistent kernel selection and numerical fidelity. While this source of error is subtle, it is measurable: for \textit{DiffPure} with $t^*=0.1$ and $dt=0.001$, we observed raw and relative gradient discrepancies of $1e-4$ and $7.28e-5$ when all other issues have been addressed. With this fix, the discrepancy dropped to exactly zero. Although less severe than other sources of error (e.g., time inconsistency or missing gradient paths), this level of precision is critical in setups where gradients are small to prevent sign changes that can affect optimization trajectories. Our inspection of prior work's code reveals this drift was not previously addressed.

\textbf{Noise Sampler for Guidance Robustness.} While the contribution of the stochastic component of $d\hat{\vx}$ often cancels analytically in vanilla \textbf{VP-SDE}-style purification (see~\S\ref{sec:Background}), meaning it does not affect the gradient and therefore is not required to be accurately reproduced during checkpointing, this is not guaranteed in all variants. In particular, many guidance-based schemes---see~\S\ref{subsec:precise_grads}---integrate stochastic terms within the guidance function itself. That is, the guidance function may incorporate stochasticity. When this occurs, gradients become sensitive to noise realizations, and failure to preserve them could break correctness. Yet, standard checkpointing only stores the intermediate $\hat{\vx}(t)$s. Hence, $\hat{\vx}(t\!+\!dt)$ will differ between the propagation phases as $d\hat{\vx}$ is computed via different random variables.

We restructure the logic computing $d\hat{\vx}$: We define a function $\boldsymbol{calc\_dx}$ that accepts the noise as an external parameter and utilize a \textit{Noise Sampler $\boldsymbol{NS}$}, initialized upon every forward propagation. For each $t$, $\boldsymbol{NS}$ returns all the necessary random noise vectors to pass to $\boldsymbol{calc\_dx}$. Instead of $\hat{\vx}(0)$ only, forward propagation also outputs $\boldsymbol{NS}$ and a state $\boldsymbol{S}$ containing all $\hat{\vx}(t)$s. These objects are used to restore the path in backpropagation. This design also future-proofs \libname{} for emerging guided or hybrid purifiers: any randomness on which the guidance depends is captured by $\boldsymbol{NS}$ and replayed exactly during differentiation.

\emph{The memory cost of our design is $\mathcal{O}(\frac{t^*}{|dt|})$, storing only $\boldsymbol{NS}$, and all $\hat{\vx}(t)$s, each with a negligible footprint ($\mathcal{O}(1)$) compared to graph dependencies~\citep{kang2024diffattack}.}

 \subsubsection{Guidance Gradients}
\label{app:guidance}
In schemes that involve guidance, it is typically obtained by applying a function $\boldsymbol{g_{fn}}$ to $\hat{\vx}(t)$ at each step (note that $\boldsymbol{g_{fn}}$ may involve stochasticity---see \aref{app:stoc}---as is often the case in \textit{GDMP}~\cite{gmdp}). For instance, in \textit{Guided}-\textbf{DDPM}~\citep{gmdp}, the original sample $\vx$ is used to influence the reverse procedure to retain key semantic information, allowing for an increased budget $t^*$ to better counteract adversarial perturbations. Effectively, it modifies \hyperref[eq:ddpm]{\textit{eq.~(8)}} describing the reverse pass of \textbf{DDPM} as:
\begin{equation*}
d\hat{\vx}= \frac{1}{\sqrt{1-\beta_i}}((1 - \sqrt{1-\beta_i}) \hat{\vx}_i + \beta_i \ws_\theta (\hat{\vx}_i, i)) - s \beta_i \nabla_{\hat{\vx}_i} \boldsymbol{GC} (\hat{\vx}_i,\; \vx) + \sqrt{\beta_i} \vz_i
\end{equation*}
where $\boldsymbol{GC}$ is a guidance condition (typically a distance metric), each step minimizes by moving in the opposite direction of its gradient, while the scale $s$ controls the guidance's influence. 

That is, in the specific case of \textit{Guided}-\textbf{DDPM}, $\boldsymbol{g_{fn}}(\hat{\vx}(t))\!\equiv\!-\nabla_{\hat{\vx}(t)} \boldsymbol{GC}(\hat{\vx}(t),\; \vx)$, where $\boldsymbol{GC}$ is a distance metric. Nonetheless,  other choices for $\boldsymbol{g_{fn}}$ may be employed in general. Typically, as the goal of the guidance is to ensure key information from the original sample $\vx$ is retained, $\boldsymbol{g\_fn}$ will also \textit{directly} involve this $\vx$ in addition to $\hat{\vx}(t)$ (e.g., $\boldsymbol{GC}$ above).  Yet, a naive implementation would back-propagate the gradients to $\vx$ by only considering the path through $\hat{\vx}(t)$. Yet, when $\boldsymbol{g\_fn}$ relies on a $\boldsymbol{guide}$ constructed from $\vx$ to influence $\hat{\vx}(t)$ (e.g., $\boldsymbol{guide}\!\equiv\!\vx$ in \textit{Guided}-\textbf{DDPM}, it creates additional paths from $\vx$ to the loss through this $\boldsymbol{guide}$ at each step $t$. Accordingly, \sysname{} augments the process to include the gradients due to these paths. In the general case, this $\boldsymbol{guide}$ may not be identical to $\vx$ but can rather be computed based on $\vx$ or even completely independent (in which case no guidance gradients are collected). \sysname{} captures this nuance through an abstract function $\boldsymbol{g\_aux}$ that, given $\vx$, outputs the $\boldsymbol{guide}$. Similar to \hyperref[eq:recurse_grad]{\textit{eq.~(6)}}, we have that for each $t$, the gradient of any function $F$ applied to $\hat{\vx}(0)$ w.r.t. $\boldsymbol{guide}$ due to the path from $\hat{\vx}(t+dt)$ to $\hat{\vx}(0)$ is given by:
\begin{equation}
\nabla_{\boldsymbol{guide}}^t {F} = \nabla_{\boldsymbol{guide}} \langle \hat{\vx}(t+dt), \nabla_{\hat{\vx}(t+dt)} {F}\rangle
    \label{eq:guide_grad_0}
\end{equation}
as $\hat{\vx}(t+dt)$ is a function of $\boldsymbol{guide}$. Recall that we are interested in $F$, which is the loss function over the classifier's output on $\hat{\vx}(0)$. The gradient of $F$ w.r.t. $\boldsymbol{guide}$ is a superposition of all these paths' gradients, given as:
\begin{equation} 
\nabla_{\boldsymbol{guide}} {F} =    \underset{t}{\sum} \nabla_{\boldsymbol{guide}}^t {F}
    \label{eq:guide_grad}
\end{equation}
By the chain rule, $F$'s gradient w.r.t $\vx$ due to the guidance paths, which we denote as $\nabla_{\vx}^{g} F$, is:
\begin{equation}
    \nabla_{\vx}^{g} F = \nabla_{\boldsymbol{\vx}} \langle\boldsymbol{guide}, \nabla_{\boldsymbol{guide}} {F}\rangle
    \label{eq:guide_grad_x}
\end{equation}
As a convention, for an unguided process or when $\boldsymbol{guide}$ and $\vx$ are not related, we define the gradient returned from \hyperref[eq:guide_grad_x]{\textit{eq.~(11)}} as $\nabla_{\vx}^{g} F\!\equiv\!\boldsymbol{0}$, preserving correctness in general. Since $\vx$ in this guided scenario traverses both the guidance and standard purification paths, the final gradient is the sum of both components.

Finally, automatic differentiation engines, by default, generate gradients without retaining dependencies on the inputs that produced them. Thus, when the guidance itself is in the form of a gradient as in the case of \textit{Guided}-\textbf{DDPM} or other potential alternatives, its effects will not be back-propagated to $\vx$ through any of the two paths described above, despite our proposed extensions. \sysname{} alters this behavior, retaining the dependencies of such gradient-based guidance metrics as well.

\ActivateWarningFilters[hpr]
\subsubsection{Proper Use of The Surrogate Method~\cite{lee2023robust}}
\DeactivateWarningFilters[hpr]
\label{app:diffhammer_surrogate}
As noted in~\S\ref{subsec:precise_grads}, \textbf{DiffHammer}~\cite{diffhammer} adopts the \textbf{surrogate} method (originally proposed by \citet{lee2023robust}) for gradient computation, as acknowledged in their Appendix C.1.1. This approach approximates \textit{DBP}’s gradients by replacing the standard fine-grained reverse process (e.g., $dt = 0.001$) with a coarser one (e.g., $\widebar{dt} = 0.01$), thereby enabling memory-efficient differentiation using standard autodiff. For instance, when using a diffusion time $t^*=0.1$ for \textit{DBP} (see~\S\ref{sec:Background}), the fine-grained process would require $100$ steps, which is computationally infeasible (memory-exhaustive) for gradient computation without checkpointing. In contrast, the coarse-grained process would only require $10$ steps, which is possible. 

Specifically, the forward pass (diffusion) is computed using the closed-form solution from \hyperref[eq:closed]{eq.~(2)}, while the reverse pass uses $\widebar{dt}$, enabling efficient gradient computation. Notably, this replacement only occurs for the purpose of updating the attack sample (i.e., computing the gradients) and happens during both forward and backpropagation so that the two phases match. In contrast, when evaluating the generated \textit{AE}, it is purified via the standard fine-grained process that uses $dt$, as this is the true (full) procedure the model owner (defender) would execute. Although the surrogate process yields approximate gradients for a slightly different process, \citet{lee2023robust} found it effective and more accurate than other approximations like the \textbf{adjoint} method~\cite{adjoint} (despite remaining slightly suboptimal---see \aref{app:one_shot}).

However, a manual inspection of \textbf{DiffHammer}'s code\footnote{\url{https://github.com/Ka1b0/DiffHammer} (accessed in May 2025)} reveals that their implementation of this surrogate method contains a critical issue. Rather than computing the reverse process with the coarse $\widebar{dt}$ as intended for attack optimization during both forward and backpropagation, they first run the reverse pass during forward propagation with the standard $dt$, storing intermediate states. Then, during backpropagation, they attempt to backpropagate gradients using checkpoints spaced at $\widebar{dt}$ intervals, but instead reuse the stored $dt$-based forward states. This leads to a mismatch: gradients intended for steps of size $\widebar{dt}$ are applied to states computed with $dt$, effectively disconnecting the computation graph and introducing significant gradient errors. In practice, this behaves like a hybrid between checkpointing and \textbf{BPDA}~\cite{obfuscated}, where gradients are approximated around fixed anchor points without correctly tracking the reverse dynamics.

As we show in~\S\ref{subsec:exp_mv}, this implementation leads to significant attack performance degradation compared to the full gradients. It illustrates the fragility of gradient computation in diffusion attacks and further motivates the need for our reliable \sysname{} module.

\subsection{Demonstrating The Effects of Incorrect Backpropagation}
\label{app:practical_examples}
\begin{figure}[H]
    \centering
    \begin{subfigure}[b]{0.48\textwidth}
        \includegraphics[width=\linewidth,height=4cm]{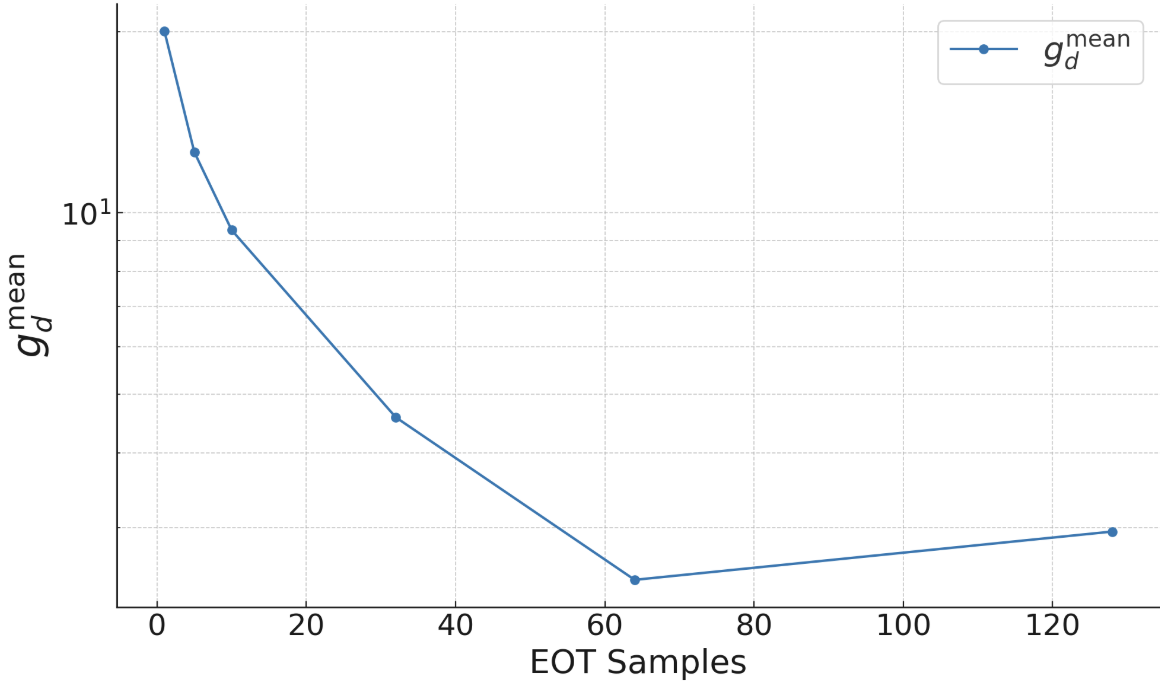}
        \caption{Effect of insufficient EOT samples}
        \label{fig:eot}
    \end{subfigure}%
    \hspace{\fill}
    \begin{subfigure}[b]{0.48\textwidth}
        \includegraphics[width=\linewidth,height=4cm]{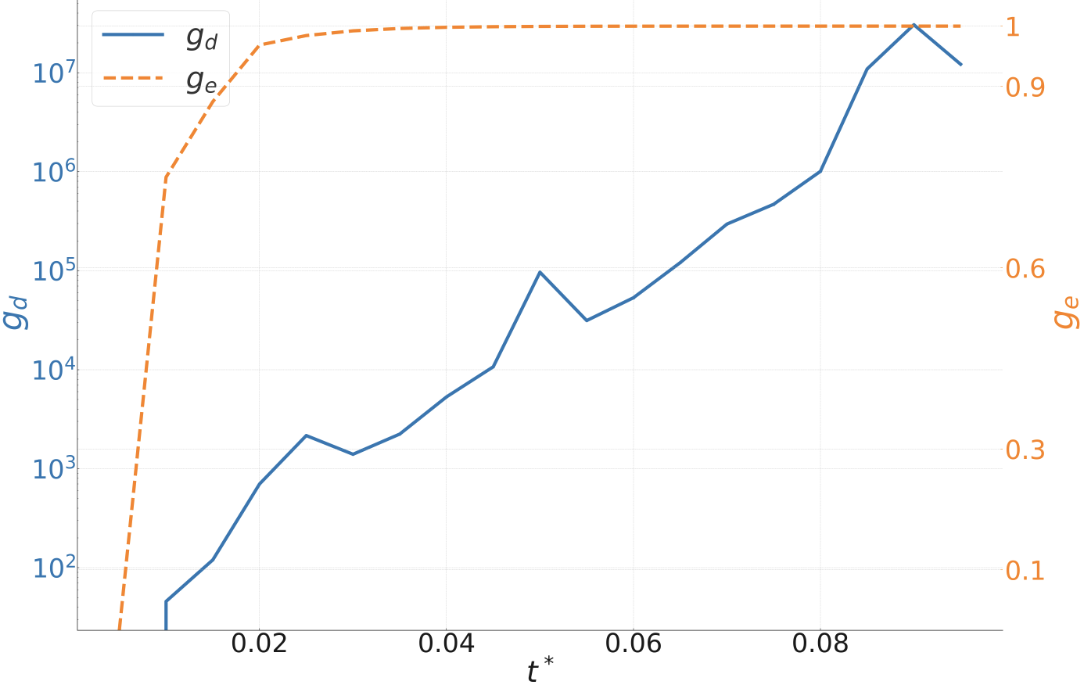}
        \caption{Effect of time inconsistency}
        \label{fig:time_inconsistency}
    \end{subfigure}%

    \vspace{0.4cm}

    \begin{subfigure}[b]{0.48\textwidth}
        \includegraphics[width=\linewidth,height=4cm]{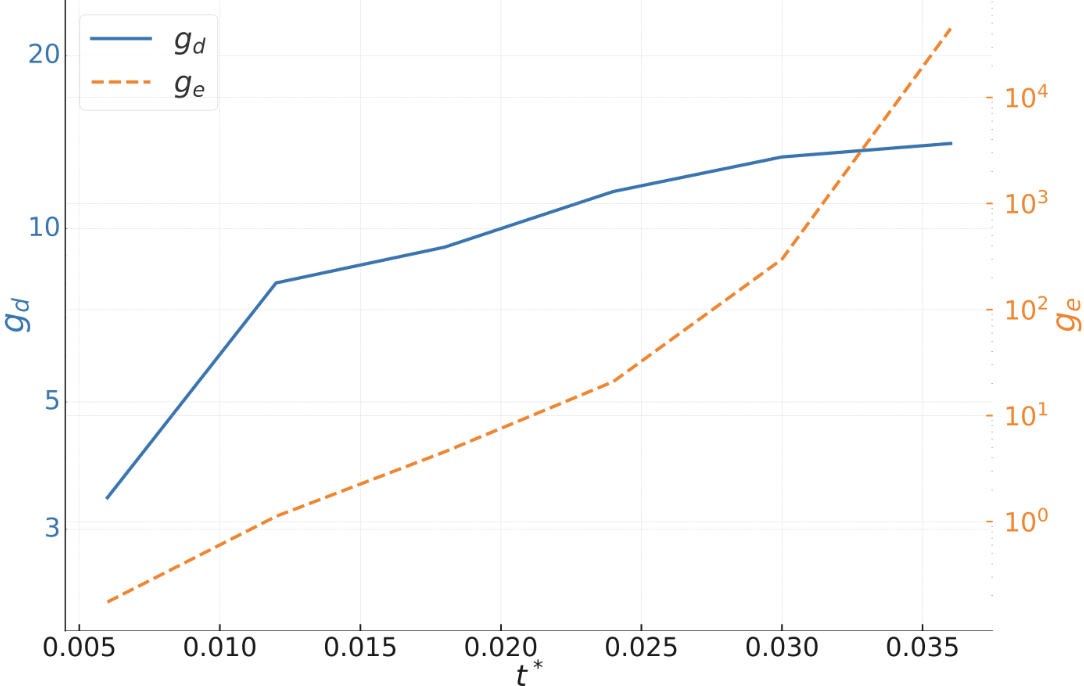}
        \caption{Effect of ignoring guidance gradients}
        \label{fig:guidance}
    \end{subfigure}%
    \hspace{\fill}
    \begin{subfigure}[b]{0.48\textwidth}
        \includegraphics[width=\linewidth,height=4cm]{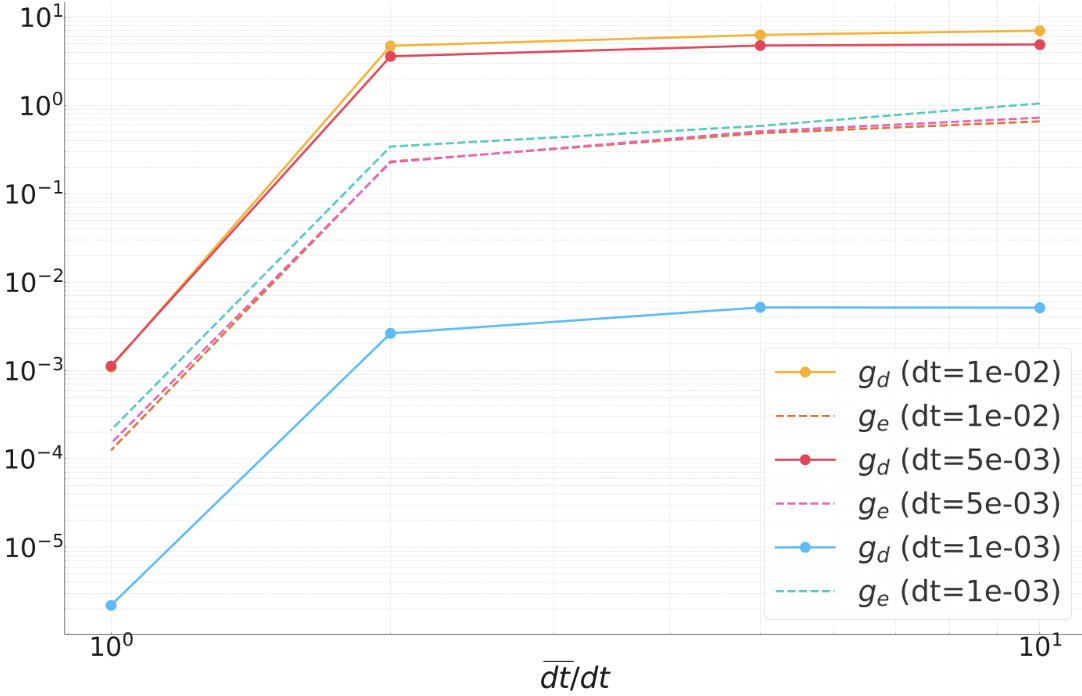}
        \caption{Effect of incorrect surrogate gradients}
        \label{fig:surrogate}
    \end{subfigure}
    
    \caption{Effects of the identified issues in \textit{DBP}'s backpropagation. Each subfigure visualizes a specific error source.}
    \label{fig:error_srcs}
\end{figure}

We empirically demonstrate the impact of the identified issues on \textit{DBP}'s gradients in Fig.~\ref{fig:error_srcs} and briefly explain each result below. As numerical results for CuDNN-induced nondeterminism (i.e., unhandled stochasticity) are already presented in \aref{app:stoc}, we focus here on the remaining four issues detailed in \aref{app:prev_issues}:

\textbf{(a) Effect of insufficient EOT samples.} Fig.\ref{fig:eot} illustrates the impact of using too few EOT samples when computing gradients—see\S\ref{subsec:precise_grads} and \aref{app:high_variance}. While prior works~\cite{diffhammer,kang2024diffattack,lee2023robust,liu2024towards,diffpure} typically use 10–20 samples, such low counts, while somewhat effective, may introduce significant variance, resulting in noisy gradients and suboptimal attacks.

To quantify this, we run the following experiment on a \textbf{\textit{CIFAR-10}} sample $\vx$ purified with \textit{DiffPure} ($t^*{=}0.1$, $dt{=}10^{-3}$). For each EOT count $N$ (shown on the x-axis), we generate $N$ purified copies, compute the average loss, and backpropagate to obtain the EOT gradient. This is repeated 20 times to yield 20 gradients ${g_1^N,\dots,g_{20}^N}$ for each $N$. We then define:
\begin{equation*}
    g_d^{mean}(N)=\frac{1}{20}\sum_{i=1}^{20}{\underset{j}{max}{\;\|\vg^N_i-\vg^N_j\|_2}}
\end{equation*}

This metric reflects the worst-case deviation between gradients under repeated EOT sampling. By the central limit theorem (see \aref{app:proof}), the variance of EOT gradients decays with $N$, and thus $g_d^{\text{mean}}(N) \to 0$ as $N \to \infty$. The curve in Fig.~\ref{fig:eot} captures this decay and lets us identify a threshold where variance becomes acceptably low.

The plot confirms that gradient variance decreases sharply with the number of EOT samples. While $N=10$—the typical choice in prior work—is somewhat effective and viable under resource constraints, it remains suboptimal. We observe that variance continues to drop until around $N=64$, after which it plateaus. This suggests that although larger $N$ values like $128$ yield marginally better stability, the bulk of the benefit is realized by $N=32$–$64$. Thus, $N=10$ is a practical baseline, but $N \ge 64$ should be preferred when accuracy is critical and compute allows, coroborating our claims in~\S\ref{subsec:precise_grads}.

\textbf{(b) Effect of time inconsistency.} 
As detailed in~\S\ref{subsec:precise_grads} and \aref{app:time_inconsistency}, rounding inconsistencies can emerge when using checkpointing with \textit{torchsde} solvers. These stem from how the solver internally handles time as a \textit{PyTorch} tensor, which may diverge subtly from floating-point representations if not carefully synchronized. While such discrepancies are negligible for short purification paths (small $t^*$), they accumulate significantly over longer trajectories—particularly at standard \textit{DBP} settings like $t^*{=}0.1$ with $dt{=}10^{-3}$ (i.e., $100$ reverse steps), as in \textit{DiffPure} on \textbf{\textit{CIFAR-10}}.

To quantify this effect, we purify a fixed \textbf{\textit{CIFAR-10}} sample $\vx$ using values of $t^*$ from $0.005$ to $0.1$ in steps of $0.005$. For each $t^*$, we compute gradients using \sysname{} under two conditions: (1) without correcting the inconsistency (yielding $\vg_{nf}$), and (2) with the issue fixed (yielding $\vg_f$), both using the same stochastic path. We then compute two metrics: (1) $g_d = \|\vg_f - \vg_{nf}\|_2$ (absolute error), and
(2) $g_e = \frac{\|\vg_f - \vg_{nf}\|_2}{\|\vg_f\|_2}$ (relative error).
These quantify the divergence introduced by uncorrected rounding—both in magnitude and in proportion to the true gradient—and illustrate its increasing severity with longer reverse trajectories.

Fig.~\ref{fig:time_inconsistency} confirms that even minute rounding discrepancies in time handling, if uncorrected, can completely corrupt gradients over longer diffusion trajectories. Both the absolute error $g_d$ (blue) and the relative error $g_e$ (orange) grow rapidly with $t^*$. By $t^* \geq 0.03$, $g_e$ exceeds $90\%$, and for the typical setting of $t^*=0.1$, the relative error reaches $100\%$, indicating that the gradients are nearly orthogonal to the correct direction. This validates our claim that unpatched checkpointing introduces severe errors and demonstrates that our \sysname{} fix is essential for correct gradient computation in practical \textit{DBP} setups.

\textbf{(c) Effect of ignored guidance gradients.} We evaluate the impact of neglecting guidance gradients, a critical problem in current attacks against \textit{Guided} \textit{DBP} defenses (e.g., \textit{GDMP}, \textit{MimicDiffusion})—see~\S\ref{subsec:precise_grads} and \aref{app:guidance}. As explained earlier, guidance introduces additional gradient paths from the input $\vx$ to $\hat{\vx}(t)$, typically requiring second-order derivatives to capture. Existing approaches ignore these paths, leading to incomplete gradients and suboptimal attacks. To our knowledge, \sysname{} is the first system to correctly handle them.

To quantify this effect, we purify a fixed \textbf{\textit{CIFAR-10}} sample $\vx$ using \textit{GDMP} (setup in \aref{app:systems}) while varying the purification horizon $t^* \in \{0.006, 0.012, \ldots, 0.036\}$. For each $t^*$, we compute gradients using \sysname{} with and without accounting for guidance—yielding $\vg_f$ and $\vg_{nf}$, respectively—under the same stochastic path. We then compute the absolute and relative gradient error ($g_d$, $g_e$) as in previous experiments.

The findings in Fig.~\ref{fig:guidance} clearly demonstrate the severe impact of neglecting guidance gradients in \textit{Guided} \textit{DBP} setups. Both the raw error ($g_d$) and the relative error ($g_e$) increase steeply with $t^*$, confirming that as the purification path grows longer, the omitted gradient paths—stemming from guidance dependencies—dominate the overall gradient signal. Most notably, $g_e$ exceeds $10^4$ at $t^*=0.036$, indicating that the gradient used in prior works is essentially meaningless for optimization, as it no longer aligns with the true direction of steepest descent.

These findings directly explain the poor attack success rates previously observed on guided defenses like \textit{GDMP} and \textit{MimicDiffusion}, and reinforce why our attacks, which correctly account for these guidance gradients via \sysname{}, achieve drastically superior performance---see~\S\ref{subsec:exp_mv} and~\S\ref{sec:counter}. In short, the guidance mechanism, when improperly handled, not only fails to help but actively hinders attack effectiveness by yielding incorrect gradient signals.

\textbf{(d) Effect of incorrect surrogate gradients.} 
As explained in~\S\ref{subsec:precise_grads} and \aref{app:diffhammer_surrogate}, prior work~\cite{diffhammer} uses the \textbf{surrogate} method~\cite{lee2023robust} to enable efficient gradient computation for \textit{DBP}. However, we identify a critical divergence in \textbf{DiffHammer}’s~\cite{diffhammer} implementation of this method that substantially compromises attack performance. This not only leads to inflated robustness estimates for \textit{DBP}, but also to misleading conclusions about the relative effectiveness of the standard gradient-based attacks versus \textbf{DiffHammer}'s proposed enhancements---see~\S\ref{subsec:exp_mv}. Specifically, as detailed in \aref{app:diffhammer_surrogate}, \textbf{DiffHammer} mismatches the time discretization used in the forward and backward passes: the forward pass computes intermediate denoised states $\hat{\vx}(t)$ using a fine step size $dt$, while the backward pass attempts to propagate gradients at a coarser granularity $\widebar{dt} > dt$ using those same (misaligned) forward states. This inconsistency leads to gradients being applied to states generated under a different dynamics, corrupting the computation and severely degrading the attack.

To validate this, we conduct a final experiment: For three different values of $dt \in \{0.001,\ 0.005,\ 0.01\}$ and fixed $t^* = 0.1$, we purify a fixed \textbf{\textit{CIFAR-10}} sample $\vx$. For each $dt$, we evaluate four surrogate variants using $\widebar{dt} \in \{dt,\ 2dt,\ 5dt,\ 10dt\}$. For every configuration, we compute $\vg_{nf}$ using \textbf{DiffHammer}’s surrogate implementation and $\vg_f$ using the correct implementation. As before, we report both $g_d$ and $g_e$ to quantify the absolute and relative gradient error due to the mismatch between $dt$ and $\widebar{dt}$. Each experiment is repeated 10 times, and we report the mean values.

The results in Fig.~\ref{fig:surrogate} clearly validate our claim: the gradient errors induced by \textbf{DiffHammer}’s surrogate implementation grow substantially with increasing $\widebar{dt}/dt$, saturating quickly even for modest mismatches. Both absolute error $g_d$ and relative error $g_e$ consistently worsen across all $dt$ configurations, indicating that gradients are being applied to the wrong points in the computation graph.

Crucially, \textbf{DiffHammer} uses $dt = 0.01$ and $\widebar{dt} = 2dt = 0.02$, a setup which already produces substantial degradation ($g_d \approx 3$, $g_e \approx 1$), confirming that their reported results rely on gradients that diverge significantly from the correct ones. These errors propagate into the attack, weakening its effectiveness and misleadingly suggesting that first-order attacks are inherently inferior. This experiment conclusively demonstrates that \textbf{DiffHammer}'s surrogate misuse is not a minor detail, but a critical bug that undermines their central claims.

We note that \sysname{} also provides a correct implementation of the \textbf{surrogate} method (despite the full correct gradient being the main method considered in the paper).

\subsection{Pseudo-Code for Our Memory-Efficient Gradient-Enabled Purification with \sysname{}}
\label{app:mem_efficient_algs}
\ActivateWarningFilters[fontsz]
\begin{algorithm}
\caption{Differentiable Purification with \textit{\sysname{}} — Forward Propagation}
\label{alg:DiffGrad_forward}
\begin{algorithmic}[1] 
    \REQUIRE Sample $\vx$, Score model $\ws_\theta$, Optimal diffusion time $t^*$, step size $dt$, Noise scheduler $\beta$, Reverse diffusion function $\mathbf{calc\_dx}$, {\color{red}Noise sampler initializer $\mathbf{init\_noise\_sampler}$}, {\color{blue}Guidance condition $\mathbf{g_{fn}}$, Guidance scale $s$, Auxiliary guidance extractor $\mathbf{g\_aux}$}

    \STATE $steps \gets \left|\frac{t^*}{dt}\right|$, {\color{blue}$\mathbf{guide} \gets \mathbf{g\_aux}(\vx)$} \CCOMMENT[black]{Calc. \#steps {\color{blue}and init. guide}}
    \STATE {\color{red}$\mathbf{disable\_dependencies}()$} 
    \MULTILINECOMMENT[red]{Dependencies enabled during forward \\ propagation. Disable them.}
    \STATE {\color{red}$\mathbf{S} \gets []$}
    \MULTILINECOMMENT[red]{Saved state (will eventually hold all \\ intermediate reverse steps' outputs).}
    \STATE {\color{red}$seed \gets \mathbf{random\_seed}()$} \CCOMMENT[red]{Seed to initialize noise path}
    \STATE {\color{red}$\mathbf{NS} \gets \mathbf{init\_noise\_sampler}(seed)$} \CCOMMENT[red]{\textbf{Reproducible} sampler}
    \STATE $\alpha \gets \mathbf{calc\_alpha}(\beta)$ \CCOMMENT[black]{Calculate $\alpha$ factors from \hyperref[eq:closed]{\textit{eq.~(2)}}}
    \STATE Draw $\boldsymbol{\epsilon} \sim \mathcal{N}(\boldsymbol{0}, \boldsymbol{I}_d)$
    \STATE $\hat{\vx} \gets \sqrt{\alpha(t^*)} \vx + \sqrt{1-\alpha(t^*)} \boldsymbol{\epsilon}$ \CCOMMENT[black]{Diffuse according to \hyperref[eq:closed]{\textit{eq.~(2)}}}
    
    \FOR{$i \gets steps, steps-1, ..., 1$}
        \STATE {\color{red}$\mathbf{S.append}(\hat{\vx})$} \CCOMMENT[red]{Set $\mathbf{S}[i] = \hat{\vx}(t)$.}
        \STATE {\color{red}$\mathbf{step\_noise} \gets \mathbf{NS.sample}(i)$}
        \MULTILINECOMMENT[red]{Sample the random noise used to \\ calculate $d\hat{\vx}$ at step $i$.}
        \STATE $d\hat{\vx} \gets \mathbf{calc\_dx}(\hat{\vx}, \ws_\theta, i, dt, \beta,$ \hfill \textcolor{black}{/* Calc. $d\hat{\vx}$ according to}
        \STATE \hspace{2.5em}${\color{red}\mathbf{step\_noise}}, {\color{blue}\mathbf{g_{fn}}, s, \mathbf{guide}})$ \hfill \textcolor{black}{\hyperref[eq:reverse_sde]{\textit{eq.~(4)}} */}
        \STATE $\hat{\vx} \gets \hat{\vx} + d\hat{\vx}$ \CCOMMENT[black]{Update $\hat{\vx} = \hat{\vx}(t+dt)$.}
    \ENDFOR
    
    \STATE {\color{red}$\mathbf{enable\_dependencies}()$} \CCOMMENT[red]{Re-enable dependencies.}
    \RETURN $\hat{\vx}, {\color{red}\mathbf{S}, \mathbf{NS}}, {\color{red}\mathbf{guide}}$
\end{algorithmic}
\end{algorithm}
\DeactivateWarningFilters[fontsz]
\paragraph{Forward Propagation.} \sysname{}'s forward propagation logic is in Algorithm~\ref{alg:DiffGrad_forward}. The code in blue is optional, pertaining to the use of guidance. We highlight in red the portions that differ from standard forward propagation. First, we generate the guidance $\boldsymbol{guide}$ from $\vx$ (line 1) and disable all graph dependency storage (line 2), enabling our code to run efficiently without attempting to store graphs that will lead to memory failures. Afterward (lines 3-5), we initialize $\boldsymbol{S}$ as an empty list and draw a random seed that is then used to invoke the abstract function $\boldsymbol{init\_noise\_sampler}$, which returns a noise sampler that provides a reproducible random path for the backpropagation phase (see~\S\ref{subsec:precise_grads}). After the input is diffused (lines 6-8) via \hyperref[eq:closed]{\textit{eq.~(2)}}, lines 9-15 correspond to the reverse pass: At each step $t$ (effectively $i$), 
$\hat{\vx}$ (that now represents $\hat{\vx}(t)$) is first appended to $\boldsymbol{S}$, which will eventually contain all such intermediate outputs (line 10). The noise provided by $\boldsymbol{NS}$ for the current step $i$ is then retrieved (line 11) and used to compute $d\hat{\vx}$ (line 12). $d\hat{\vx}$ is then added to $\hat{\vx}$ so that its current value becomes $\hat{\vx}(t+dt)$. This repeats until $\hat{\vx}=\hat{\vx}(0)$. Unlike the naive implementation, we only store the intermediate results. For efficiency, we also avoid saving the random noise for each step $i$ but utilize $\boldsymbol{NS}$ to reproduce those variables on demand. Before termination, we re-enable dependency storage (line 16) to ensure our code does not interfere with the execution of any other modules. Finally, $\hat{\vx}(0)$ is returned together with the state $\boldsymbol{S}$ and the sampler $\boldsymbol{NS}$, which are stored internally for reproducibility during backpropagation.

\textit{\underline{DBP Parallelism.}} Although Algorithm~\ref{alg:DiffGrad_forward} is presented for a single purification path, \sysname{} natively supports parallel purification of $N$ stochastic copies to potentially obtain a significant speedup when computing higher-quality EOT gradients---see~\S\ref{subsec:precise_grads}. For simplicity, we abstracted away the batch logic in the pseudo-code. In practice, prior to line 1, the input $\vx$ is replicated $N$ times ($N$ is an additional argument that can be provided to Algorithm~\ref{alg:DiffGrad_forward}). Line 4 returns $N$ random seeds, and the resulting sampler $\mathbf{NS}$ manages $N$ reproducible random paths. Line 7 draws $N$ distinct noise samples to generate $N$ diffused versions of $\vx$, and all subsequent steps operate copy-wise in parallel across these $N$ instances.

\paragraph{Backpropagation.} \sysname{}'s backpropagation logic is in Algorithm~\ref{alg:DiffGrad_backward}. Similar to earlier, red text refers to operations that deviate from traditional backpropagation, while blue lines are optional (guidance-related). In addition to the usual gradient $\boldsymbol{grad}$ w.r.t. $\hat{\vx}(0)$, the inputs include multiple parameters normally exclusive to forward propagation, as they are required to re-calculate the dependencies. Additionally, the algorithm accepts the saved state $\boldsymbol{S}$, and the same noise sampler $\boldsymbol{NS}$ to retrieve the stochastic path of the forward propagation. Before providing details, we note that by definition $\forall \boldsymbol{A},\boldsymbol{B}\in \mathbb{R}^d$, it holds that:
\begin{equation*}
    \langle \boldsymbol{A}, \boldsymbol{B} \rangle = \displaystyle\sum_d \boldsymbol{A} \odot \boldsymbol{B}
\end{equation*}
where $\odot$ denotes the element-wise product. Therefore, in order to calculate the gradients w.r.t. $\hat{\vx}(t)$ and $\boldsymbol{guide}$ as described in \hyperref[eq:recurse_grad]{\textit{eq.~(6)}} and \hyperref[eq:guide_grad_0]{\textit{eq.~(9)}}, we may define an objective at each step $t$ as:
\begin{equation}
    {Obj}_t = \displaystyle\sum_d (\hat{\vx}(t+dt) \odot \nabla_{\hat{\vx}(t+dt)} {F})
\label{eq:obj_t}
\end{equation}
and take its gradient w.r.t. the two elements of interest above, which explains the steps in our pseudo-code in
Algorithm~\ref{alg:DiffGrad_backward}.

\ActivateWarningFilters[fontsz]
\tikzmark{left}
\begin{algorithm}
\caption{Differentiable Purification with \textit{\sysname{}} — Backpropagation}
\label{alg:DiffGrad_backward}
\begin{algorithmic}[1] 
    \REQUIRE Loss gradient $\boldsymbol{grad}$ w.r.t $\hat{\vx}_0$, {\color{red}Sample $\vx$, Score model $\ws_\theta$, Optimal diffusion time $t^*$, step size $dt$, Noise scheduler $\beta$, Reverse diffusion function $\mathbf{calc\_dx}$, State $\mathbf{S} = \{\hat{\vx}(dt*i) | i \in [\![ |\frac{t^*}{dt}| ]\!] \}$, Noise sampler $\mathbf{NS}$,} {\color{blue}Auxiliary guidance input $\mathbf{guide}$, Guidance function $\mathbf{g_{fn}}$, Guidance scale $s$}

    \STATE $steps \gets \left|\frac{t^*}{dt}\right|$ 
    \STATE {\color{blue}$\mathbf{g\_grad} \gets \boldsymbol{0}$} \CCOMMENT[blue]{Init. gradient w.r.t guidance input}
    \FOR{$i \gets 1, 2, ..., steps$}
        \STATE {\color{red}$\hat{\vx} \gets \mathbf{S}[i]$} \CCOMMENT[red]{Set $\hat{\vx} = \hat{\vx}(t)$}
        \STATE {\color{red}$\mathbf{step\_noise} \gets \mathbf{NS.sample}(i)$} \CCOMMENT[red]{Retrieve noise for step $i$}
        \STATE {\color{red}$\mathbf{enable\_dependencies}()$}
        \STATE {\color{red}$d\hat{\vx} \gets \mathbf{calc\_dx}(\hat{\vx}, \ws_\theta, i, dt, \beta,$}
        \STATE \hspace{2.5em}{\color{red}$\mathbf{step\_noise}$, \color{blue}$\mathbf{g_{fn}}, s,$ $\mathbf{guide}$})
        \STATE {\color{red}$\hat{\vx}_{+dt} \gets \hat{\vx} + d\hat{\vx}$} \CCOMMENT[red]{$\hat{\vx}_{+dt}=\hat{\vx}(t+dt)$}
        \STATE ${Obj}_t \gets \sum (\hat{\vx}_{+dt} \odot \boldsymbol{grad})$ \CCOMMENT[black]{Objective due to \hyperref[eq:obj_t]{\textit{eq.~(12)}}}
        \STATE {\color{red}$\mathbf{disable\_dependencies}()$}
        \STATE $\boldsymbol{grad} \gets \nabla_{\hat{\vx}} {Obj}_t$ \CCOMMENT[black]{Update gradient w.r.t $\hat{\vx}(t)$ (\hyperref[eq:recurse_grad]{\textit{eq.~(6)}})}
        \STATE {\color{blue}$\mathbf{g\_grad} \gets \mathbf{g\_grad} + \nabla_{\mathbf{guide}} {Obj}_t$} \CCOMMENT[blue]{Update $\mathbf{guide}$ gradient (\hyperref[eq:guide_grad]{\textit{eq.~(10)}})}
    \ENDFOR
    \STATE $\alpha \gets \mathbf{calc\_alpha}(\beta)$
    \STATE $\boldsymbol{grad} \gets \boldsymbol{grad} * \sqrt{\alpha(t^*)}$ \CCOMMENT[black]{Loss gradient w.r.t $\vx$ (\hyperref[eq:closed]{\textit{eq.~(2)}})}
    \STATE {\color{blue}$\mathbf{g\_grad} \gets \nabla_{\vx} \sum (\mathbf{guide} \odot \mathbf{g\_grad})$} \CCOMMENT[blue]{Guidance gradient w.r.t $\vx$ (\hyperref[eq:guide_grad_x]{\textit{eq.~(11)}})}
    \STATE {\color{blue}$\boldsymbol{grad} \gets \boldsymbol{grad} + \mathbf{g\_grad}$} \CCOMMENT[blue]{Merge loss and guidance gradients}
    \RETURN $\boldsymbol{grad}$
\end{algorithmic}
\end{algorithm}
\DeactivateWarningFilters[fontsz]

The procedure begins by creating a variable $\boldsymbol{g\_grad}$ and setting it to $\boldsymbol{0}$ (line 2). This will later be used to store the guidance gradients (see \aref{app:guidance}). For each time step $t$ (i.e., step $i$), starting from $t^\prime=-dt$ ($i=1$), the process (lines 3-14) first retrieves $\hat{\vx}(t)$ from the saved state $S$ (line 4) and the corresponding random noise for that step used during forward propagation (line 5) and computes $\hat{\vx}(t+dt)$, denoted as $\hat{\vx}_{+dt}$ (lines 7-9). Importantly, these computations are performed while storing graph dependencies (enabled on line 6 and re-disabled on line 11 to restore the normal execution state). Specifically, during the first step, we calculate $\hat{\vx}(0)$ from $\hat{\vx}(-dt)$. Afterward, we compute the objective ${Obj}_t$ (line 10) following \hyperref[eq:obj_t]{\textit{eq.~(12)}} that allows us to back-propagate the gradient from $\hat{\vx}(0)$ to $\hat{\vx}(-dt)$ and $\boldsymbol{guide}$ using the stored dependencies, as per \hyperref[eq:recurse_grad]{\textit{eq.~(6)}} and \hyperref[eq:guide_grad_0]{\textit{eq.~(9)}}. $\boldsymbol{grad}$ is then updated to hold the gradient of the loss function w.r.t. $\hat{\vx}(-dt)$ as desired (line 12), and the gradient of $\boldsymbol{guide}$ due to this guidance path (i.e., from $\boldsymbol{guide}$ to the loss due to $\boldsymbol{guide}$ participating directly in the calculation of $\hat{\vx}(t+dt)$--- see \aref{app:guidance}) is added to $\boldsymbol{g\_grad}$ (line 13). This process repeats until $\boldsymbol{grad}$ finally holds the gradient w.r.t. $\hat{\vx}(t^*)$ and $\boldsymbol{g\_grad}$ holds the sum of gradients due to all guidance paths w.r.t. $\boldsymbol{guide}$ (\hyperref[eq:guide_grad]{\textit{eq.~(10)}}). Note that after the required gradients w.r.t. $\hat{\vx}(t)$ and $\boldsymbol{guide}$ are obtained at each step, the dependencies are no longer needed and can be discarded. This is where our approach differs from traditional backpropagation algorithms, enabling memory-efficient gradient calculations (at the cost of an additional forward propagation in total). At this point (line 14), we have the gradient $\nabla_{\hat{\vx}(t^*)} F$ and all is required is to use it to calculate $\nabla_{\vx} F$, which is trivial due to the chain rule since the closed-form solution for $\hat{\vx}(t^*) \equiv \vx(t^*)$ from \hyperref[eq:closed]{\textit{eq.~(2)}} indicates that this is equivalent to $\nabla_{\vx} F=\sqrt{\alpha(t^*)}*(\nabla_{\vx(t^*)} F)$ as we compute on line 16. We then calculate the guidance paths' gradient w.r.t $\vx$ following \hyperref[eq:guide_grad_x]{\textit{eq.~(11)}} on line 17. Finally, we sum both components, returning the precise full gradient w.r.t. $\vx$.

\textit{\underline{DBP Parallelism.}} Algorithm~\ref{alg:DiffGrad_backward} is written for a single purified copy but in practice operates over a batch of $N$ stochastic instances like Algorithm~\ref{alg:DiffGrad_forward}. As the forward pass stores $N$ trajectories, the backward pass propagates gradients independently for each. The operations in the pseudo-code are thus applied copy-wise in parallel across all $N$ instances. Finally, after line 18, the $N$ gradients are averaged, yielding the EOT gradient.

\subsection{Verifying the Correctness of \sysname{}}
\label{app:manual_verification}
To ensure the correctness of \sysname{}, all that is required is to verify that each $\hat{\vx}(t+dt)$ computed during backpropagation (line 9 in Algorithm~\ref{alg:DiffGrad_backward}) exactly matches the corresponding forward-computed $\hat{\vx}(t+dt)$ (line 14 in Algorithm~\ref{alg:DiffGrad_forward}). This equality guarantees that the reconstructed computation graph faithfully mirrors the one produced by standard autodiff engines during their normal (non-checkpointed) operation, ensuring exact gradient recovery since we use them to perform the necessary backpropagation between each $\hat{\vx}(t+dt)$ and $\hat{\vx}(t)$.

Since our forward pass explicitly stores all intermediate states in $\boldsymbol{S}$, we compare each recomputed $\hat{\vx}(t+dt)$ with its forward counterpart during backpropagation. An exact match confirms that the system is computing precise gradients. We manually validated this for all timesteps.

For guidance gradients, correctness follows from the derivations in \aref{app:guidance}. Once incorporated, the same matching procedure ensures their correctness as well.

\section{Additional Details on Systems \& Models}
\label{app:systems}
WideResNet-28-10 and WideResNet-70-16~\cite{zagoruyko2016wide} are used for \textit{\textbf{CIFAR-10}}, and ResNet-50~\cite{he2016deep}, WideResNet-50-2, and DeiT-S~\cite{dosovitskiy2020image} for \textit{\textbf{ImageNet}}, similar to~\cite{diffpure,kang2024diffattack}. For \textbf{VP-SDE} \textit{DBP} (\textit{DiffPure})~\cite{diffpure}, the \textit{DM}s~\cite{DBLP:conf/nips/DhariwalN21,dmint} are those from the original work. We also experiment with the \textit{Guided}-\textbf{DDPM} (see~\S\ref{subsec:precise_grads}), \textit{GDMP}~\cite{gmdp}, due to its \textit{SOTA} robustness, using the author-evaluated \textit{DM}s~\cite{DBLP:conf/nips/DhariwalN21,ho2020denoising}. The settings match the original optimal setup~\cite{gmdp,diffpure}: For \textit{Diffpure},  $t^*\!\!=\!\!0.1$ for \textbf{\textbf{\textit{CIFAR-10}}} and $t^*\!\!=\!\!0.15$ for \textbf{\textit{ImageNet}}. For \textit{GDMP}, a \textbf{\textbf{\textit{CIFAR-10}}} sample is purified $m\!\!=\!\!4$ times, with each iteration running for $36$ steps ($t^*\!\!=\!\!0.036$), using \textit{MSE} guidance~\cite{gmdp}. \textit{\textbf{ImageNet}} uses $45$ steps ($m\!\!=\!\!1$) under \textbf{DDPM}-acceleration~\cite{DBLP:conf/nips/DhariwalN21} with \textit{SSIM} guidance~\cite{wang2004image}.

\section{One-Shot \textit{DBP} Baseline Comparisons Against \sysname{}'s Accurate Gradients}
\label{app:one_shot}
As noted in~\S\ref{sec:theory}, \textit{DBP}'s robustness stems from two sources: inaccurate gradients and improper evaluation. As our work offers enhancements on both fronts, we evaluate each factor separately. Here, we isolate the gradient issue by re-running prior experiments under the same one-shot evaluation protocol, but with accurate gradients via our \textbf{DiffGrad} module and compare the results to those from the literature. We restrict our attacks to $10$ optimization steps (with \textit{AA}-$\ell_\infty$), while prior works often use up to $100$, giving them a clear advantage. We evaluate on \textbf{\textit{CIFAR-10}} with WideResNet-28-10, using $N=128$ EOT samples as justified in~\S\ref{subsec:precise_grads}. For papers reporting several results, we chose their best.

\underline{\textbf{Baselines.}} To demonstrate the efficacy of exact full gradients, we compare our \textbf{DiffGrad} module against prior gradient approaches. For \textit{DiffPure}, the \textbf{adjoint} method was originally used~\cite{diffpure}, while \textit{GDMP} was initially evaluated using \textbf{BPDA}~\cite{obfuscated} and a \textbf{blind} variant (ignoring the defense entirely) in \citet{gmdp}. These approximations were later criticized for poor attack performance~\cite{diffhammer,liu2024towards,kang2024diffattack,lee2023robust}. More recently,~\citet{lee2023robust} proposed the \textbf{surrogate} process to approximate the gradients, performing the reverse pass with fewer steps during backpropagation to reduce memory usage, enabling approximate gradients via standard autodiff tools. \textbf{DiffAttack}~\cite{kang2024diffattack}, \textbf{DiffHammer}~\cite{diffhammer}, and~\citet{liu2024towards} proposed checkpointing for memory-efficient full-gradient backpropagation. However, \textbf{DiffHammer} avoids the standard one-shot evaluation (1-evaluation) protocol and reports no results under it, while \textbf{DiffAttack} does not evaluate \textit{GDMP} and continues to use the \textbf{adjoint} method for \textit{DiffPure}. We focus here on existing results under the 1-evaluation protocol and defer the discussion of these works' conceptual issues (see~\S\ref{sec:intro}) to \S\ref{subsec:exp_mv}.

\noindent
\underline{\textbf{Metric.}} Following prior works, we report \textit{robust accuracy} (\textit{Rob-Acc}): the fraction of samples correctly classified after the attack completes and the final adversarial example is purified and evaluated (once).

\underline{\textbf{Results.}} Table~\ref{tab:first_exp} shows our comparison. All methods achieve similar clean accuracy (Cl-Acc) without attacks, with minor variation due to sample selection. Thus, robust accuracy (Rob-Acc) differences reflect the effect of gradient methods. \textbf{Full} denotes the standard $AA$ attack that uses the full exact gradients (i.e., via checkpointing).

\begin{wraptable}{r}{0.65\textwidth}
    \vspace{-4mm}
    \centering
    \caption{One-shot \textit{AA}-$\ell_{\infty}$ comparison on \textbf{\textbf{\textit{CIFAR-10}}} ($\epsilon_{\infty}\!\!=\!\!8/255$). $\dag$ indicates strategy is \textit{PGD}.}
    \label{tab:single_cifar_auto}
\resizebox{0.95\hsize}{1.7cm}{
    \begin{tabular}{ccl|cc}
    \toprule
    Models &  Pur. & Gradient Method & Cl-Acc \% & Rob-Acc \%\\
    \midrule
     \multirow{9}{*}{WideResNet-28-10} & \multirow{5}{*}{\textit{DiffPure}~\cite{diffpure}} & {\textbf{Adjoint} (\citet{diffpure})} & 89.02 & 70.64 \\
         & & \textbf{DiffAttack} (\citet{kang2024diffattack})  & 89.02 & 46.88 \\
         & & \textbf{Surrogate} (\citet{lee2023robust})$\dag$ &  90.07 & 48.28  \\
         & & \textbf{Full} (\citet{liu2024towards}) &  89.26 & 62.11  \\
          & & \textbf{Full-}\sysname{} (Ours) & 89.46 & \textbf{48.05} \\
    \cline{2-5}
      & \multirow{4}{*}{\textit{GDMP}~\cite{gmdp}} & \textbf{Blind} (\citet{gmdp}) & 93.50 & 90.06 \\
          & & \textbf{BPDA} (\citet{lee2023robust}) &   89.96 &  75.59 \\
          & & \textbf{Surrogate} (\citet{lee2023robust})$\dag$ &   89.96 &  24.53 \\
          & & \textbf{Full-}\sysname{} (Ours) & 93.36 & \textbf{19.53} \\
    \bottomrule
    \end{tabular}
}
\label{tab:first_exp}
\end{wraptable}

Our approach significantly outperforms gradient approximations such as \textbf{Adjoint}, \textbf{Blind}, and \textbf{BPDA}, reaffirming their known weaknesses. More notably, despite using only $10$ optimization steps, our method reduces \textit{DiffPure}'s Rob-Acc by $14.06\%$ compared to~\citet{liu2024towards}, who also use exact gradients, confirming their backpropagation mismatches (see~\S\ref{subsec:precise_grads}). 

While \textbf{DiffAttack} remains slightly stronger, the difference is extremely small ($1.4\%$) and can be safely attributed to differing evaluation samples. Importantly, the original gap reported in~\citet{kang2024diffattack} between \textbf{DiffAttack} and the standard \textit{AA}-$\ell_\infty$ dropped by $23.76\%$, undermining its claimed advantage due to the per-step deviated reconstruction losses and aligning with our theoretical findings in~\S\ref{subsec:path}. As \textbf{DiffAttack} involves broader architectural changes beyond gradient logic, we provide detailed comparisons in~\S\ref{subsec:exp_mv}, where we clearly showcase its inferiority.

Our method also slightly outperforms \textbf{Surrogate} for \textit{DiffPure} ($0.23\%$), but we caution against overinterpreting this small gap: under the improper one-shot evaluation, several purification paths can still cause misclassification even if the majority yield correct labels. As such, good approximation methods like \textbf{Surrogate} may appear closer in performance than they truly are. To verify the superiority of our \sysname{}'s full gradients over the \textbf{surrogate} approximation against \textit{DiffPure}, we thus further compare the two under the realistic \textit{MV} protocol studied in~\S\ref{subsec:exp_mv}, executing \textit{AA}-$\ell_\infty$ with the \textbf{surrogate} process to obtain the gradients when attacking the same \textbf{\textit{CIFAR-10}} classifier considered in~\S\ref{subsec:exp_mv} (i.e., WideResNet-70-16) and using the same number of samples ($N=10$) over which the majority vote is taken. We find that the \textbf{surrogate} process brings MV.Rob to $43.75\%$ only, whereas our \textbf{Full}-\sysname{} lowers this number to $39.45\%$, achieving a considerable improvement of $4.3\%$ and unequivocally proving the advantage of exact gradient computations. Finally, for \textit{GDMP}, our method outperforms \textbf{Surrogate} by $5\%$ even in the one-shot evaluation protocol (see Table~\ref{tab:first_exp}), largely due to our incorporation of guidance gradients—entirely absent in all prior approaches—highlighting the unique strength of \textbf{DiffGrad}.

Note that the \textbf{surrogate} method shortens the trajectory and backpropagates through this shorter horizon but retains the full graph over that path. Hence, one may argue that the attack success degradation is an acceptable tradeoff when using the \textbf{surrogate} method given the potential speedup. Yet, in \textit{DBP}, memory---not FLOPs---is the bottleneck; Checkpointing (e.g., our \textbf{DiffGrad}) recomputes individual path segments, yielding exact gradients with lower peak memory. The \textbf{surrogate} may improve speed but not memory pressure, and its truncation can miss dependencies, weakening gradients as shown above. In practice, attackers precompute adversarial examples offline since \textit{DBP}’s latency makes real-time adversarial generation infeasible either way, so runtime savings alone do not translate to a practical advantage.

All in all, the results demonstrate the fragility of \textit{DBP} in the face of accurate gradients and highlight the issues in previous works' backpropagation. Nonetheless, this one-shot evaluation protocol remains problematic, as previously stated, leading to an inflated robustness estimate. In fact, under more realistic settings (see~\S\ref{subsec:exp_mv}), we demonstrate that this gradient-based attack almost entirely defeats \textit{DBP} when only one sample is used to predict the label (i.e., Wor.Rob).

\ActivateWarningFilters[hpr]
\section{Evaluations with~\citet{liu2024towards}'s Fixed \textit{AutoAttack}}
\DeactivateWarningFilters[hpr]
\label{app:liu} 
\citet{liu2024towards}, like \textbf{DiffHammer}~\cite{diffhammer} and our own analysis, note that evaluating a single purification at attack termination inflates robustness scores. \citet{liu2024towards} address this by evaluating $20$ replicas of the final \textit{AE}, declaring success if any is misclassified. While similar in spirit to the Wor.Rob metric we consider \begin{wraptable}{r}{0.7\textwidth}
\vspace{-2mm}
    \centering
    \caption{\textit{AA}-$\ell_{\infty}$ comparison on \textbf{\textbf{\textit{CIFAR-10}} under~\citet{liu2024towards}'s protocol (i.e., Fixed \textit{AutoAttack})} ($\epsilon_{\infty}\!\!=\!\!8/255$).}
    \label{tab:20_eval}
\resizebox{0.95\hsize}{1.2cm}{
    \begin{tabular}{ccl|cc}
    \toprule
    Models &  Pur. & Gradient Method & Cl-Acc \% & Rob-Acc \%\\
    \midrule
     \multirow{4}{*}{WideResNet-28-10} & \multirow{2}{*}{\textit{DiffPure}~\cite{diffpure}} & \textbf{Full} (\citet{liu2024towards}) &   89.26 & 56.25 \\
          & & \textbf{Full-}\sysname{} & 89.46 & \textbf{30.86} \\
    \cline{2-5}
      & \multirow{2}{*}{\textit{GDMP}~\cite{gmdp}} & \textbf{Full}  (\citet{liu2024towards}) &    91.80 &  40.97  \\
          & & \textbf{Full-}\sysname{} & 93.36 & \textbf{10.55} \\
    \hline
    \multirow{2}{*}{WideResNet-70-16} & \multirow{1}{*}{\textit{DiffPure}~\cite{diffpure}} & \textbf{Full-}\sysname{} & 89.06 & \textbf{35.16}  \\
    \cline{2-5}
    & \textit{GDMP}~\cite{gmdp} & \textbf{Full-}\sysname{} & 91.8 & \textbf{8.59} \\
    \bottomrule
    \end{tabular}
}
\vspace{-4mm}
\end{wraptable} 
(see~\S\ref{subsec:flaws}), this protocol is more limited: it evaluates only at the final step, whereas Wor.Rob evaluates $N$ copies at each attack iteration. Accordingly,~\citet{liu2024towards} group their method with one-shot evaluations, providing a slightly more realistic assessment that leads to results similar to those attained via \textit{PGD}~\cite{pgd}.

We replicate their setup using \textbf{DiffGrad} (Table~\ref{tab:20_eval}), running $20$ iterations with $N=10$ EOT samples per step and evaluating $20$ final replicas. Under this protocol, our improvements are stark. On WideResNet-28-10, we reduce Rob-Acc by $25.39\%$ and $30.42\%$ for \textit{DiffPure} and \textit{GDMP}, bringing final accuracy to $30.86\%$ and $10.55\%$, respectively. These results confirm the superiority of \textbf{DiffGrad}'s gradients and expose \textit{DBP}’s realistic vulnerability. We observe similar results on WideResNet-70-16 (not evaluated in~\cite{liu2024towards}).

\section{Ablation Study with Different Numbers of Samples for Label Prediction}
\label{app:ablation}
To assess the impact of the number of evaluation samples $N$, we test both single-purification and majority-vote (MV) settings with $N \in \{1, 10, 128\}$. Our goal is to highlight the brittleness of \textit{DBP}'s standard deployment, which classifies based on a single purified copy. As explained in~\S\ref{subsec:flaws}, in stateless setups (e.g., phishing, spam, CSAM), adversaries can resubmit identical queries indefinitely. Since randomized defenses like \textit{DBP} can fail along certain stochastic paths, any non-negligible misclassification probability compounds with repeated attempts---see~\S\ref{subsec:flaws}. Although \textit{DBP} assumes this probability is negligible, our proof in~\S\ref{subsec:path}, results in~\S\ref{sec:exp}, and the ablation experiments here contradict this. Furthermore, we wish to showcase the advantages of our proposed majority-vote \textit{(MV)} setting that strives to mitigate this vulnerability by, instead, classifying based on expectation of the randomized defense.

We evaluate \textit{DiffPure} and \textit{GDMP} on \textbf{\textit{CIFAR-10}} using WideResNet-70-16 and (our) \textit{AA}-$\ell_\infty$ (with $\epsilon=8/255$ and $100$ iterations). For $N=1$ and $N=10$, we use $10$ EOT samples; for $N=128$, we reuse the same $128$ samples for both EOT and evaluation. This improves gradient quality, thus strengthening attacks, yet still leads to higher \textit{MV} robustness, proving the superiority of our proposed method despite the use of more accurate gradients.

For Wor.Rob, clean accuracy (Cl-Acc) remains constant across all $N$ values since we always compute it using a single purified copy ($N=1$), independent of the number $N$ of samples used in the attack. This reflects the actual standard single-purification deployment of \textit{DBP}, where the defender classifies a single output, while the attacker may retry multiple times. Hence, for Rob-Acc we consider a batch of multiple samples (the corresponding $N$ in that row) and then declare attack success if a single misclassification occurs, revealing the gap between measured and effective robustness. In contrast, MV.Rob clean accuracy varies with $N$, as the prediction always aggregates over $N$ purified samples, consistent with our proposed deployment.

\begin{wraptable}{r}{0.6\textwidth}
    \vspace{-2mm}
    \centering
    \caption{\textit{AA}-$\ell_\infty$ Performance under various evaluation sample counts}
    \label{tab:batch_auto_128}
\resizebox{0.95\hsize}{1.5cm}{
    \begin{tabular}{cc|c|c}
    \toprule
    Pur. & \#Samples & \begin{tabular}{cc}
         \multicolumn{2}{c}{Wor.Rob \%}  \\
        Cl-Acc & Rob-Acc
    \end{tabular} & \begin{tabular}{cc}
         \multicolumn{2}{c}{MV.Rob \%}  \\
        Cl-Acc & Rob-Acc
    \end{tabular}\\
    \midrule
        \multirow{3}{*}{\textit{DiffPure}~\cite{diffpure}} & 
         {\textbf{N=1}} & \begin{tabular}{lc}
        \multirow{3}{*}{89.06} & \phantom{0000000}35.16
    \end{tabular} & \begin{tabular}{lc}
        89.06 & \phantom{0000000}35.16
    \end{tabular} \\
          & \textbf{N=10} & \begin{tabular}{lc}
         & \phantom{000000000000}17.19
    \end{tabular} & \begin{tabular}{lc}
       91.02 & \phantom{0000000}39.45
    \end{tabular} \\
          & \textbf{N=128} & \begin{tabular}{lc}
          & \phantom{000000000000}17.58
    \end{tabular} & \begin{tabular}{lc}
        92.19  & \phantom{0000000}47.72
    \end{tabular} \\
    \cline{1-4}
    \multirow{3}{*}{\textit{GDMP}~\cite{gmdp}} & 
         {\textbf{N=1}} & \begin{tabular}{lc}
        \multirow{3}{*}{91.8} & \phantom{000000000}8.59
    \end{tabular} & \begin{tabular}{lc}
        91.8 & \phantom{00000000}8.59
    \end{tabular} \\
          & \textbf{N=10} & \begin{tabular}{lc}
         & \phantom{0000000000000}7.03
    \end{tabular} & \begin{tabular}{lc}
       92.19  & \phantom{0000000}16.8
    \end{tabular} \\
          & \textbf{N=128} & \begin{tabular}{lc}
         & \phantom{0000000000000}5.47
    \end{tabular} & \begin{tabular}{lc}
        \phantom{0}92.19 & \phantom{0000000}32.81
    \end{tabular} \\
    \bottomrule
    \end{tabular}
}
\vspace{-4mm}
\end{wraptable}

Table~\ref{tab:batch_auto_128} confirms that Wor.Rob consistently declines with $N$, revealing the illusion of robustness under single evaluation. For instance, \textit{DiffPure} shows a drop from $35.16\%$ (at $N=1$) to $17.58\%$ at $N=128$ (far lower than the inflated numbers reported in previous studies---see \aref{app:one_shot}), confirming that many stochastic paths yield incorrect predictions, as our properly implemented gradient-based attack (see~\S\ref{subsec:precise_grads}) lowers the expected classification confidence, making such failure modes more likely. For \textit{GDMP}, we observe a similar trend.

In contrast, MV.Rob improves with $N$, rising from $35.16\%$ to $47.72\%$ for \textit{DiffPure}, and from $8.59\%$ to $32.81\%$ for \textit{GDMP}. This affirms \textit{MV} as a more stable and accurate evaluation method that must be adopted as the de facto standard for \textit{DBP} evaluations in the future. 

Yet, despite \textit{MV}’s benefits, our gradient-based \textit{AA}-$\ell_\infty$ attack still degrades robustness, which never exceeds $50\%$, validating our theoretical finding from~\S\ref{subsec:path} that such attacks repurpose \textit{DBP} into an adversarial distribution generator.

\paragraph{Computational Cost.} While larger $N$ increases computational overhead, we identify $N=128$ as the max batch size fitting in a single A100 GPU run requiring $\sim 26.53s$ for inference, with $N=10$ offering a practical middle ground ($6.54s$ vs. $5.29s$ for $N=1$). Hence, for practicality and if throughput is critical, one should opt for $N=10$. However, in security-critical systems where latency is not crucial, a larger $N$ is favorable. Yet, additional tests (not shown) suggest MV.Rob plateaus near $N=128$. Note that these latencies refer to purification inference alone (classification excluded). During attacks (not standard deployment), there is also the cost of backpropagation. When accounting for the cost of classification and backpropagation, the latency becomes $\sim 17$s for $N=1$ and $N=10$ compared to $\sim 53$s for $N=128$. As explained in~\S\ref{subsec:precise_grads}, our design allows purifying multiple stochastic copies of the same sample in one step, in contrast to previous work. This reduces the runtime per EOT gradient from up to $N\times 17$s---where $N$ stochastic purifications are run serially—to just $T_N$s, where $T_N$ is the latency incurred by purifying $N$ samples in a batch, yielding a speedup of up to $N\times17/T_N$ for single-sample attacks, which amounts to $41.06\times$ for $N=128$ allowing for a considerable speedup when the objective is to utilize many EOT samples to obtain accurate gradient estimates. For batch attacks, prior methods purify one stochastic copy per sample per EOT step and require all samples to converge before proceeding to the next batch of inputs, blocking early termination. In contrast, \sysname{} purifies $N$ copies of a single sample in parallel, allowing samples to terminate independently. This greatly improves overall throughput by freeing compute sooner, especially when convergence varies across inputs.

\underline{\textit{\textbf{ImageNet} timings:}} While we limit our ablation studies to \textbf{\textit{CIFAR-10}} for feasibility, we provide \textbf{\textit{ImageNet}} timings for completeness. At \textbf{\textit{ImageNet}} resolution on a 40 GB A100 GPU, \textit{DBP} inference (forward propagation only) requires $\sim 7$s for $N=1$ and $\sim 25$s for $N=8$. A single attack iteration (i.e., forward and backward propagation) lasts $\sim 16$s ($N=1$) and $\sim 75$s ($N=8$). Thus, batching EOT copies yields a significant speedup over serial purification for both attacks and inference, but models and activations dominate cost at this scale, so batching returns diminish faster than on \textbf{\textit{CIFAR-10}}.

\paragraph{Recommendations for Choosing The Batch Size $N$.} We found $N=128$ for \textit{\textbf{CIFAR-10}} and $N=8$ for \textbf{\textit{ImageNet}} to be the largest feasible batch sizes to fit into the 40 GB A100 GPU memory, making larger batches impractical for large-scale evaluations. That said, control experiments with larger $N$ values for both datasets (split into batches) showed negligible robustness or attack performance gains, indicating saturation. 

Following the above discussion, we recommend $N=10$ for \textit{\textbf{CIFAR-10}} as a practical default and $N=64-128$ for security-critical inference (\textit{MV}) and highly accurate gradients when evaluating new attack methods such as our \textit{LF} strategy---see~\S\ref{subsec:low_freq}. For \textit{\textbf{ImageNet}}, we recommend $N=8$ for similar reasons; larger $N$ values remain preferable when added latency is acceptable.

\section{Details on Our Low-Frequency (\textit{LF}) Adversarial Optimization Strategy}
\subsection{Understanding Optimizable Filters}
\label{app:low_freq}

In practice, \textit{OFs} extend an advanced class of filters, namely \textbf{\textit{guided}} filters, that improve upon the basic filters discussed in~\S\ref{subsec:low_freq}. \textit{\textbf{Guided}} filters employ additional per-pixel \textit{color kernels} that modulate the distortion at critical points: Since filters interpolate each pixel with its neighbors, they are destructive at edges (intersections between different objects in the image), while the values of non-edge pixels are similar to their neighbors, making such operations of little effect on them. Depending on a permissiveness $\sigma$, \textit{\textbf{guided}} filters construct, \textbf{for each pixel {\small \mbox{$(i,\; j)$}}} a \textit{color kernel} $\text{\tiny \mbox{$\boldsymbol{\mathcal{C}}_{\boldsymbol{\vx},\sigma\!_{_c}}^{i,j}$}}$ of the same dimensionality {\small \mbox{$M \! {\times} \! N$}} as {\small \mbox{$\boldsymbol{\mathcal{K}}$}} that assigns a multiplier for each of {\small \mbox{$(i,\; j)$}}'s neighbors, that decays with the neighbor's difference in value from {\small \mbox{$(i,\; j)$}}'s. The output at {\small \mbox{$(i,\; j)$}} involves calculating the effective kernel ${\text{\tiny \mbox{$\boldsymbol{\mathcal{V}}_{\boldsymbol{\vx},\boldsymbol{\mathcal{K}},\sigma\!_{_c}}^{i,j}$}}} = \text{\tiny \mbox{$\boldsymbol{\mathcal{C}}_{\boldsymbol{\vx},\sigma\!_{_c}}^{i,j} \odot \boldsymbol{\mathcal{K}}$}}$ (normalized) which then multiplies {\small \mbox{$(i,\; j)$}}'s vicinity, taking the sum of this product. Thus, contributions from neighbors whose values differ significantly are diminished, better preserving information.

\textit{\textbf{Guided}} filters still employ the same {\small \mbox{$\boldsymbol{\mathcal{K}}$}} for all pixels, changing only the \textit{color kernel} that is computed similarly for all pixels. Thus, to incur sufficient changes, they would also require destructive parameters despite them still potentially performing better compared to their pristine counterparts. Their parameters are also predetermined,  making it impossible to optimize them for a specific purpose. The \textit{OFs} by~\citet{unmarker} build upon \textbf{\textit{guided}} filters but differ in two ways: First, instead of using the same {\small \mbox{$\boldsymbol{\mathcal{K}}$}}, they allow each pixel to have its own kernel {\small \mbox{$\boldsymbol{\mathcal{K}}^{i,j}$}} to better control the filtering effects at each point, ensuring visual constraints are enforced based on each pixel's visual importance. In this setting, {\small \mbox{$\boldsymbol{\mathcal{K^*}}$}} denotes the set including all the per-pixel kernels {\small \mbox{$\boldsymbol{\mathcal{K}}^{i,j}$}}. Second, the parameters {\small \mbox{$\boldsymbol{\theta}_{\boldsymbol{\mathcal{K}}^*}$}} of each filter are learnable using feedback from a perceptual metric (\textit{\textbf{lpips}})~\citep{lpips} that models the human vision, leading to an optimal assignment that ensures similarity while maximizing the destruction at visually non-critical regions. To further guarantee visual similarity, they also include \textit{color kernels} similar to \textit{\textbf{guided}} filters (see original paper for details~\citep{unmarker}).

\subsection{Attack Hyperparameters}
\label{app:cifar_network_unmarker}
Through experimentation, we found that the loss balancing constants $c=10^8$ for \textit{\textbf{CIFAR-10}} and $c=10^4$ for \textit{\textbf{ImageNet}} lead to the fastest convergence rates and selected these values accordingly (although other choices are also possible). \textbf{\textit{UnMarker}}'s filter network architecture for \textit{\textbf{ImageNet}} is identical to that from the original paper~\cite{unmarker}. For \textit{\textbf{CIFAR-10}}, since the images are much smaller, we found the original architecture unnecessarily costly and often prevents convergence since larger filters group pixels from distant regions together in this case, easily violating visual constraints upon each update, resulting in the \textit{\textbf{lpips}} condition being violated. Thus, we opt for a more compact network that includes filters with smaller dimensions, which was chosen based on similar considerations to~\cite{unmarker}, allowing us to explore several interpolation options. The chosen architecture for \textit{\textbf{CIFAR-10}} includes $4$ filters, whose dimensions are: $(5,5),\; (7,7),\; (5,5),\; (3,3)$. We use fixed learning rates of $0.008$ for the direct modifier $\boldsymbol{\delta}$ and $0.05$ for the filters' weights, optimized using $\boldsymbol{Adam}$. The remaining hyperparameters were left unchanged compared to~\cite{unmarker}.

\subsection{Pseudo-Code}
\label{app:pseudo_c}
The pseudo-code for our low-frequency (\textit{LF}) strategy (see~\S\ref{subsec:low_freq}) is in Algorithm~\ref{alg:Unmarker}. Importantly, as each {\small \mbox{$\boldsymbol{\mathcal{K}}_b^{i,j}$}}'s values should be non-negative and sum to $1$, the values for each such per-pixel kernel are effectively obtained by \textit{\textbf{softmax}}ing the learned weights. Initially, the modifier $\boldsymbol{\hat{\delta}}$ is initialized to $\boldsymbol{0}$ and the weights $\{\widehat{\boldsymbol{\theta}}_{\boldsymbol{\mathcal{K}}^*\!\!\!_{_b}}\}$ are selected s.t. the filters perform the identity function (line 1). As a result, the attack starts with $\vx_{adv}=\vx$ that is iteratively optimized. Similar to \textit{C\&W}~\citep{cw}, we directly optimize $\vx_{adv}$ (i.e., using the modifier $\boldsymbol{\hat{\delta}}$) in the $\boldsymbol{arctanh}$ space, meaning we transform the sample first to this space by applying $\boldsymbol{arctanh}$ (after scaling $\vx_{adv}$ to $\boldsymbol{arctanh}$'s valid range $[-1,\; 1]$) where $\boldsymbol{\hat{\delta}}$ is added and then restore the outcome to the original problem space (i.e., $[min\_val,\; max\_val]$, which is typically $[0,\; 1]$) via the $\boldsymbol{tanh}$ operation. Further details on this method and its benefits can be found in~\citep{cw}. All other steps correspond to the description brought in~\S\ref{subsec:low_freq}. Unlike~\S\ref{subsec:path}, we assume the classifier $\mathcal{M}$ outputs the logit vector rather than the probabilities (i.e., we omit the $\boldsymbol{softmax}$ layer over its output, which me or may not be re-introduced by the loss $\ell$), as is traditionally done for a variety of adversarial optimization strategies (e.g., \textit{C\&W}~\citep{cw}) to avoid gradient vanishing. We use the known \textbf{\textit{max-margin}} loss~\citep{cw}--- \mbox{\small$\ell(\boldsymbol{logits}, y) \!=\! \boldsymbol{logits}[,:y] -\! \mathrel{\raisebox{2pt}{$\underset{j\neq y}{max}$}} \{\boldsymbol{logits}[,:j]\}$}.

\ActivateWarningFilters[fontsz]
\begin{algorithm}
\caption{Low-Frequency (\textit{LF}) Adversarial Optimization}
\label{alg:Unmarker}
\begin{algorithmic}[1] 
    \REQUIRE Sample $\vx$, Model (classifier) $\mathcal{M}$, \textit{DBP} pipeline $D$, Loss function $\ell$, True label $y$ of $\vx$, Perceptual loss $\boldsymbol{lpips}$, $\boldsymbol{lpips}$ threshold $\tau_p$, Filter architecture $\text{\mbox{\tiny $\overset{B}{\underset{\boldsymbol{OF}}{\boldsymbol{\prod}}}$}}$, Balancing constant $c$, Iterations $max\_iters$, Success condition $\boldsymbol{Cond}$, Filter weights learning rate ${lr}_{_{\boldsymbol{OF}}}$, Modifier learning rate ${lr}_{\boldsymbol{\delta}}$, Number of purified copies $n$, Number of \textbf{\textit{EoT}} $eot\_steps$, Input range limits ($min\_val, max\_val$)

    \STATE $\{\widehat{\boldsymbol{\theta}}_{\boldsymbol{\mathcal{K}}^*\!\!\!_{_b}}\} \gets \boldsymbol{identity\_weights}(\text{\mbox{\tiny $\overset{B}{\underset{\boldsymbol{OF}}{\boldsymbol{\prod}}}$}})$, $\boldsymbol{\hat{\delta}} \gets \boldsymbol{0}$ \CCOMMENT[black]{Initialize attack parameters.}
    \STATE $\boldsymbol{Optim} \gets \boldsymbol{Adam}([\{\widehat{\boldsymbol{\theta}}_{\boldsymbol{\mathcal{K}}^*\!\!\!_{_b}}\}, \boldsymbol{\hat{\delta}}], [{lr}_{_{\boldsymbol{OF}}},{lr}_{\boldsymbol{\delta}}])$
    \STATE $\vx_{inv} \gets \boldsymbol{inv\_scale\_and\_arctanh}(\vx, min\_val, max\_val)$ \CCOMMENT[black]{Scale $\vx$ to $[-1, 1]$ and take $\boldsymbol{arctanh}$}

    \FOR{$i \gets 1$ \textbf{to} $max\_iters \cdot eot\_steps$}
        \STATE $\vx_{adv} \gets \text{\mbox{\tiny $\overset{B}{\underset{\boldsymbol{OF}}{\boldsymbol{\prod}}}$}}(\boldsymbol{tanh\_and\_scale}(\vx_{inv} + \boldsymbol{\hat{\delta}}, min\_val, max\_val))$
        \MULTILINECOMMENT[black]{Generate adversarial input using \\ new $\{\widehat{\boldsymbol{\theta}}_{\boldsymbol{\mathcal{K}}^*\!\!\!_{_b}}\}$ and $\boldsymbol{\hat{\delta}}$ via (\hyperref[eq:struct]{\textit{eq.~(7)}}) \\ scaled to $[min\_val, max\_val]$.}

        \STATE $dist \gets \boldsymbol{lpips}(\vx, \vx_{adv})$ \CCOMMENT[black]{Calculate perceptual distance.}
        \STATE ${\hat{\vx}_{adv}}^{0} \gets D(\boldsymbol{repeat}(\vx_{adv}, n))$ \CCOMMENT[black]{Get purified outputs.}
        \STATE $\boldsymbol{logits} \gets \mathcal{M}({\hat{\vx}_{adv}}^{0})$ \CCOMMENT[black]{Compute model output.}
        \IF{$\boldsymbol{Cond}(\boldsymbol{logits}, y)$ \AND $dist \leq \tau_p$}
            \RETURN $\boldsymbol{\vx_{adv}}$ \CCOMMENT[black]{Success. Return $\vx_{adv}$.}
        \ENDIF
        \STATE $Objective \gets \ell(\boldsymbol{logits}, y) + c \cdot max(dist - \tau_p, 0)$ \CCOMMENT[black]{Compute loss (\hyperref[eq:struct]{\textit{eq.~(7)}}).}
        \STATE $Objective.\boldsymbol{backward()}$ \CCOMMENT[black]{Get gradients for parameters.}
        \IF{$i \bmod eot\_steps = 0$}
            \STATE $\boldsymbol{Optim.step()}$ \CCOMMENT[black]{Update parameters.}
            \STATE $\boldsymbol{Optim.zero\_grad()}$ \CCOMMENT[black]{Reset gradients.}
        \ENDIF
    \ENDFOR
    \RETURN $\boldsymbol{\vx}$ \CCOMMENT[black]{Failure. Return original $\vx$.}
\end{algorithmic}
\end{algorithm}
\DeactivateWarningFilters[fontsz]

To average the gradients over multiple ($N$) paths as per the adaptive attack's requirements from~\S\ref{subsec:path}, we generate several purified copies by repeating the sample $\vx_{adv}$ under optimization $n$ times before feeding it into the \textit{DBP} pipeline (line 9). Here, $n$ corresponds to the maximum number of copies we can fit into the \textit{GPU}'s memory during a single run. However, as this $n$ may be smaller than the desired $N$ from~\S\ref{subsec:path} (i.e., number of \textit{EoT} samples) that allows us to sufficiently eliminate the error in the computed gradient, we use gradient accumulation by only making updates to the optimizable parameters (and then resetting their gradients) every $eot\_steps$ iterations (lines 16-19). By doing so, the effective number of used copies becomes $n*eot\_steps$, which can represent any $N$ of choice that is divisible by $n$. Note that if $n$ is not a divisor of $N$, we can always increase $N$ until this condition is met, as a larger $N$ can only enhance the accuracy). This also explains why the algorithm runs for $max\_iters*eot\_steps$ (line 5). 

Finally, the condition $\boldsymbol{Cond}$ captures the threat model (either \textit{SP} or \textit{MV}---see~\S\ref{sec:threat_model}): When the logits for the batch of $n$ copies are available together with the target label $y$, $\boldsymbol{Cond}$ outputs a success decision based on whether we seek misclassification for the majority of these purified copies or a single copy only. Note that, as explained in~\S\ref{subsec:exp_mv}, we take the majority vote over the maximum number of copies we can fit into the \textit{GPU} (i.e., $n$) for \textit{MV}. As this choice was only made for practical considerations, one may desire to experiment with different configurations wherein another number of copies is used. Yet, this is easily achievable by simply modifying $\boldsymbol{Cond}$ accordingly: For instance, we may augment it with a history that saves the output logits over all $eot\_steps$ (during which $\vx_{adv}$ is not updated). Then, the majority vote can be taken over all copies in this window. Note that by increasing the number of $eot\_steps$, we can use as many copies for the majority vote decision as desired in this case. That said, the attack will become significantly slower.

In addition to the precise gradient module \sysname{}, our \libname{} toolkit provides the implementation of our \textit{LF} strategy as well as various other common methods (e.g., \textit{AA} and \textit{StAdv}), to enable robust and reliable evaluations of \textit{DBP}. All strategies are optimized for performance to speed up computations via various techniques such as just-in-time (\textit{JIT}) compilation. Our code is available at \codeurl{}.

\section{Example Attack Images}
\label{app:attack_images}
In \aref{app:F1}-\ref{app:F10}, we provide a variety of successful attack images that cause misclassification in the rigorous \textit{MV} setting, generated using our low-frequency (\textit{LF}) strategy against all systems considered in~\S\ref{subsec:exp_lf}. For configurations that were also evaluated against \textit{AA}-$\ell_\infty$ under the \textbf{same} \textit{MV} setting (i.e., using the same sample counts $N$ as used for \textit{LF} in~\S\ref{subsec:exp_lf}), we include successful \textit{AE}s generated with this method for direct comparison. Specifically, for \textbf{\textit{ImageNet}}, both the \textit{LF} attack in~\S\ref{subsec:exp_lf} and the \textit{AA} attack in~\S\ref{subsec:exp_mv} were evaluated under \textit{MV} with $N=8$; hence, we include \textit{AA} samples for \textbf{\textit{ImageNet}} directly from~\S\ref{subsec:exp_mv}. For \textbf{\textit{CIFAR-10}}, the \textit{AA} experiments in~\S\ref{subsec:exp_mv} use the more permissive $N=10$ setting. Therefore, we instead draw samples from the corresponding $N=128$ \textit{AA} experiments reported in~\aref{app:ablation} to ensure a fair comparison under equal majority-vote conditions. Note that all samples are crafted using the parameters listed in~\S\ref{subsec:exp_mv} and~\S\ref{subsec:exp_lf}. That is, $\tau_p\!=\!0.05$ for \textit{LF} and $\epsilon_{\infty}\!=\!8/255$ for \textit{AA} against \textit{\textbf{CIFAR-10}} and $\epsilon_{\infty}\!=\!4/255$ against \textit{\textbf{ImageNet}}. 

For the configurations that were evaluated against both strategies, we provide two sets of samples: 1) Three triplets containing the original image, the \textit{AE} generated using \textit{AA}, and the \textit{AE} crafted using \textit{LF}. Importantly, all original samples in this set are inputs for which both methods can generate successful \textit{AE}s, and we provide these to allow for a direct comparison between the two strategies' output quality on a sample-by-sample basis. Yet, as \textit{AA} is inferior to our approach (\textit{LF}), resulting in the systems retaining robustness on many inputs for which it fails to generate successful \textit{AE}s under \textit{MV}, it is essential to inspect \textit{LF}'s outputs on such more challenging samples to demonstrate that it still preserves quality despite its ability to fool the target classifiers. Thus, we include a second set of 2) Three successful \textit{AE}s generated with \textit{LF} from inputs on which \textit{AA} fails under \textit{MV}. For the remaining configurations that were not evaluated against \textit{AA} under the same \textit{MV} sample counts from~\S\ref{subsec:exp_lf}, we provide six successful \textit{AE}s generated with \textit{LF}.

In \aref{app:F11}, we present attack images generated by the non-norm-bounded \textit{StAdv}~\cite{stadv} strategy under \textit{MV} (with the above sample counts $N$). This method has demonstrated superior performance to norm-based techniques in the past against \textit{DBP}~\cite{diffpure} even in the absence of the correct exact gradients, indicating it could be a viable attack strategy with our gradient computation fixes, thereby making our \textit{LF} approach unnecessary. Yet, previous evaluations only considered \textit{StAdv} against \textit{DBP} for \textit{\textbf{CIFAR-10}}~\cite{diffpure}. While we find \textit{StAdv} capable of defeating all systems (for both \textit{\textbf{CIFAR-10}} and \textit{\textbf{ImageNet}}), it leads to severe quality degradation when used to attack \textit{DBP}-protected classifiers for high-resolution inputs (i.e., \textbf{\textit{ImageNet}}), leaving the \textit{AE}s of no utility. Thus, we deem it unsuitable, excluding it from the main body of the paper accordingly. Further details are in \aref{app:F11}.

All samples below are originally (without adversarial perturbations) correctly classified.

\subsection{Attack Samples Generated Against \textbf{\textit{CIFAR-10}}'s WideResNet-70-16 with \textit{GDMP} Purification}
\label{app:F1}
\begin{figure}[H]
        \centering
        \begin{subfigure}[b]{0.32\textwidth}
                \includegraphics[width=\linewidth]{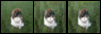}
        \end{subfigure}%
        \hspace{\fill}
        \begin{subfigure}[b]{0.32\textwidth}
                \includegraphics[width=\linewidth]{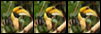}
        \end{subfigure}%
        \hspace{\fill}
        \begin{subfigure}[b]{0.32\textwidth}
                \includegraphics[width=\linewidth]{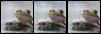}
        \end{subfigure}%
        \hspace{\fill}
        \caption{Successful attacks generated by \textit{LF} and \textit{AA}-$\ell_{\infty}$. Left -original image. Middle - \textit{AA}. Right - \textit{LF}.}\vspace{3mm}
        \begin{subfigure}[b]{0.32\textwidth}
                \includegraphics[width=\linewidth]{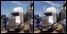}
        \end{subfigure}
        \hspace{\fill}
        \begin{subfigure}[b]{0.32\textwidth}
                \includegraphics[width=\linewidth]{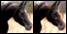}
        \end{subfigure}
        \hspace{\fill}
        \begin{subfigure}[b]{0.32\textwidth}
                \includegraphics[width=\linewidth]{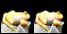}
        \end{subfigure}
        \caption{Successful \textit{LF} attacks on inputs for which \textit{AA}-$\ell_{\infty}$ fails. Left - original image. Right - \textit{LF}.}
\end{figure}

\subsection{Attack Samples Generated Against \textbf{\textit{CIFAR-10}}'s WideResNet-28-10 with \textit{GDMP} Purification}
\label{app:F2}
\emph{Note: While WideResNet-28-10 was not evaluated against \textit{AA} under \textit{MV} with $N=128$ in \aref{app:ablation}, we include \textit{AA} samples here to complement our CIFAR-10 analyses—our primary benchmark for \textit{AA}-based evaluation. Among the two main purification paradigms considered (\textit{GDMP} and \textit{DiffPure}), we randomly selected \textit{GDMP} for this illustrative example.}

\begin{figure}[H]
        \centering
        \begin{subfigure}[b]{0.32\textwidth}
                \includegraphics[width=\linewidth]{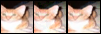}
        \end{subfigure}%
        \hspace{\fill}
        \begin{subfigure}[b]{0.32\textwidth}
                \includegraphics[width=\linewidth]{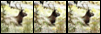}
        \end{subfigure}%
        \hspace{\fill}
        \begin{subfigure}[b]{0.32\textwidth}
                \includegraphics[width=\linewidth]{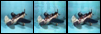}
        \end{subfigure}%
        \hspace{\fill}
        \caption{Successful attacks generated by \textit{LF} and \textit{AA}-$\ell_{\infty}$. Left -original image. Middle - \textit{AA}. Right - \textit{LF}.}\vspace{3mm}
        \begin{subfigure}[b]{0.32\textwidth}
                \includegraphics[width=\linewidth]{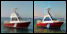}
        \end{subfigure}
        \hspace{\fill}
        \begin{subfigure}[b]{0.32\textwidth}
                \includegraphics[width=\linewidth]{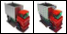}
        \end{subfigure}
        \hspace{\fill}
        \begin{subfigure}[b]{0.32\textwidth}
                \includegraphics[width=\linewidth]{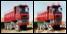}
        \end{subfigure}
        \caption{Successful \textit{LF} attacks on inputs for which \textit{AA}-$\ell_{\infty}$ fails. Left - original image. Right - \textit{LF}.}
\end{figure}

\subsection{Attack Samples Generated Against \textbf{\textit{CIFAR-10}}'s WideResNet-70-16 with \textit{DiffPure} Purification}
\label{app:F3}
\begin{figure}[H]
        \centering
        \begin{subfigure}[b]{0.32\textwidth}
                \includegraphics[width=\linewidth]{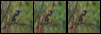}
        \end{subfigure}%
        \hspace{\fill}
        \begin{subfigure}[b]{0.32\textwidth}
                \includegraphics[width=\linewidth]{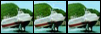}
        \end{subfigure}%
        \hspace{\fill}
        \begin{subfigure}[b]{0.32\textwidth}
                \includegraphics[width=\linewidth]{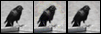}
        \end{subfigure}%
        \hspace{\fill}
        \caption{Successful attacks generated by \textit{LF} and \textit{AA}-$\ell_{\infty}$. Left -original image. Middle - \textit{AA}. Right - \textit{LF}.}\vspace{3mm}
        \begin{subfigure}[b]{0.32\textwidth}
                \includegraphics[width=\linewidth]{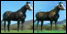}
        \end{subfigure}
        \hspace{\fill}
        \begin{subfigure}[b]{0.32\textwidth}
                \includegraphics[width=\linewidth]{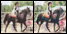}
        \end{subfigure}
        \hspace{\fill}
        \begin{subfigure}[b]{0.32\textwidth}
                \includegraphics[width=\linewidth]{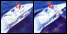}
        \end{subfigure}
        \caption{Successful \textit{LF} attacks on inputs for which \textit{AA}-$\ell_{\infty}$ fails. Left - original image. Right - \textit{LF}.}
\end{figure}

\subsection{Attack Samples Generated Against \textbf{\textit{CIFAR-10}}'s WideResNet-28-10 with \textit{DiffPure} Purification}
\label{app:F4}
\begin{figure}[H]
        \centering
        \begin{subfigure}[b]{0.32\textwidth}
                \includegraphics[width=\linewidth]{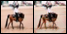}
        \end{subfigure}%
        \hspace{\fill}
        \begin{subfigure}[b]{0.32\textwidth}
                \includegraphics[width=\linewidth]{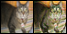}
        \end{subfigure}%
        \hspace{\fill}
        \begin{subfigure}[b]{0.32\textwidth}
                \includegraphics[width=\linewidth]{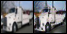}
        \end{subfigure}%
        \hspace{\fill}
        \begin{subfigure}[b]{0.32\textwidth}
                \includegraphics[width=\linewidth]{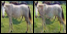}
        \end{subfigure}
        \hspace{\fill}
        \begin{subfigure}[b]{0.32\textwidth}
                \includegraphics[width=\linewidth]{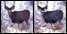}
        \end{subfigure}
        \hspace{\fill}
        \begin{subfigure}[b]{0.32\textwidth}
                \includegraphics[width=\linewidth]{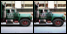}
        \end{subfigure}
        \caption{Successful attacks generated with \textit{LF}. Left -original image. Right - \textit{LF}.}
\end{figure}

\subsection{Attack Samples Generated Against \textbf{\textit{ImageNet}}'s DeiT-S with \textit{GDMP} Purification}
\label{app:F5}
\begin{figure}[H]
        \centering
        \begin{subfigure}[b]{0.9\textwidth}
                \includegraphics[width=\linewidth,height=3cm]{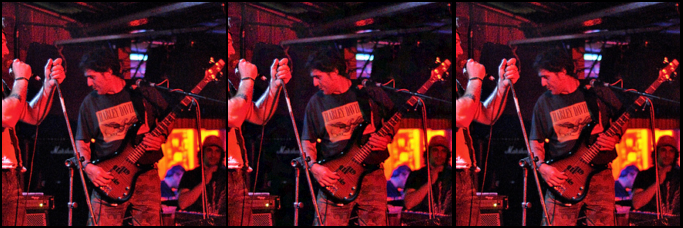}
        \end{subfigure}%
        \hspace{\fill}
        \begin{subfigure}[b]{0.9\textwidth}
        \includegraphics[width=\linewidth,height=3cm]{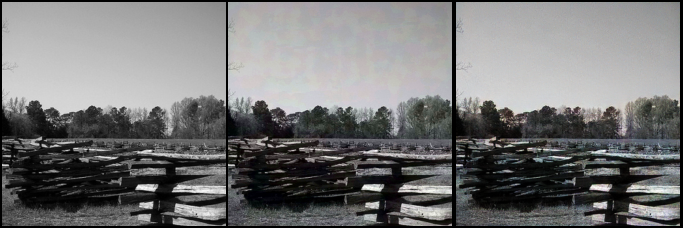}
        \end{subfigure}%
        \hspace{\fill}
        \begin{subfigure}[b]{0.9\textwidth}
                \includegraphics[width=\linewidth,height=3cm]{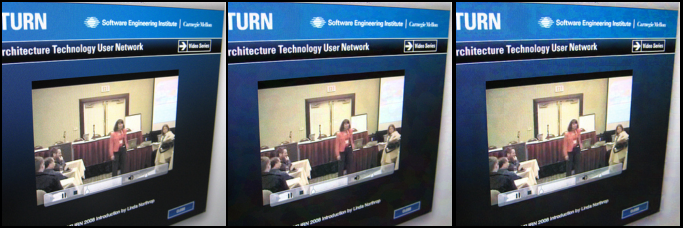}
        \end{subfigure}
        \caption{Successful attacks generated by \textit{LF} and \textit{AA}-$\ell_{\infty}$. Left -original image. Middle - \textit{AA}. Right - \textit{LF}.}
\end{figure}
\begin{figure}[H]
        \centering
        \begin{subfigure}[b]{0.66\textwidth}
                \includegraphics[width=\linewidth,height=3cm]{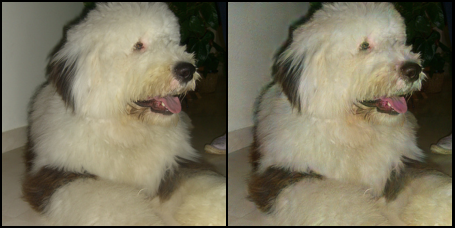}
        \end{subfigure}%
        \hspace{\fill}
        \begin{subfigure}[b]{0.66\textwidth}
                \includegraphics[width=\linewidth,height=3cm]{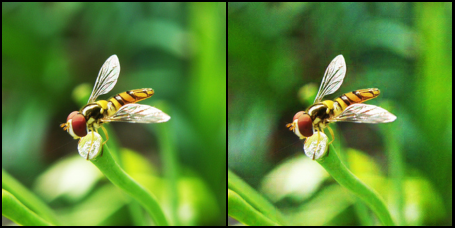}
        \end{subfigure}%
        \hspace{\fill}
        \begin{subfigure}[b]{0.66\textwidth}
                \includegraphics[width=\linewidth,height=3cm]{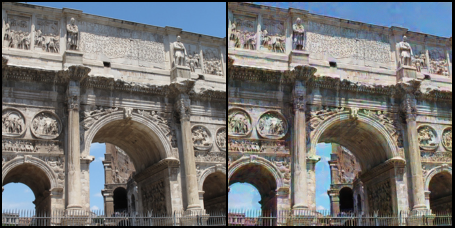}
        \end{subfigure}
        \caption{Successful \textit{LF} attacks on inputs for which \textit{AA}-$\ell_{\infty}$ fails. Left - original image. Right - \textit{LF}.}
\end{figure}

\subsection{Attack Samples Generated Against \textbf{\textit{ImageNet}}'s WideResNet-50-2 with \textit{GDMP} Purification}
\label{app:F6}
\begin{figure}[H]
        \centering
        \begin{subfigure}[b]{0.48\textwidth}
                \includegraphics[width=\linewidth,height=3cm]{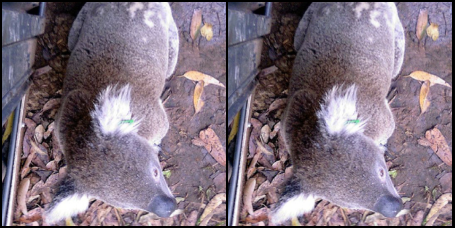}
        \end{subfigure}%
        \hspace{\fill}
        \begin{subfigure}[b]{0.48\textwidth}
                \includegraphics[width=\linewidth,height=3cm]{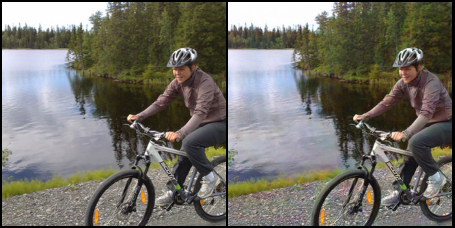}
        \end{subfigure}%
        \hspace{\fill}
        \begin{subfigure}[b]{0.48\textwidth}
                \includegraphics[width=\linewidth,height=3cm]{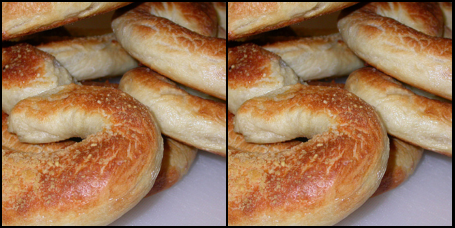}
        \end{subfigure}
        \hspace{\fill}
        \begin{subfigure}[b]{0.48\textwidth}
                \includegraphics[width=\linewidth,height=3cm]{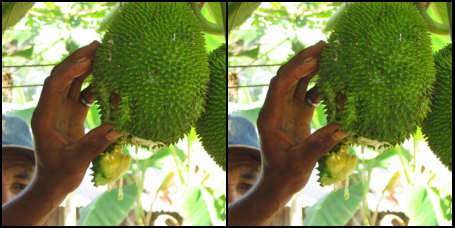}
        \end{subfigure}
        \hspace{\fill}
        \begin{subfigure}[b]{0.48\textwidth}
                \includegraphics[width=\linewidth,height=3cm]{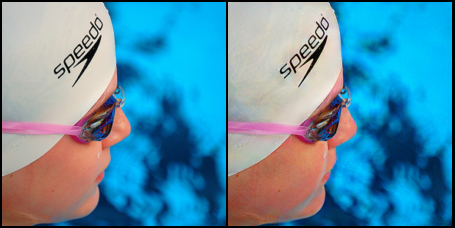}
        \end{subfigure}
        \hspace{\fill}
        \begin{subfigure}[b]{0.48\textwidth}
                \includegraphics[width=\linewidth,height=3cm]{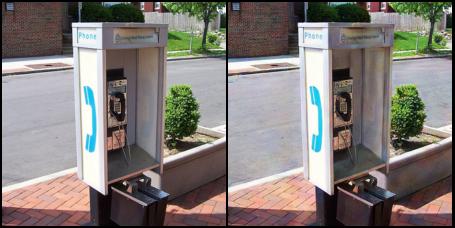}
        \end{subfigure}
        \caption{Successful attacks generated with \textit{LF}. Left -original image. Right - \textit{LF}.}
\end{figure}

\subsection{Attack Samples Generated Against \textbf{\textit{ImageNet}}'s ResNet-50 with \textit{GDMP} Purification}
\label{app:F7}
\begin{figure}[H]
        \centering
        \begin{subfigure}[b]{0.48\textwidth}
                \includegraphics[width=\linewidth,height=3cm]{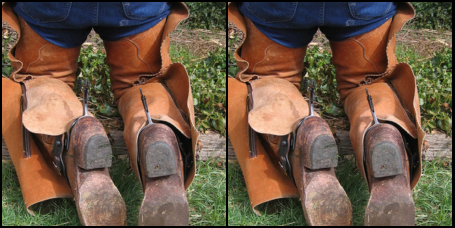}
        \end{subfigure}%
        \hspace{\fill}
        \begin{subfigure}[b]{0.48\textwidth}
                \includegraphics[width=\linewidth,height=3cm]{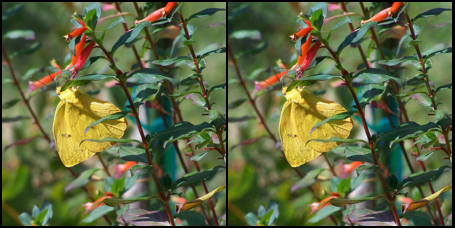}
        \end{subfigure}%
        \hspace{\fill}
        \begin{subfigure}[b]{0.48\textwidth}
                \includegraphics[width=\linewidth,height=3cm]{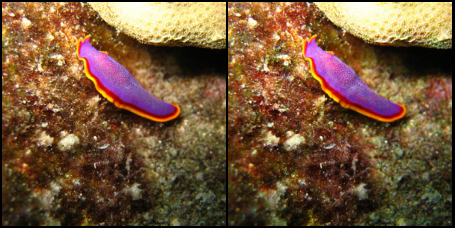}
        \end{subfigure}
        \hspace{\fill}
        \begin{subfigure}[b]{0.48\textwidth}
                \includegraphics[width=\linewidth,height=3cm]{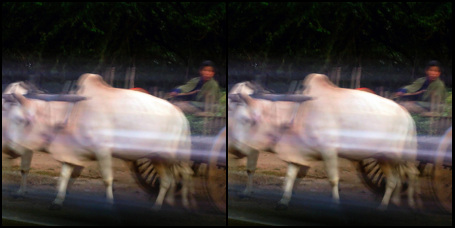}
        \end{subfigure}
        \hspace{\fill}
        \begin{subfigure}[b]{0.48\textwidth}
                \includegraphics[width=\linewidth,height=3cm]{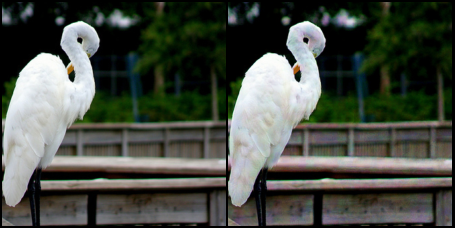}
        \end{subfigure}
        \hspace{\fill}
        \begin{subfigure}[b]{0.48\textwidth}
                \includegraphics[width=\linewidth,height=3cm]{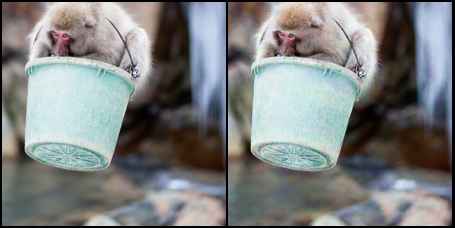}
        \end{subfigure}
        \caption{Successful attacks generated with \textit{LF}. Left -original image. Right - \textit{LF}.}
\end{figure}

\subsection{Attack Samples Generated Against \textbf{\textit{ImageNet}}'s DeiT-S with \textit{DiffPure} Purification}
\label{app:F8}
\begin{figure}[H]
        \centering
        \begin{subfigure}[b]{0.9\textwidth}
                \includegraphics[width=\linewidth,height=3cm]{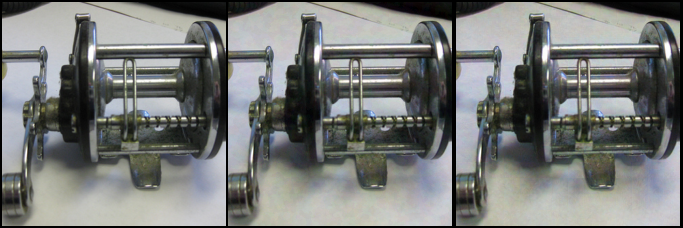}
        \end{subfigure}%
        \hspace{\fill}
        \begin{subfigure}[b]{0.9\textwidth}
                \includegraphics[width=\linewidth,height=3cm]{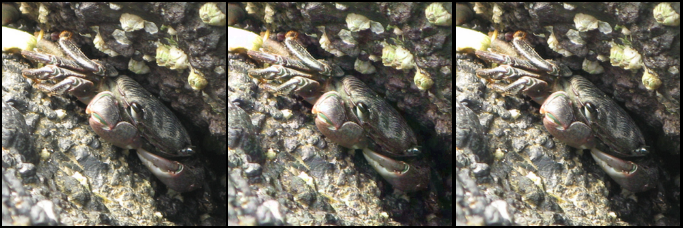}
        \end{subfigure}%
        \hspace{\fill}
        \begin{subfigure}[b]{0.9\textwidth}
                \includegraphics[width=\linewidth,height=3cm]{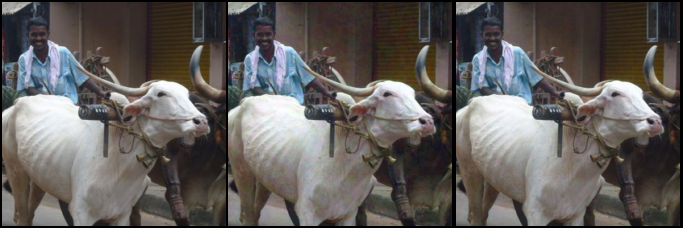}
        \end{subfigure}
        \caption{Successful attacks generated by \textit{LF} and \textit{AA}-$\ell_{\infty}$. Left -original image. Middle - \textit{AA}. Right - \textit{LF}.}
\end{figure}
\begin{figure}[H]
        \centering
        \begin{subfigure}[b]{0.66\textwidth}
                \includegraphics[width=\linewidth,height=3cm]{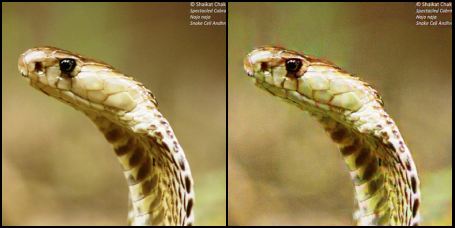}
        \end{subfigure}%
        \hspace{\fill}
        \begin{subfigure}[b]{0.66\textwidth}
                \includegraphics[width=\linewidth,height=3cm]{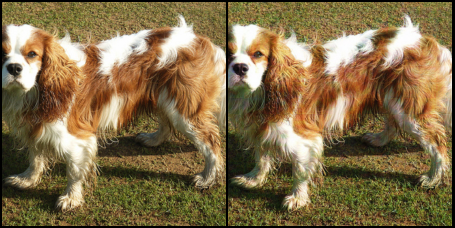}
        \end{subfigure}%
        \hspace{\fill}
        \begin{subfigure}[b]{0.66\textwidth}
                \includegraphics[width=\linewidth,height=3cm]{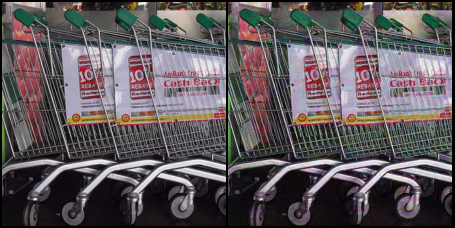}
        \end{subfigure}%
        \caption{Successful \textit{LF} attacks on inputs for which \textit{AA}-$\ell_{\infty}$ fails. Left - original image. Right - \textit{LF}.}
\end{figure}

\subsection{Attack Samples Generated Against \textbf{\textit{ImageNet}}'s WideResNet-50-2 with \textit{DiffPure} Purification}
\label{app:F9}
\begin{figure}[H]
        \centering
        \begin{subfigure}[b]{0.9\textwidth}
                \includegraphics[width=\linewidth,height=3.2cm]{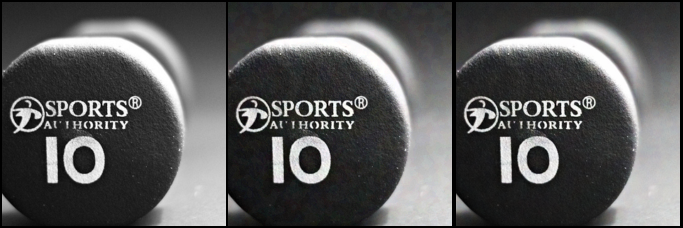}
        \end{subfigure}%
        \hspace{\fill}
        \begin{subfigure}[b]{0.9\textwidth}
                \includegraphics[width=\linewidth,height=3.2cm]{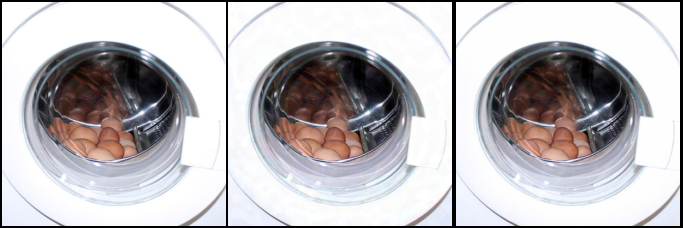}
        \end{subfigure}%
        \hspace{\fill}
        \begin{subfigure}[b]{0.9\textwidth}
                \includegraphics[width=\linewidth,height=3.2cm]{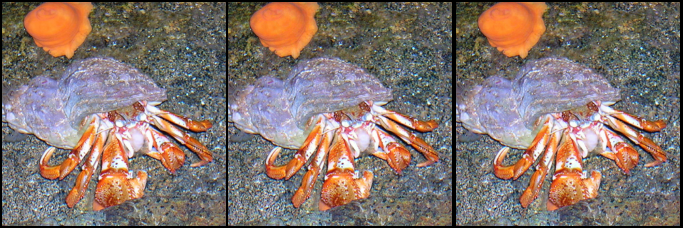}
        \end{subfigure}
        \caption{Successful attacks generated by \textit{LF} and \textit{AA}-$\ell_{\infty}$. Left -original image. Middle - \textit{AA}. Right - \textit{LF}.}
\end{figure}
\begin{figure}[H]
        \centering
        \begin{subfigure}[b]{0.66\textwidth}
                \includegraphics[width=\linewidth,height=3cm]{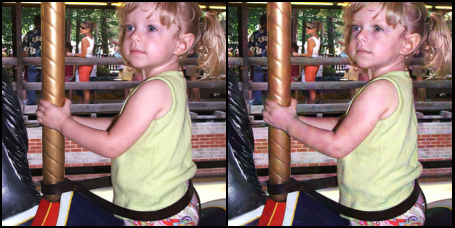}
        \end{subfigure}%
        \hspace{\fill}
        \begin{subfigure}[b]{0.66\textwidth}
                \includegraphics[width=\linewidth,height=3cm]{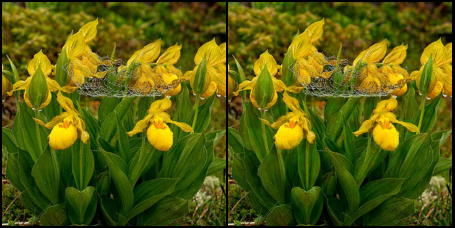}
        \end{subfigure}%
        \hspace{\fill}
        \begin{subfigure}[b]{0.66\textwidth}
                \includegraphics[width=\linewidth,height=3cm]{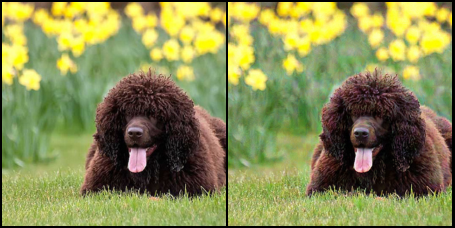}
        \end{subfigure}%
        \caption{Successful \textit{LF} attacks on inputs for which \textit{AA}-$\ell_{\infty}$ fails. Left - original image. Right - \textit{LF}.}
\end{figure}

\subsection{Attack Samples Generated Against \textbf{\textit{ImageNet}}'s ResNet-50 with \textit{DiffPure} Purification}
\label{app:F10}
\begin{figure}[H]
        \centering
        \begin{subfigure}[b]{0.48\textwidth}
                \includegraphics[width=\linewidth,height=3cm]{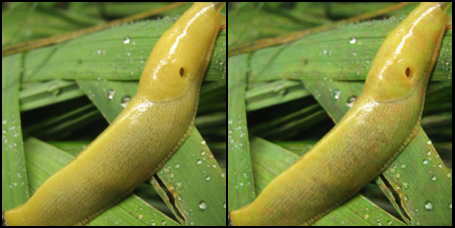}
        \end{subfigure}%
        \hspace{\fill}
        \begin{subfigure}[b]{0.48\textwidth}
                \includegraphics[width=\linewidth,height=3cm]{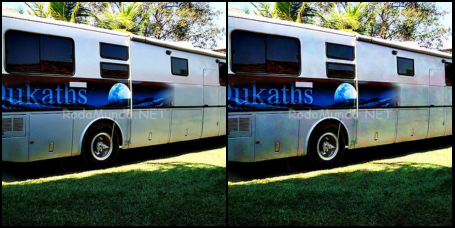}
        \end{subfigure}%
        \hspace{\fill}
        \begin{subfigure}[b]{0.48\textwidth}
                \includegraphics[width=\linewidth,height=3cm]{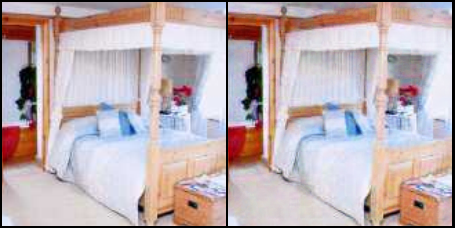}
        \end{subfigure}
        \hspace{\fill}
        \begin{subfigure}[b]{0.48\textwidth}
                \includegraphics[width=\linewidth,height=3cm]{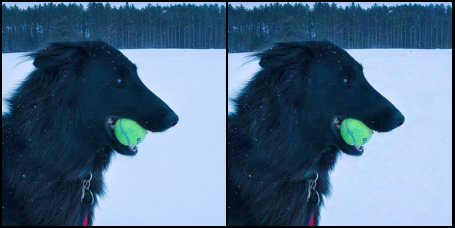}
        \end{subfigure}
        \hspace{\fill}
        \begin{subfigure}[b]{0.48\textwidth}
                \includegraphics[width=\linewidth,height=3cm]{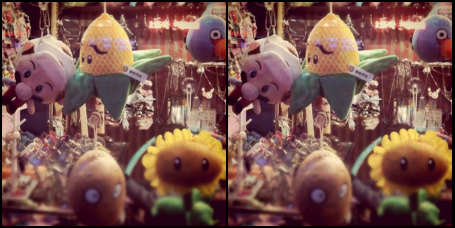}
        \end{subfigure}
        \hspace{\fill}
        \begin{subfigure}[b]{0.48\textwidth}
                \includegraphics[width=\linewidth,height=3cm]{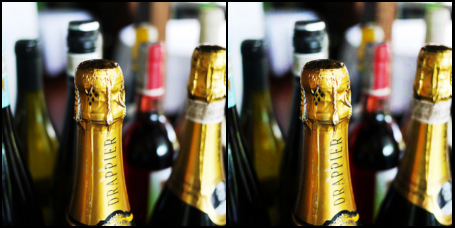}
        \end{subfigure}
        \caption{Successful attacks generated with \textit{LF}. Left -original image. Right - \textit{LF}.}
\end{figure}

\subsection{Quality Comparison with \textit{StAdv}}
\label{app:F11}
We found \textit{StAdv} capable of generating outputs that defeat \textit{DBP} even under \textit{MV}. However, it is not suitable for targeting \textit{DBP}-defended classifiers that operate on high-resolution images. The reason is that \textit{StAdv} performs spatial transformations that relocate the different pixels. Thus, its changes quickly become visible when applied excessively. Due to the considerable stochasticity of \textit{DBP} (see~\S\ref{subsec:exp_mv}), the required displacements (especially in the \textit{MV} setting) are significant, which in turn can severely impact the quality. For low-resolution inputs (e.g., \textit{\textbf{CIFAR-10}}), \textit{StAdv} can still be effective, with the quality degradation remaining unnoticeable due to the size of the images that renders them blurry by default, masking \textit{StAdv}'s effects. For high-resolution inputs, the degradation is substantial, leaving the outputs useless as stealthiness is a key requirement from practical \textit{AE}s~\cite{unmarker}. \textit{StAdv}'s successfully misclassified samples (under \textit{MV}) below prove these claims. We use \textbf{Full-}\sysname{} for backpropagation and run \textit{StAdv} with its default parameters~\cite{stadv}. When the parameters are changed to better retain quality, \textit{StAdv} ceases to converge for \textit{\textbf{ImageNet}}, making it of no use. All provided samples are originally correctly classified.

\begin{figure}[H]
        \centering
        \begin{subfigure}[b]{0.32\textwidth}
                \includegraphics[width=\linewidth]{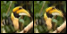}
        \end{subfigure}%
        \hspace{\fill}
        \begin{subfigure}[b]{0.32\textwidth}
                \includegraphics[width=\linewidth]{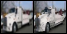}
        \end{subfigure}%
        \hspace{\fill}
        \begin{subfigure}[b]{0.32\textwidth}
                \includegraphics[width=\linewidth]{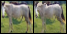}
        \end{subfigure}%
        \hspace{\fill}
        \begin{subfigure}[b]{0.32\textwidth}
                \includegraphics[width=\linewidth]{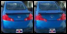}
        \end{subfigure}
        \hspace{\fill}
        \begin{subfigure}[b]{0.32\textwidth}
                \includegraphics[width=\linewidth]{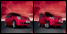}
        \end{subfigure}
        \hspace{\fill}
        \begin{subfigure}[b]{0.32\textwidth}
                \includegraphics[width=\linewidth]{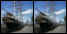}
        \end{subfigure}
        \caption{\textit{StAdv} attacks against \textbf{\textit{CIFAR-10}}'s WideResNet-70-16 with \textit{GDMP} purification. Left - original image. Right - \textit{StAdv}.}
\end{figure}
\begin{figure}[H]
        \centering
        \begin{subfigure}[b]{0.48\textwidth}
                \includegraphics[width=\linewidth,height=3cm]{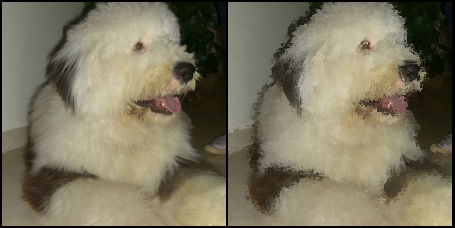}
        \end{subfigure}%
        \hspace{\fill}
        \begin{subfigure}[b]{0.48\textwidth}
                \includegraphics[width=\linewidth,height=3cm]{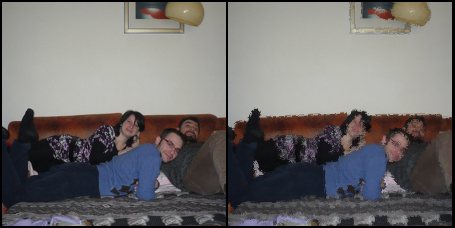}
        \end{subfigure}%
        \hspace{\fill}
        \begin{subfigure}[b]{0.48\textwidth}
                \includegraphics[width=\linewidth,height=3cm]{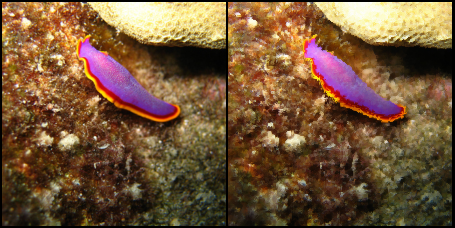}
        \end{subfigure}
        \hspace{\fill}
        \begin{subfigure}[b]{0.48\textwidth}
                \includegraphics[width=\linewidth,height=3cm]{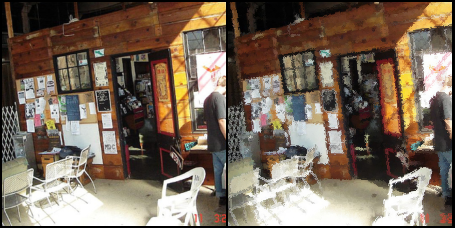}
        \end{subfigure}
        \hspace{\fill}
        \begin{subfigure}[b]{0.48\textwidth}
                \includegraphics[width=\linewidth,height=3cm]{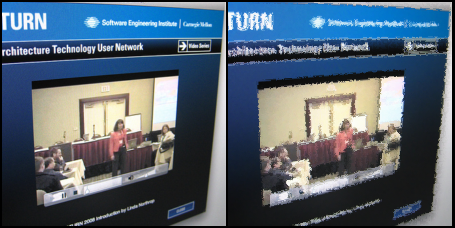}
        \end{subfigure}
        \hspace{\fill}
        \begin{subfigure}[b]{0.48\textwidth}
                \includegraphics[width=\linewidth,height=3cm]{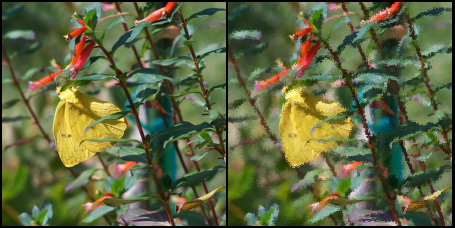}
        \end{subfigure}
        \caption{\textit{StAdv} attacks against \textbf{\textit{ImageNet}}'s DeiT-S with \textit{DiffPure} purification. Left - original image. Right - \textit{StAdv}.}
\end{figure}

\end{document}